\documentclass{jpp}
\usepackage{epsfig}

\usepackage{graphicx}
\usepackage{epstopdf, epsfig}

\usepackage{amsmath}
\usepackage{amssymb}
\usepackage{mathabx}

\usepackage{natbib}
\usepackage{color}
\usepackage{enumitem} 
\usepackage{tabu,multirow}

\usepackage{color}

\newcommand\beq{\begin{equation}}
\newcommand\eeq{\end{equation}}

\newcommand{\ben}{\begin{eqnarray}}
\newcommand{\een}{\end{eqnarray}}
\newcommand{\benn}{\begin{eqnarray*}}
\newcommand{\eenn}{\end{eqnarray*}}

\newcommand{\apar}{ A_{\parallel}}

\newcommand{\pa}{\partial}

\newcommand{\lapp}{\Delta_\perp}

\newcommand{\gpar}{\nabla_{\parallel}}

\allowdisplaybreaks

\title{Imbalanced kinetic Alfv\'en wave turbulence: from weak turbulence theory to nonlinear diffusion models for the strong regime}

\author{T. Passot,  P.L. Sulem}

\affiliation{Universit\'e C\^ote d'Azur, Observatoire de la C\^ote d'Azur,  CNRS, Laboratoire J.L. Lagrange, Boulevard de l'Observatoire, CS  34229, 06304 Nice Cedex 4, France}
\begin{document}

\maketitle

\begin{abstract}
A two-field Hamiltonian gyrofluid model for kinetic Alfv\'en waves retaining ion finite Larmor radius corrections, parallel magnetic field fluctuations and electron inertia,  is used to study turbulent cascades from the MHD to the sub-ion scales. Special attention is paid to  the case of imbalance between waves propagating along or opposite to the ambient magnetic field. For  weak turbulence in the absence of electron inertia,  kinetic equations for the spectral density of the conserved quantities (total energy and generalized cross-helicity) are obtained. They provide a global description, matching between the regimes of reduced MHD at large scales and electron reduced MHD  at small scales, previously considered  in the literature. In the limit of ultra-local interactions, Leith-type nonlinear diffusion equations in the Fourier space are derived and heuristically extended to the strong turbulence regime by  modifying the transfer time appropriately. Relations with existing phenomenological models for imbalanced MHD and balanced sub-ion turbulence are discussed. \textcolor{black}{It turns out that in the presence of dispersive effects, the dynamics is sensitive on the way turbulence is maintained in a steady state. Furthermore, the total energy spectrum  at sub-ion scales becomes steeper as the generalized cross-helicity flux is increased.}
\end{abstract}


\section{Introduction}
In addition to its intrinsic astrophysical interest and its relevance  for space weather, solar wind is  often viewed,  thanks to the high-quality of in situ data obtained by Earth orbiting spacescraft, as a natural laboratory for collisionless plasma turbulence \citep{Bruno13, BC16}.
\textcolor{black}{It appears that Alfv\'en waves play a dominant role both at large scales \citep{Belcher71} and at sub-ion scales, where they are referred to kinetic Alfv\'en waves \citep{SGRK09,Alexandrova09,Salem12, Podesta13}.} Turbulence mostly results from the nonlinear interactions between counter-propagating Alfv\'en waves. Since the waves are   emitted at the Sun's surface, reflection on density gradients is often invoked for  generating backward-propagating waves \citep{Perez13}. Furthermore, solar-wind Alfvenic turbulence is  usually in an imbalanced regime characterized by an excess of energy carried by outward propagating waves \citep{Tu89,Lucek98,Wicks13}. The degree of imbalance is  greatly dependent on the type of wind \citep{Tu90} and also on the distance from the Sun \citep{Roberts87,Marsch-Tu90}. 

Imbalanced turbulence was the object of theoretical and numerical  studies, mostly at MHD scales where the flow is almost incompressible, with negligible dispersive and kinetic effects. 
In the context of strong turbulence, various phenomenological models have been proposed \citep{Lithwick07, Chandran08, Beresnyak08, Perez-Boldyrev09, Podesta10}, but at this date no real consensus has been reached (see \cite{Chen16} for review).  For example, while \citet{Lithwick03} predict $-5/3$ power laws for the energy spectra of both outer and inner propagating waves, the model of \citet{Chandran08} only prescribes an entanglement relation for the spectral exponents. These two models also lead to very different scalings for the ratio of  the transfer rates with  the outer-scale energy ratio. An intermediate phenomenology by \citet{Beresnyak08}  seems, on this point, in better agreement with numerical simulations \citep{Beresnyak09}. \textcolor{black}{Nevertheless,  significant imbalance makes the  numerics delicate, in part because it reduces the strength of the nonlinear interactions and thus requires both long integration times and high spatial resolution}. Using reduced MHD (RMHD) equations with regular (rather than hyper) viscosity and diffusivity, \citet{Perez-Boldyrev09} found, independently of the level of the velocity magnetic field correlation, a $-3/2$ power law for both types of waves. This exponent is interpreted by taking into account a dynamical alignment between the velocity and magnetic field, \textcolor{black}{an effect which depletes the strength of the nonlinear coupling. This correction however leads to a theoretical prediction for the variation of the ratio of transfer rates with the ratio of the energies at a given scale which is not supported by the numerics \citep{Beresnyak10}. Even in the absence of imbalance, a regime where  numerical simulations at higher resolutions have recently been performed, the MHD spectral exponents are in fact still debated, simulations by  \citet{Perez12} showing a $-3/2$ energy spectrum, and those of \citet{Beresnyak14} an exponent close to $-5/3$. At the level of theoretical analysis, \citet{Mallet17}, retaining the  disruption by magnetic reconnection of the sheet-like structures formed by dynamically aligned Alfvenic turbulence, recently predicts a steepening at small scales of a -3/2 large-scale spectrum}.  Furthermore, transition from MHD to sub-ion scales in imbalanced turbulence was addressed by \citet{Voitenko16}, assuming that the energy fluxes associated to parallel and anti-parallel propagating waves are   scale-independent,  in spite of wave dispersion.  

The above contributions concern the strong turbulent regime for which no systematic theory is available. Differently, in the case of weak turbulence, kinetic equations can be systematically derived for the densities of the quadratic invariants. Weak Alfvenic turbulence was studied in the framework of incompressible MHD \citep{Ng96,GS97,Galtier02} and later on in the context of Hall-MHD \citep{Galtier06}. Weak turbulence at the sub-ion scales was addressed using electron-MHD (EMHD) by \citet{Galtier03,Lyutikov13} and \citet{Galtier15}. Kinetic equations for imbalanced weak Alfv\'en waves turbulence in the MHD range are analyzed and simulated in \citet{Lithwick03}. 

The aim of the present paper is to provide a framework for a uniform description of imbalanced Alfv\'en-wave turbulence from the MHD  to the sub-ion scales, neglecting the coupling to other types of waves. For this purpose, a two-field gyrofluid derived in \citet{PST18} as a reduction of a more general model of \citet{Bri92} is used. It involves a strong spectral anisotropy, consistent with solar wind observations \citep{MacBride08,Sahraoui10}. This reduced model, which displays an Hamiltonian structure, retains ion finite Larmor radius (FLR) corrections, parallel magnetic field fluctuations and electron inertia. It describes Alfv\'en waves and dispersive Alfv\'en waves (DAWS) at MHD scales, kinetic Alfv\'en waves (KAWs) at sub-ion scales and also inertial kinetic Alfv\'en waves (IKAWs) at scales smaller than the electron skin depth $d_e$ \citep{PST17,Chen-Bold17}. This last regime was recently observed in numerical simulations \citep{Royter18}.

Imbalanced Alfvenic turbulence, as described by the two-field gyrofluid, is amenable to a systematic approach  in the weak-turbulence regime. In addition to their own interest, weak turbulence kinetic equations can be used as a starting point for a rigorous derivation, \textcolor{black}{under the assumption of strong spectral locality}, of nonlinear diffusion equations in Fourier space. This model can be heuristically extended to strong turbulence, as the cascade phenomenology of the two regimes  mostly differs by the characteristic transfer time. Special attention is payed to the condition for the existence, in the imbalanced regime, of stationary cascades resulting from a large-scale driving. In the framework of weak MHD turbulence, a complete solution of the problem cannot be found \textcolor{black}{without prescribing additional conditions} \citep{Lithwick07}. It turns out that in the presence of diffusive effects, the phenomenon of pinning of the  spectra at the dissipation scales, first reported by \citet{Grappin83} in the framework of EDQNM closure for strong imbalanced turbulence, permits the determination of both the energy fluxes and the spectral indices which are found to be non-universal. The pinning effect was later observed in numerical simulations of kinetic equations for weak visco-diffusive MHD turbulence \citep{Lithwick03} and in direct integration of  three-dimensional diffusive MHD and RMHD equations \citep{Perez12}. An interesting question is whether dissipation and pinning still play a crucial role in the presence of dispersive effects. In this context, it is worth mentioning that pinning was recently observed in direct simulations of three-dimensional EMHD displaying a direct energy cascade and an inverse cascade of magnetic helicity \citep{Cho16}.

The paper is organized as follows. In Section \ref{Gyrofluid}, the two-field gyrofluid model is presented, together with the quadratic invariants (energy and generalized cross helicity). The model is also written in terms of the linear eigenmodes, which conveniently leads to dynamical equations for the Alfv\'en-wave amplitudes. In Section \ref{weak_turbulence}, the weak turbulence formalism is implemented, leading to kinetic equations for the spectral density tensor of the Alfv\'en modes. \textcolor{black}{The equations for the energy and generalized cross-helicity spectra} are specified in the case of negligible electron inertia in Section \ref{kin-eq-zeroinertia}. In Section \ref{local_model}, a simplified model is derived, based on the assumption of strongly-local interactions. It is governed by nonlinear diffusion equations in the spectral space, in the spirit of Leith's phenomenological model for hydrodynamic turbulence and \citet{Zhou90} model  for MHD flows. Section \ref{strong_turbulence} provides a phenomenological extension of this model to the  strong turbulence regime, by heuristically adjusting the transfer time. Section \ref{transverse-spectra} is devoted to the study of the energy and generalized cross helicity cascades in weak and strong turbulence, obtained as stationary solutions of the corresponding diffusion models.  Section \ref{conclusion} is the Conclusion.
 
\section{A two-field gyrofluid model} \label{Gyrofluid}
 The plasma dynamics results from perturbations of a homogeneous equilibrium state
 characterized by a  density $n_0$, isotropic ion and electron temperatures $T_{0i}$ 
 and $T_{0e}$, and subject to a strong ambient magnetic field of amplitude $B_0$
 along the $z$-direction. The various characteristic  scales are conveniently measured in terms of the  
 sonic Larmor radius $\rho_s  = c_s/\Omega_i$, where $c_s = \sqrt {T_{0e}/m_i}$ is the sound speed
 and $\Omega_i= eB_0/(mc)$ the ion gyrofrequency, in the form
 \begin{equation}
 d_i = \sqrt{\frac{2}{\beta_e}} \rho_s , \quad
 d_e= \sqrt{\frac{2}{\beta_e}}\delta \rho_s , \quad
 \rho_i = \sqrt{2\tau} \rho_s , \qquad  \rho_e= \sqrt{2} \delta \rho_s,
 \end{equation}
 where $\beta_e= 8 \pi n_0 T_{0e}/B_0^2$, $\delta^2= m_e/m_i$ is
 the electron to ion mass ratio and $\tau = T_{0i}/T_{0e}$. We have here defined the particle Larmor
 radii ($r=i$ for the ions, $r=e$ for the electrons) by $\rho_r =v_{th\,r}/\Omega_r$ where the particle thermal velocities are given by 
 $v_{th \, r}=(2T_r/m_r)^{1/2}$ and the inertial lengths by $d_r=v_A/\Omega_r$ where $v_A=B_0/(4\pi n_0 m_i)^{1/2}
 =c_s \sqrt{2/\beta_e}$ is the Alfv\'en velocity.

 The two-field gyrokinetic model derived in \cite{PST18} as a restriction of the  model of \citet{Bri92}, isolates the  Alfv\'en wave dynamics. \textcolor{black}{By construction, gyrofluids involve the gyrokinetic scaling (see e.g. \citet{HCD06}). Assuming a strong ambient  field, it prescribes an anisotropic dynamics with transverse scales much smaller than the parallel ones. As only low frequencies compared to the ion gyrofrequency are retained, fast magnetosonic waves average out. Furthermore, in the two-fluid model,  slow waves are decoupled as a result of the prescribed asymptotics which assumes small $\beta_e$. For larger $\beta_e$, Landau damping can act more efficiently on slow rather than on Alfv\'en waves, justifying a two-field description. Similar reductions were presented both in the context of compressible Hall-MHD \citep{Bian09} and at sub-ion scales \citep{BHXP13}.}
In addition to ion FLR corrections and parallel magnetic fluctuations, \textcolor{black}{the model} retains electron inertia as well as an electron FLR contribution which becomes relevant when the ion-electron temperature ratio $\tau$ is comparable to or larger than the inverse electron beta  $1/\beta_e$. This model, which displays a Hamiltonian structure, covers a spectral
 range extending from the MHD scales (large compared to the ion inertial length
 $d_i$)  to scales comparable to the electron skin depth $d_e$.  The considered scales are nevertheless taken large compared to the electron Larmor radius $\rho_e$ in order to prevent the full FLR electron corrections to be relevant. This requires that  $\rho_e/d_e=\beta_e^{1/2}$ be small enough, a regime where Landau damping can efficiently  homogenize electron temperatures along the magnetic field lines. The present model thus assumes isothermal electrons, which is a good approximation when neglecting dissipation phenomena \citep{TSP16,Sulem2016}.

 \subsection{Formulation of the model}
The model \textcolor{black}{(which retains no dissipation process)} is written as equations for the electron gyrocenter number density $N_e$ and the parallel component of the magnetic potential $A_\|$, in the form 
\begin{eqnarray}
&&\partial_t N_e +[\varphi,N_e]-[B_z,N_e]+\frac{2}{\beta_e}\nabla_\| \Delta_\perp A_\|=0\label{eq:gyro-2fields-Ne} \label{eq:Ne}\\
  &&\partial_t (1-\frac{2\delta^2}{\beta_e}\Delta_\perp)A_\| -[\varphi,\frac{2\delta^2}{\beta_e}\Delta_\perp A_\|]
  +[B_z,\frac{2\delta^2}{\beta_e}\Delta_\perp A_\|] 
 + \nabla_\| (\varphi-N_e-B_z)=0\label{eq:gyro-2fields-A},\label{eq:A}
\end{eqnarray}
with the parallel magnetic fluctuations $B_z$ and the electrostatic potential $\varphi$ given by 
\begin{eqnarray}
&&\left (\frac{2}{\beta_e}  +(1+2\tau)(\Gamma_0 - \Gamma_1) \right ) B_z=
  \left ( 1 -(\frac{\Gamma_0-1}{\tau}) -\Gamma_0
  +\Gamma_1  \right)\varphi  \label{gyro:Bzphi}\\
&&N_e=\left ( (\frac{\Gamma_0-1}{\tau}) +\delta^2\Delta_\perp\right )\varphi
-(1-\Gamma_0+\Gamma_1) B_z.\label{eq:gyro-Ne-phi}
\end{eqnarray}

Here, \textcolor{black}{$\Delta_\perp = \partial_{xx} + \partial_{yy}$ is the Laplacian in the plane transverse to the ambient field and  $[f,g]= \partial_x f \partial_y g-\partial_y f \partial_x g$}  the canonical bracket of two scalar functions $f$ and $g$.
Furthermore, $\Gamma_n$ denotes the (non-local) operator 
$\Gamma_n(-\tau \Delta_\perp)$ associated with  the Fourier multiplier  $\Gamma_n(\tau k_\perp^2)$, defined by  $\Gamma_n(x) = I_n(x) e^{-x}$ where $I_n$ is the modified Bessel function of first type of order n. For a scalar function $f$, the parallel gradient operator $\gpar$ is defined by
\beq
\gpar f=-[\apar , f]+\frac{\pa f}{\pa z}.
\eeq
The equations are written in a nondimensional form, using the following units: $\Omega_i^{-1}$ for time, $\rho_s$ for the space coordinates (and thus 
$\rho_s^{-1}$ for the wavenumber components), the ambient magnetic field $B_0$ for the parallel magnetic fluctuations $B_z$, the equilibrium density
$n_0$ for the electron gyrocenter density $N_e$, $T_e/e$ for the electric potential $\varphi$ and $B_0\rho_s$ for the parallel magnetic potential $A_\|$.
The ion and electron particle number densities $n_i$ and $n_e$ are given by
\begin{equation}
n_i=n_e=N_e+B_z,
\end{equation}
\textcolor{black}{the latter equality being only valid in the absence of electron inertia and FLR contributions.}
 An advantage of the present formulation is that it provides a continuous transition between the MHD and the sub-ion ranges.
 
Introducing the operators 
\begin{eqnarray}
&&L_1 = \frac{2}{\beta_e}  +(1+2\tau)(\Gamma_0-\Gamma_1)\\
&&L_2 = 1 +\frac{1-\Gamma_0}{\tau} - \Gamma_0 +\Gamma_1 \\
&&L_3 = \frac{1-\Gamma_0}{\tau} -\delta^2\Delta_\perp\\
&&L_4 = 1-\Gamma_0+\Gamma_1\\
  &&L_e = 1-\frac{2 \delta^2}{\beta_e} \lapp, 
\end{eqnarray}
one  writes $ B_z = M_1 \varphi$, with $M_1 =  L_1^{-1} L_2 $,
and $N_e = -M_2 \varphi$, where $M_2 = L_3 + L_4L_1^{-1}L_2$ is positive 
definite, as numerically seen on its Fourier transform.
Thus, $B_z$ and $\varphi$  can be expressed in terms of $N_e$.

At the level of the linear approximation, one easily checks 
that the phase velocity $v_{ph}= \omega/k_z$ is given by the dispersion relation 
\begin{equation}
v_{ph}^2\equiv\left(\frac{\omega}{k_z}\right)^2 = 
\frac{2}{\beta_e} \frac{k_\perp^2}{1 + \frac{2\delta^2 k_\perp^2}{\beta_e}} \frac{1 -{\widehat M}_1 + {\widehat M}_2} {{\widehat M}_2},
\end{equation}
\textcolor{black}{where the caret refers to the Fourier symbol of the operator.}
The associated operator $V_{ph}$ is given by 
\begin{equation}
	V_{ph}= s (-\lapp)^{1/2} (1 -M_1 + M_2)^{1/2}M_2^{-1/2}L_e^{-1/2},
\end{equation}
where  $s= (2/\beta_e)^{1/2}$ is the equilibrium Alfv\'en velocity in sound speed units.
In physical space, the  eigenmodes satisfy the conditions
\begin{equation}
\textcolor{black}{\pm V_{ph} M_2 \varphi - s^2 \lapp A_\|= 0}
\end{equation}
or 
\begin{equation}
\Lambda \varphi \pm s A_\| =0,
\end{equation}
where
\begin{equation}
\Lambda = D_e^{-1} (1+M_2-M_1)^{1/2} M_2^{1/2} 
\end{equation}
and $D_e^2 = (-\lapp) L_e$. 
Defining $M_3=1+M_2-M_1$, Eqs. (\ref{eq:Ne})-(\ref{eq:A})  rewrite
\begin{eqnarray}
&&\partial_t M_2\varphi-s^2\partial_z\Delta_\perp A_\| +[M_3\varphi,M_2\varphi]+s^2[A_\|,\Delta_\perp A_\|]=0\\
&&\partial_t L_e A_\| +\partial_z M_3\varphi+[(1-M_1)\varphi,L_e A_\|]-[A_\|,M_2\varphi]=0.
\end{eqnarray}
Applying the operator $\Lambda/M_2$ on the first equation and the operator $sL_e^{-1}$ on the second one,  adding and  subtracting the resulting equations  and introducing the generalized Elsasser potentials (which identify with the linear eigenmodes)
\begin{equation}
\mu^\pm = \Lambda \varphi \pm s A_\| ,
\end{equation}  leads to
\begin{eqnarray}
&&\partial_t \mu^\pm \pm V_{ph}\partial_z \mu^\pm +\frac{\Lambda}{M_2}\left \{[M_3\varphi,M_2\varphi]+s^2[A_\|,\Delta_\perp A_\|]\right\}\nonumber\\
&&\pm s L_e^{-1}\left\{[(1-M_1)\varphi,L_e A_\|]-[A_\|,M_2\varphi]\right\}=0,
\end{eqnarray}
which rewrites,  using 
$\Lambda \varphi = \frac{1}{2}(\mu^+ + \mu^-)$ and $A_\| = \frac{1}{2s}(\mu^+ -\mu^-)$,
\begin{eqnarray}
&&\partial_t \mu^\pm \pm V_{ph} \partial_z \mu^\pm \nonumber \\
&& +\frac{1}{4} \Lambda^{-1} D_e^{-2} M_3\Big \{ [\Lambda^{-1} M_3 (\mu^++\mu^-) , \Lambda^{-1}M_2 (\mu^+ + \mu^-)]
+[(\mu^+-\mu^-), \lapp (\mu^+-\mu^-)] \Big \} \nonumber \\
&&\pm \frac{1}{4} D_e^{-2}\lapp \Big \{
[(\mu^+ -\mu^-), \Lambda^{-1} M_2 (\mu^++\mu^-)]
+ [L_e (\mu^+ -\mu^-), \Lambda^{-1} (1-M_1)(\mu^+ +\mu^-)]
\Big \}=0.\nonumber \\ 
&&\label{Dtmu}
\end{eqnarray}

From Eq. (\ref{Dtmu}), it is easily seen that, \textcolor{black}{in the MHD limit (where $M_2= -\Delta_\perp$ and subdominant terms associated e.g. to the Hall effect are neglected), only counter-propagating waves are interacting, as expected. In this regime, Eq. (\ref{Dtmu}) indeed reduces to the RMHD equations  \citep{Schekochihin09}
\begin{equation}
\partial_t \Delta_\perp\mu^\pm \pm V_{ph} \partial_z \Delta_\perp\mu^\pm + \frac{1}{2}\left \{ [\mu^+, \Delta_\perp\mu^-] + [\mu^-, \Delta_\perp\mu^+]
\mp \Delta_\perp [\mu^+, \mu^-]\right \}=0.
\end{equation}	
}
\textcolor{black}{\textbf{\textit{Remark:}}
In the MHD range, $\mu^\pm =\varphi \pm s A_\|$ where the potentials $\varphi$ and $A_\|$ are related to the transverse ion velocity ${\boldsymbol u}_{\perp i}$ and the transverse magnetic field ${\boldsymbol B}_\perp$ by ${\boldsymbol u}_{\perp i}= {\widehat{\boldsymbol z}} \times \bnabla \varphi$	and ${\boldsymbol B}_\perp= -{\widehat{\boldsymbol z}} \times \bnabla A_\|$, with ${\widehat{\boldsymbol z}}$ denoting the unit vector in the direction of the ambient magnetic field. It follows that the transverse Elsasser variables ${\boldsymbol z}^\pm = {\boldsymbol u}_{\perp i} \pm s {\boldsymbol B}_\perp$ are given by
${\boldsymbol z}^\pm = {\widehat{\boldsymbol z}} \times \bnabla \mu^\mp$. 
}
\textcolor{black}{
\subsection{Limiting forms of the model}}

{\small
	\begin{table*}
		\begin{center}
			\begin{tabu}{|c||c|c|c|c|}
				\hline  & & & &  \\
				&DAWs: 2-field & Reconnection & KAWs: ERMHD & IKAWs \\
				& HRMHD & model & &\\ & & & &\\
				& $\tau k^2_\perp\ll 1$ & $\tau k^2_\perp\ll 1$ & $\tau k^2_\perp\gg 1$ & $\tau k^2_\perp\gg 1$ \\
				& $\beta_e\lesssim 1$ and $\delta=0$ & $\beta_e\ll 1$ and $\delta\ne 0$ & $\beta_e\lesssim 1$ and $\delta=0$ & $\beta_e\ll 1$ and $\delta\ne 0$ \\
				& $\tau\ll 1$ & $\tau\ll 1$ & $\tau \sim  1$ &  $\tau \gg 1$ \\
				& & & &  \\ \hline\hline  & & & &  \\
				${\widehat M_1}$	& $(1+\frac{2}{\beta_e})^{-1}k^2_\perp$ & 0& $\frac{\beta_e}{2}(1+\frac{1}{\tau})$ & $\frac{\beta_e}{2}$ \\
				& & & &  \\\hline& & & &  \\
				${\widehat M_2}$	& $k^2_\perp$ & $k^2_\perp$ & $\frac{\beta_e}{2}(1+\frac{1}{\tau}+\frac{2}{\beta_i})$& $\frac{\beta_e}{2}(1+\frac{2}{\beta_i}+\frac{2\delta^2}{\beta_e}k^2_\perp)$ \\
				& & & &  \\\hline& & & &  \\
				${\widehat \Lambda}$	& $\sqrt{1+\frac{\frac{2}{\beta_e}}{1+\frac{2}{\beta_e}}k^2_\perp}$ & $\sqrt{\frac{1+k^2_\perp}{1+\frac{2\delta^2}{\beta_e}k^2_\perp}}$ & $\frac{1}{k_\perp}\sqrt{\frac{1}{\tau}+\frac{1}{\tau^2}+\frac{\beta_e}{2}(1+\frac{1}{\tau})^2}$  & $\frac{\sqrt{\frac{\beta_e}{2}(1+\delta^2k^2_\perp)(1+\frac{2}{\beta_i}+\frac{2\delta^2k^2_\perp}{\beta_e})}}{k_\perp\sqrt{1+\frac{2\delta^2}{\beta_e}k^2_\perp}}$ \\
				& & & &  \\\hline& & & &  \\
				${\widehat V}_{ph}$	& $s\sqrt{ 1+\frac{\frac{2}{\beta_e}}{1+\frac{2}{\beta_e}}k^2_\perp}$ & $s\sqrt{\frac{1+k^2_\perp}{1+\frac{2\delta^2}{\beta_e}k^2_\perp}}$ & $s k_\perp\sqrt{ \frac{1+\tau}{1+\frac{\beta_e}{2}(1+\tau)}}$& $\frac{s^2k_\perp }{\sqrt{(1+\frac{2}{\beta_i}+\frac{2\delta^2}{\beta_e}k^2_\perp)(1+\frac{2\delta^2}{\beta_e}k^2_\perp))}}$ \\
				& & & &  \\\hline& & & &  \\
				Eqs.	& (\ref{eq:HRMHD1})-(\ref{eq:HRMHD2}) &  (\ref{eq:rec-mod1})-(\ref{eq:rec-mod2})  &  (\ref{eq:ERMHD1})-(\ref{eq:ERMHD2}) &  (\ref{eq:IKAW1})-(\ref{eq:IKAW2}) \\
				& & & &  \\\hline 
			\end{tabu} 
			\caption{Asymptotic forms of the Fourier symbols (indicated by a caret) of the operators ${M_1}$, ${M_2}$, ${\Lambda}$ and ${V}_{ph}$, together with the corresponding equation numbers, in four different regimes.}
		\end{center}
	\end{table*}
}
\textcolor{black}{Equations (\ref{eq:Ne})-(\ref{eq:A}) reduce to several classical systems found in the literature when restricted to large or small scales with respect to the ion gyroradius,  with and without electron inertia.
Table 1 summarizes the asymptotic forms of the various operators in four limiting cases discussed below.
}
\textcolor{black}{	
The equations, whose numbers are given in the table, all admit an Hamiltonian structure and are given by:}
\begin{itemize}
	\item \textcolor{black}{Two-field HRMHD for DAWs}
\textcolor{black}{	\begin{align}
	&\partial_t A_\| + \nabla_\| \left ( \varphi- \frac{2}{\beta_e} \frac{1}{1+\frac{2}{\beta_e}} \Delta_\perp\varphi     \right )=0 \label{eq:HRMHD1}\\
	&\partial_t \Delta_\perp\varphi  + [\varphi,\Delta_\perp \varphi] + \frac{2}{\beta_e} \nabla_\|\Delta_\perp A_\| =0.\label{eq:HRMHD2}
	\end{align}
}
\textcolor{black}{These equations, which extend to three dimensions the model of \citet{Grasso99}, identify with the HRMHD equations (E19)-(E20) of \citet{Schekochihin09} when neglecting the parallel ion velocity $u_i$ and assuming that $B_z$ is "slaved" to the potential $\varphi$. At small $\beta_e$, HRMHD is often further simplified by taking $M_1=0$ (this  supresses the second term in the parenthesis of Eq. (\ref{eq:HRMHD1}) which accounts for the Hall effect), leading to the two-field RMHD equations.}

	\item \textcolor{black}{Two-field reconnection model (\citet{Schep94}; see also Eqs. (143)-(144) of \cite{Tassi17} and  references therein)}
\textcolor{black}{	\begin{align}
	&\partial_t (1-\frac{2\delta^2}{\beta_e}\Delta_\perp) A_\| -[\varphi, \frac{2\delta^2}{\beta_e}\Delta_\perp A_\| ]+ \nabla_\| \left ( \varphi-   \Delta_\perp\varphi     \right )=0 \label{eq:rec-mod1}\\
	&\partial_t \Delta_\perp\varphi + [\varphi,\Delta_\perp \varphi]+ \frac{2}{\beta_e} \nabla_\|\Delta_\perp A_\| =0. \label{eq:rec-mod2}
	\end{align}
}	
	
	\item \textcolor{black}{ERMHD for KAWs (\cite{Schekochihin09, BHXP13})}
\textcolor{black}{	\begin{align}
	&\partial_t A_\| + \nabla_\| (1+\frac{1}{\tau}) \varphi=0.  \label{eq:ERMHD1}\\
	&\partial_t \varphi - \frac{\frac{2\tau}{\beta_e}}{1+\frac{\beta_e}{2}(1+\tau)} \nabla_\|\Delta_\perp A_\| =0. \label{eq:ERMHD2}
	\end{align}
}	
	\item \textcolor{black}{IKAW model (\cite{Chen-Bold17, PST17})}
\textcolor{black}{	\begin{align}
	&\partial_t (1-\frac{2\delta^2}{\beta_e}\Delta_\perp) A_\| -[\varphi, \frac{2\delta^2}{\beta_e}\Delta_\perp A_\| ]+ \nabla_\|  \varphi=0 \label{eq:IKAW1}\\
	&\partial_t \left (1+\frac{2}{\beta_i} -\frac{2\delta^2}{\beta_e}\Delta_\perp \right)\varphi - [\varphi,\frac{2\delta^2}{\beta_e}\Delta_\perp \varphi]- \frac{4}{\beta_e^2} \nabla_\|\Delta_\perp A_\| =0, \label{eq:IKAW2}
	\end{align}
}	
\textcolor{black}{where $\beta_i=\tau\beta_e$ refers to the ion beta parameter.} 
\end{itemize}

\subsection{Quadratic invariants}
The system (\ref{eq:Ne})-(\ref{eq:A})  preserves the energy ${\mathcal E}$ and the generalized cross-helicity ${\mathcal C}$, defined as 
\begin{eqnarray}
&&{\mathcal E} = \frac{1}{2} \int \Big ( \frac{2}{\beta_e} |\bnabla_\perp A_\| |^2 
+ \frac{4\delta^2}{\beta_e^2}|\Delta_\perp A_\| |^2 
- N_e(\varphi -N_e-B_z) \Big ) d^3 {x}, \label{energy}\\
&&{\mathcal C} =-\int N_e \Big ( 1 - \frac{2\delta^2}{\beta_e}\Delta_\perp \Big) A_\|d^3 {x}.\label{defC}
\end{eqnarray}
Note that ${\mathcal C}$ was defined with an opposite sign in \citet{PST18}.

For proving the conservation of energy and generalized cross helicity, one
notices that $N_e$, $\varphi -N_e -B_z$ and $\varphi-B_z$
result from the action of Hermitian operators on $\varphi$
and that $M_2$ and $M_3=1+M_2-M_1$ are positive definite.
One can thus write
\begin{equation}
\int \partial_t \{ (\varphi -N_e -B_z) N_e \} d{\boldsymbol x}= 2 \int (\varphi -N_e -B_z)\partial_t N_e d^3x
\end{equation}
and 
\begin{equation}
\int N_e \partial_z (\varphi-N_e-B_z) d^3x = 0.
\end{equation} 

In terms of the previously defined operators and of the generalized Elsasser potentials, the two invariants rewrite
\begin{align}
{\mathcal E}  &=  \frac{1}{2} \int \left \{s^2 (D_e A_\|)^2 + (D_e \Lambda \varphi)^2\right \} d^3x \label{Energy2}\\
&=\frac{1}{4} \int \left \{ (D_e \mu^+)^2 + (D_e \mu^-)^2 \right \} d^3x\\
{\mathcal C} &=  \int M_2 \Lambda^{-1} \left ( \frac{\mu^+ +\mu^-}{2}\right )
L_e \left ( \frac{\mu^+ -\mu^-}{2s} \right ) d^3x.
\end{align}

Using $\displaystyle{(1/s)M_2 \Lambda^{-1} L_e = D_e^2 V_{ph}^{-1}}$,
and the Hermitian character of the involved (commutating) operators, one also finds
\begin{eqnarray}
{\mathcal C} &=& \int \left ( \frac{\mu^+ +\mu^-}{2}\right )D_e^2 V_{ph}^{-1} \left ( \frac{\mu^+ -\mu^-}{2}\right )d^3x \nonumber \\
&=& \frac{1}{4} \int \left \{\left ( V_{ph}^{-1/2}D_e \mu^+\right)^2 - \left ( V_{ph}^{-1/2}D_e \mu^-\right)^2 \right\} d^3x, \label{eq:Cmu}
\end{eqnarray}

\textcolor{black}{\subsubsection{Limiting forms of the invariants}}
\textcolor{black}{The first term in the rhs of  Eq. (\ref{energy}) corresponds to the magnetic energy, the second one to the parallel electron kinetic energy, while the third one includes both internal energy and a term which, at MHD scales,  is dominated by the perpendicular ion kinetic energy. The latter becomes subdominant at small scales due to ion FLR effects, as a result of the $\Lambda$ operator in the second term of the rhs of Eq. (\ref{Energy2}). It then reduces to the perpendicular electron kinetic energy modified by electron FLR. These latter contributions are here retained only through the $\delta^2$ term in Eq. (\ref{eq:gyro-Ne-phi}) which is relevant when $\beta_i$ is of order unity or larger.} The complete form of the energy in the sub-$d_e$ range is given by Eq. (5.13) of \citet{PST17}.

It is also of interest to discuss the forms taken by the generalized cross-helicity when considered  in the large or small-scale limit.
\begin{itemize}
	\item \quad In the MHD range, $N_e=\Delta_\perp\varphi$ which also corresponds to the ion vorticity (with a single component \textcolor{black}{$\omega_{i z}$} along the $z$ direction). The invariant ${\cal C}$ thus reads 
${\mathcal C} =-\int \omega_{i z}  A_\|d^3 {x}$ which, after integration by parts, also rewrites ${\mathcal C} =-\int {\boldsymbol u_{\perp i}}\cdot{\boldsymbol B}_\perp  d^3 {x}$, i.e. the opposite of the usual cross-helicity.
Alternatively, in terms of magnetic helicity, one has for small $\tau$ ${\mathcal C} = \left ( {2}/{\beta_e} + 1 \right) \int  B_z A_\| d^3x$.

\item \quad At sub-ions scales (leaving out the electron inertia and FLR contributions), $B_z$ and $N_e$ are proportional to $\varphi$, and thus the generalized cross-helicity is proportional to the magnetic helicity which arises in imbalanced EMHD studied by \citet{KimCho15}.
More precisely, from $N_e = - M_2 M^{-1}_1 B_z= -(L_3 L^{-1}_2 L_1 + L_4)B_z$, one has 
\begin{equation}
{\mathcal C} = \int  (L_3 L^{-1}_2 L_1 + L_4)B_z (L_e A_\|) d^3 x.
\end{equation}
At these scales, where one can take $\Gamma_0=0$ and $\Gamma_1=0$, 
\begin{eqnarray}
{\mathcal C} &=&\int \left (1 + \frac{1}{1+ 1/\tau}\left (\frac{2}{\beta_i} -\frac{2\delta^2}{\beta_e} \Delta_\perp\right)\right )B_z \left (L_e A_\|\right ) d^3 x\\
&\approx& \int (L_eB_z)(L_e A_\|) d^3x + \frac{2}{\beta_e + \beta_i} \int B_z (L_e A_\|) d^3x. \label{C-magn-hel}
\end{eqnarray}
In Eq. (\ref{C-magn-hel}), $\delta^2\Delta_\perp\ll 1$ has been assumed. It is interesting to note that in the incompressible limit where the beta parameters tend to infinity, ${\mathcal C} =\int (L_eB_z)(L_e A_\|) d^3x$, thus recovering, in the quasi-transverse limit, the  generalized magnetic helicities of EMHD \citep{Biskamp99}, or of extended MHD when the ion velocity is taken to zero (Eq. (35) of \citet{Abdelhamid16}). 

The fact that the quantity ${\cal C}$ provides a generalization of the cross-helicity at sub-ion scales is easily seen  when considering Eq. (\ref{eq:Cmu}).
This expression clearly indicates the way the operator $V_{ph}$  renormalizes the usual cross-helicity, a quantity measuring the degree of imbalance between forward and backward propagating waves.
\end{itemize}

\subsection{Fourier-modes representation}

Using the same notation for the operators and their Fourier symbols, the Fourier components $c_{\boldsymbol k}^{\sigma_k} =
D_e(k_\perp) \mu_{\boldsymbol k}^{\sigma_k}$ of the fields $D_e \mu^{\pm}$, where $\sigma_k= \pm$, obey
\begin{equation}
\partial_t c_{\boldsymbol k}^{\sigma_k} + \sigma_k v_{ph}(k_\perp)\partial_z c_{\boldsymbol k}^{\sigma_k} - \int \sum_{\sigma_p, \sigma_q}{\widehat {\boldsymbol z}}\bcdot({\boldsymbol p}\times {\boldsymbol q}) L_{k_{\perp}p_{\perp}q_{\perp}}^{\sigma_k \sigma_p \sigma_q} c^{\sigma_p}_{\boldsymbol p} c^{\sigma_q}_{\boldsymbol q} \delta({\boldsymbol p}+{\boldsymbol q} - {\boldsymbol k} )
	d{\boldsymbol p} d{\boldsymbol q}=0 \label{eq:ck}
\end{equation}
with 
\begin{eqnarray}
L_{k_{\perp}p_{\perp}q_{\perp}}^{\sigma_k \sigma_p \sigma_q} &=& \frac{1}{4}\frac{1}{\Lambda(k_{\perp}) D_e(k_{\perp}) \Lambda(p_{\perp}) D_e(p_{\perp}) \Lambda(q_{\perp})D_e(q_{\perp})} \nonumber \\
&&\Big \{ 
M_3(k_{\perp}) M_3(p_{\perp}) M_2(q_{\perp}) 
-\sigma_p \sigma_q M_3(k_{\perp})\Lambda(p_{\perp})\Lambda(q_{\perp}) q_{\perp}^2 
\nonumber \\
&& - \sigma_k \sigma_p k_{\perp}^2 \Lambda(k_{\perp}) \Lambda(p_{\perp}) \Big ( M_2(q_{\perp}) + L_e(p_{\perp}) \{1-M_1(q_{\perp}) \} \Big ) 
\Big \},
\end{eqnarray}
where one also has (using $1-M_1 = M_3-M_2$)
\begin{equation}
M_2(q_\perp) + L_e(p_\perp) \{1-M_1(q_\perp) \} = M_3(q_\perp) + s^2 \delta^2 p_\perp^2\{1-M_1(q_\perp) \}= M_3(q_\perp)
 L_e(p_\perp) -M_2(q_\perp)
. \label{noLe}
\end{equation}
Furthermore, when denoting by $\alpha$, $\beta$ and $\gamma$ the interior angles of the $(k_\perp, p_\perp, q_\perp)$  triangle, one has 
\begin{equation}
{\widehat {\boldsymbol z}}\bcdot({\boldsymbol p}\times {\boldsymbol q}) =  k_\perp p_\perp q_ \perp \frac{\sin \alpha}{k_\perp} = k_\perp p_\perp q_ \perp \frac{\sin \beta}{p_\perp} = k_\perp p_\perp q_ \perp \frac{\sin \gamma }{q_\perp}.
\end{equation} 

\subsection{Detailed conservation laws}

Symmetrizing the interaction vertex in the form
\begin{equation}
V_{{\boldsymbol k}{\boldsymbol p}{\boldsymbol q}}^{\sigma_k \sigma_p \sigma_q} = \frac{1}{2} {\widehat {\boldsymbol z}}\bcdot({\boldsymbol p}\times {\boldsymbol q}) \left ( L_{k_\perp p_\perp q_\perp}^{\sigma_k \sigma_p \sigma_q}
-L_{k_\perp q_\perp p_\perp}^{\sigma_k \sigma_q \sigma_p}\right
),
\end{equation}
one has
\begin{equation}
\frac{1}{2}
\partial_t \sum_{\sigma_k}|c_{{\boldsymbol k}_\perp}^{\sigma_k}|^2 = \int \sum_{\sigma_k\sigma_p, \sigma_q} V_{-{\boldsymbol k}_\perp{\boldsymbol p}_\perp {\boldsymbol q_\perp}}^{\sigma_{-k} \sigma_p \sigma_q} c^{\sigma_p}_{\boldsymbol p} c^{\sigma_q}_{\boldsymbol q}  c^{\sigma_{k}}_{\boldsymbol k}
\delta({\boldsymbol p}+{\boldsymbol q} + {\boldsymbol k} )
d{\boldsymbol p} d{\boldsymbol q},
\end{equation}
with here $V_{-{\boldsymbol k}_\perp{\boldsymbol p}_\perp {\boldsymbol q_\perp}}^{\sigma_{-k} \sigma_p \sigma_q}=V_{\boldsymbol k_\perp{\boldsymbol p}_\perp {\boldsymbol q_\perp}}^{\sigma_k\sigma_p \sigma_q}$.
One easily checks that 
\begin{equation}
V_{\boldsymbol k_\perp{\boldsymbol p}_\perp {\boldsymbol q_\perp}}^{\sigma_k\sigma_p \sigma_q} + V_{\boldsymbol p_\perp{\boldsymbol q}_\perp {\boldsymbol k_\perp}}^{\sigma_p\sigma_q \sigma_k} +V_{\boldsymbol q_\perp{\boldsymbol k}_\perp {\boldsymbol p_\perp}}^{\sigma_q\sigma_k \sigma_p}=0.
\end{equation}
Defining 
\begin{equation}
S({\boldsymbol k}|{\boldsymbol p}, {\boldsymbol q})=\sum_{\sigma_k\sigma_p, \sigma_q} V_{{\boldsymbol k}_\perp{\boldsymbol p}_\perp {\boldsymbol q_\perp}}^{\sigma_k
\sigma_p \sigma_q} c^{\sigma_p}_{\boldsymbol p} c^{\sigma_q}_{\boldsymbol q}  c^{\sigma_k}_{\boldsymbol k}
\delta({\boldsymbol p}+{\boldsymbol q} + {\boldsymbol k} ),
\end{equation}
one thus has 
\begin{equation}
\frac{1}{2}
\partial_t \sum_{\sigma_k}|c_{{\boldsymbol k}_\perp}^{\sigma_k}|^2 = \int S({\boldsymbol k}|{\boldsymbol p}, {\boldsymbol q})
d{\boldsymbol p} d{\boldsymbol q}
\end{equation}
with 
\begin{equation}
S({\boldsymbol k}|{\boldsymbol p}, {\boldsymbol q}) + S({\boldsymbol p}|{\boldsymbol q}, {\boldsymbol k}) + S({\boldsymbol q}|{\boldsymbol k}, {\boldsymbol p})=0,
\end{equation}
which indicates that the Fourier modes associated to the triads $\pm {\boldsymbol k}, \pm {\boldsymbol p},\pm {\boldsymbol q}$ exchange energy conservatively.

The detailed conservation of the generalized cross helicity is obtained in a similar way, based on the identity
\begin{equation}
\frac{\sigma_k}{v_{ph}(k_\perp)}V_{\boldsymbol k_\perp{\boldsymbol p}_\perp {\boldsymbol q_\perp}}^{\sigma_k\sigma_p \sigma_q} + \frac{\sigma_p}{v_{ph}(p_\perp)}V_{\boldsymbol p_\perp{\boldsymbol q}_\perp {\boldsymbol k_\perp}}^{\sigma_p\sigma_q \sigma_k} +\frac{\sigma_q}{v_{ph}(q_\perp)}V_{\boldsymbol q_\perp{\boldsymbol k}_\perp {\boldsymbol p_\perp}}^{\sigma_q\sigma_k \sigma_p}=0
\end{equation}
which is established by writing 
\begin{equation}
\frac{\sigma_k}{v_{ph}(k_\perp)} = \frac{\sigma_k}{s} \frac{M_2(k_\perp)}{k_\perp^2 \Lambda(k_\perp)}=
\frac{\sigma_k}{s} \frac{\Lambda(k_\perp) M_2(k_\perp)}{k_\perp^2 \Lambda(k_\perp)^2}= \frac{\sigma_k}{s} \frac{\Lambda(k_\perp) L_e(k_\perp)}{M_3(k_\perp)}
\end{equation}
and thus, after
making the substitution $1-M_1(k_\perp) = M_3(k_\perp) -M_2(k_\perp)$, 
\begin{eqnarray}
&&\frac{\sigma_k}{v_{ph}(k_\perp)} V_{{\boldsymbol k}{\boldsymbol p}{\boldsymbol q}}^{\sigma_k \sigma_p \sigma_q}= \frac{2}{s}\Phi(k_\perp, p_\perp, q_\perp) \Big\{\sigma_k \Lambda(k_\perp) L_e(k_\perp)\left (M_3(p_\perp) M_2(q_\perp)- M_3(q_\perp)M_2(p_\perp)\right)\nonumber \\
&&-\sigma_k\sigma_p\sigma_q \Lambda(k_\perp)\Lambda(p_\perp)\Lambda(q_\perp) (q_\perp^2 -p_\perp^2)L_e(k_\perp)
\nonumber \\
&&-M_2(k_\perp) \left[\sigma_p \Lambda(p_\perp) \left(M_3(q_\perp) L_e(p_\perp)-M_2(q_\perp)\right)
-\sigma_q \Lambda(q_\perp) \left(M_3(p_\perp) L_e(q_\perp)-M_2(p_\perp)\right)
\right]\Big \},\nonumber \\
\end{eqnarray}
where 
\begin{equation}
\Phi(k_\perp, p_\perp, q_\perp)=\frac{1}{16}\frac{{\widehat {\boldsymbol z}}\bcdot({\boldsymbol p}\times {\boldsymbol q})}{\Lambda(k_{\perp}) D_e(k_{\perp}) \Lambda(p_{\perp}) D_e(p_{\perp}) \Lambda(q_{\perp})D_e(q_{\perp})}  \end{equation}
is invariant by circular permutation of the variables.

\subsection{Interaction representation}

The interaction representation is obtained by defining
\begin{equation}
a_{\boldsymbol k}^{\sigma_k} = e^{i\omega_k^{\sigma_k} t} c_{\boldsymbol k}
\end{equation}
with $\omega_{\boldsymbol k}^{\sigma_k} = \sigma_k v_{ph}(k_\perp) k_z$. Equation (\ref{eq:ck})
rewrites
\begin{equation}
\partial_t a_{\boldsymbol k}^{\sigma_k}  - \int \sum_{\sigma_p, \sigma_q} e^{i \Omega_{{\boldsymbol k};{\boldsymbol p} {\boldsymbol q}}^{\sigma_k \sigma_p \sigma_q}t}
V_{{\boldsymbol k}{\boldsymbol p}{\boldsymbol q}}^{\sigma_k \sigma_p \sigma_q} a^{\sigma_p}_{\boldsymbol p}a^{\sigma_q}_{\boldsymbol q} \delta({\boldsymbol p}+{\boldsymbol q} - {\boldsymbol k} )
d{\boldsymbol p} d{\boldsymbol q}=0, \label{eq:ak}
\end{equation}
with 
\begin{equation}
\Omega_{{\boldsymbol k};{\boldsymbol p} {\boldsymbol q}}^{\sigma_k \sigma_p \sigma_q} = \omega_{\boldsymbol k}^{\sigma_k} - \omega_{\boldsymbol p}^{\sigma_p} - \omega_{\boldsymbol q}^{\sigma_q}= \sigma_k v_{ph}(k_\perp) k_\| - \sigma_p v_{ph}(p_\perp) p_\| - \sigma_q v_{ph}(q_\perp) q_\|.  \label{Omega}
\end{equation}

\section{Weak-turbulence kinetic equations}  \label{weak_turbulence}

Beyond the possible 
presence of weak KAW turbulence in natural plasmas \citep{Moya15}, this regime, where the characteristic time of the nonlinear effects is significantly longer than the period of the waves, has the  advantage of being amenable to  a systematic theory. \textcolor{black}{It is of great interest for understanding statistical dynamics of driven and dissipating complex systems, as discussed in \citet{Newell01}.} 
As some of the properties concerning the spectral transfer of the quadratic invariants are common for weak and strong turbulence,  the study of the former can shed light on the latter.  

 We concentrate on the three-wave couplings which are dominant in \textcolor{black}{weak incompressible} MHD when retaining the interactions mediated by the $k_z=0$ modes \citep{Ng96}. As pointed out by \cite{Scheko12}, at least in the case of a finite periodic domain where continuity relatively to $k_z$ cannot be assumed, the non-dispersive MHD regime could require discriminating between the propagating Alfv\'en waves (characterized by a non-zero wavevector component $k_z$) and two-dimensional perturbations in the transverse spectral plane. These modes which mediate the interaction between counter-propagating Alfv\'en waves obey a  dynamics governed by the two-dimensional MHD equations with an oscillatory nonlinear source representing the coupling of two counter propagating Alfv\'en waves.
Influence of the couplings to purely two-dimensional modes  in numerical simulations of weak incompressible MHD turbulence is discussed in \citet{Meyrand15}. This difficulty can be avoided when retaining dispersion originating from kinetic effects. Dispersion becoming evanescent in the large-scale limit, one can suspect that an upper bound is to be assumed for the 
largest considered scales.
Note also that in contrast with whistler turbulence which remains weak at all the scales \citep{Lyutikov13,MeyrandX18}, weak KAW turbulence is expected to become strong at scales small enough for the linear and nonlinear times to become comparable \textcolor{black}{ (see \citet{GS97} for the case of MHD turbulence and also \citet{PS15} for KAWs).} Direct numerical simulations of this transition in the incompressible MHD regime are presented in \citet{Meyrand16}.

The weak-turbulence kinetic equations are obtained by repeating the
analysis done in Section 3.1 of \citet{Galtier15}. Indeed, being real, the vertex $V_{{\boldsymbol k}{\boldsymbol p}{\boldsymbol q}}^{\sigma_k \sigma_p \sigma_q}$ obeys
\begin{equation}
V_{{\boldsymbol k}{\boldsymbol p}{\boldsymbol q}}^{\sigma_k \sigma_p \sigma_q} = (V_{-{\boldsymbol k} -{\boldsymbol p} -{\boldsymbol q}}^{\sigma_k \sigma_p \sigma_q})^*,
\end{equation}
where $^*$ stands for the complex conjugate.
It is furthermore symmetric in ${\boldsymbol p}$ and ${\boldsymbol q}$ (and simultaneously in $(\sigma_p, \sigma_q)$. For ${\boldsymbol k}=0$, ${\boldsymbol p} = - {\boldsymbol q}$, thus leading to 
\begin{equation}
V_{0{\boldsymbol p}{\boldsymbol q}}^{\sigma_k \sigma_p \sigma_q}=0.
\end{equation}
One also has
\begin{equation}
V_{{\boldsymbol k}{\boldsymbol p}{\boldsymbol q}}^{\sigma_k \sigma_p \sigma_q}=
V_{{\boldsymbol k}{-\boldsymbol p}{-\boldsymbol q}}^{\sigma_k \sigma_p \sigma_q}=
V_{-{\boldsymbol k}-{\boldsymbol p}-{\boldsymbol q}}^{\sigma_k \sigma_p \sigma_q}
\end{equation}
and
\begin{equation}
V_{{\boldsymbol k}{\boldsymbol p}{\boldsymbol q}}^{\sigma_k \sigma_p \sigma_q}=
V_{{\boldsymbol k}{\boldsymbol p}{\boldsymbol q}}^{-\sigma_k -\sigma_p -\sigma_q}.
\end{equation}

Introducing the ensemble average $\langle \cdot \rangle$ and assuming statistical spatial homogeneity,  one defines the spectral density tensor $Q^{\sigma_k \sigma_{k'}}$ such that
\begin{equation}
\langle a_{\boldsymbol k}^{\sigma_k} a_{\boldsymbol k'}^{\sigma_{k'}}\rangle = Q^{\sigma_k
\sigma_{k'}}_{\boldsymbol k} \delta({\boldsymbol k} +{\boldsymbol k'}).
\end{equation}
In the context of weak turbulence, only the correlations of modes with the same $\sigma$ are resonant and contribute to the dynamics. Furthermore, for a time $t$ large compared to the period of the waves, 
\begin{equation}
\int_0^t \exp(i\Omega_{{\boldsymbol k};{\boldsymbol p} {\boldsymbol q}}^{\sigma_k \sigma_p \sigma_q}  {t'})dt'= \frac{\exp(i\Omega_{{\boldsymbol k};{\boldsymbol p} {\boldsymbol q}}^{\sigma_k \sigma_p \sigma_q} t -1)}{i\Omega_{{\boldsymbol k};{\boldsymbol p} {\boldsymbol q}}^{\sigma_k \sigma_p \sigma_q} }= \pi \delta(\Omega_{{\boldsymbol k};{\boldsymbol p} {\boldsymbol q}}^{\sigma_k \sigma_p \sigma_q} )  +i {\mathcal PV}\left (\frac{1}{\Omega_{{\boldsymbol k};{\boldsymbol p} {\boldsymbol q}}^{\sigma_k \sigma_p \sigma_q} }\right),
\end{equation}
where ${\mathcal PV}$ is the principal value of the convolution integral \textcolor{black}{(see e.g. \citet{Benney69})}.

One thus gets ($\delta_\sigma^{\sigma'}$ denoting the Kronecker symbol)
\begin{eqnarray}
&&\partial_t Q^{\sigma\sigma'}_{\boldsymbol k} \delta_\sigma^{\sigma'}= 4\pi \int \sum_{\sigma_{\pm k},\sigma_{\pm p},\sigma_{\pm q}}\delta({\boldsymbol k}- {\boldsymbol p}-{\boldsymbol q}) 
\delta (\Omega_{{\boldsymbol k};{\boldsymbol p} {\boldsymbol q}}^{\sigma_k \sigma_p \sigma_q} ) \nonumber \\
&& V_{{\boldsymbol k}{\boldsymbol p}{\boldsymbol q}}^{\sigma \sigma_p \sigma_q} \Big ( V_{{\boldsymbol p}-{\boldsymbol q} {\boldsymbol k}}^{\sigma_p \sigma_{-q} \sigma_k} Q^{\sigma_{-q} \sigma_{q}}_{\boldsymbol q} \delta_{\sigma_{-q}}^{\sigma_q} Q^{\sigma' \sigma_{k}}_{\boldsymbol k} \delta_{\sigma'}^{\sigma_k}
+ V_{{\boldsymbol q}-{\boldsymbol p} {\boldsymbol k}}^{\sigma_q \sigma_{-p} \sigma_k} Q^{\sigma_{-p} \sigma_{p}}_{\boldsymbol p} \delta_{\sigma_{-p}}^{\sigma_p} Q^{\sigma' \sigma_{k}}_{\boldsymbol k} \delta_{\sigma'}^{\sigma_k}  \nonumber \\
&& +V_{{\boldsymbol {-k}}-{\boldsymbol p} -{\boldsymbol q}}^{\sigma' \sigma_{-p} \sigma_{-q}} Q^{\sigma_{-p} \sigma_{p}}_{\boldsymbol p} \delta_{\sigma_{-p}}^{\sigma_p} Q^{\sigma_{-q} \sigma_{q}}_{\boldsymbol q} \delta_{\sigma_{-q}}^{\sigma_q} 
\Big ) d{\boldsymbol p} d{\boldsymbol q}.
\end{eqnarray}
Using the notation $Q^{\sigma}_{\boldsymbol k} = Q^{\sigma \sigma}_{\boldsymbol k}$ and $\Omega_{{\boldsymbol k}{\boldsymbol p} {\boldsymbol q}}^{\sigma_k \sigma_p \sigma_q}=\Omega_{-{\boldsymbol k};{\boldsymbol p} {\boldsymbol q}}^{\sigma_k \sigma_p \sigma_q}$,
 the above equation rewrites
\begin{eqnarray}
&&\partial_t Q^{\sigma}_{\boldsymbol k} = 4\pi \int \sum_{\sigma_{p},\sigma_{q}}\delta({\boldsymbol k}+ {\boldsymbol p}+{\boldsymbol q}) 
\delta (\Omega_{{\boldsymbol k}{\boldsymbol p} {\boldsymbol q}}^{\sigma \sigma_p \sigma_q} ) \nonumber \\
&& V_{-{\boldsymbol k}{\boldsymbol p}{\boldsymbol q}}^{\sigma \sigma_p \sigma_q} \left ( V_{{\boldsymbol p}-{\boldsymbol q} -{\boldsymbol k}}^{\sigma_p \sigma_{q} \sigma} Q^{\sigma_{q}}_{\boldsymbol q} Q^{\sigma}_{\boldsymbol k} 
+ V_{{\boldsymbol q}-{\boldsymbol p} -{\boldsymbol k}}^{\sigma_q \sigma_{p} \sigma} Q^{\sigma_{p}}_{\boldsymbol p} Q^{\sigma}_{\boldsymbol k}  +  V_{{\boldsymbol {k}}-{\boldsymbol p} -{\boldsymbol q}}^{\sigma\sigma_{p} \sigma_{q}} Q^{\sigma_{p}}_{\boldsymbol p}  Q^{\sigma_{q}}_{\boldsymbol q} 
\right ) d{\boldsymbol p} d{\boldsymbol q}.
\end{eqnarray}
As a consequence of the symmetries of the vertex, one gets the kinetic equation 
\begin{eqnarray}
	&&\partial_t Q^{\sigma}_{\boldsymbol k} = 4\pi \int \sum_{\sigma_{p},\sigma_{q}}\delta({\boldsymbol k}+ {\boldsymbol p}+{\boldsymbol q}) 
	\delta (\Omega_{{\boldsymbol k}{\boldsymbol p} {\boldsymbol q}}^{\sigma \sigma_p \sigma_q} ) \nonumber \\
	&& V_{{\boldsymbol k}{\boldsymbol p}{\boldsymbol q}}^{\sigma \sigma_p \sigma_q} \left \{\left ( V_{{\boldsymbol p}{\boldsymbol q} {\boldsymbol k}}^{\sigma_p \sigma_{q} \sigma} Q^{\sigma_{q}}_{\boldsymbol q} 
	+ V_{{\boldsymbol q}{\boldsymbol p} {\boldsymbol k}}^{\sigma_q \sigma_p \sigma} Q^{\sigma_{p}}_{\boldsymbol p} \right ) Q^{\sigma}_{\boldsymbol k}  +  V_{{\boldsymbol {k}}{\boldsymbol p} {\boldsymbol q}}^{\sigma\sigma_{p} \sigma_{q}} Q^{\sigma_{p}}_{\boldsymbol p}  Q^{\sigma_{q}}_{\boldsymbol q} \right\}
	 d{\boldsymbol p} d{\boldsymbol q}. \label{kinetic}
\end{eqnarray}
\textcolor{black}{Limiting forms of the interaction vertex $V_{{\boldsymbol k}{\boldsymbol p}{\boldsymbol q}}^{\sigma \sigma_p \sigma_q}$ in the MHD and sub-ion ranges are given in Appendix \ref{app:vertex}.}

\section{Kinetic equations for negligible electron inertia}\label{kin-eq-zeroinertia}

Concentrating in this Section on scales large  compared to the electron skin depth $d_e$ in order to make electron inertia  negligible, one first notes that 
\begin{eqnarray}
&&M_3(k_\perp) \left (M_3(p_\perp) M_2(q_\perp) -
M_3(q_\perp) M_2(p_\perp) \right ) = \nonumber \\
&& \qquad \qquad
M_3(k_\perp) M_3(p_\perp) M_3(q_\perp)
\left ( \frac{M_1(q_\perp)-1}{M_3(q_\perp)} -
\frac{M_1(p_\perp)-1}{M_3(p_\perp)}\right)
\end{eqnarray}
and
\begin{eqnarray}
&& \frac{1}{\Lambda(k_{\perp})\Lambda(p_{\perp}) \Lambda(q_{\perp})} \Big \{ -\sigma_p \sigma_q M_3(k_{\perp})\Lambda(p_{\perp})\Lambda(q_{\perp}) q_{\perp}^2 
 - \sigma_k \sigma_p k_{\perp}^2 \Lambda(k_{\perp}) \Lambda(p_{\perp})  M_3(q_\perp)\Big \} \nonumber \\
 &&- \{ p \leftrightarrow q \}  = -\sigma_p \sigma_q \frac{M_3(k_\perp)}{\Lambda(k_\perp)} (q_\perp^2 - p_\perp^2) - \sigma_k k_\perp^2 
\left (\frac{\sigma_p M_3(q_\perp)} {\Lambda(q_\perp)} - 
\frac{\sigma_q M_3(p_\perp)} {\Lambda(p_\perp)}\right)
\nonumber \\
&& = \left ( \sigma_q \frac{M_3(k_\perp) M_3(p_\perp)} {\Lambda(k_\perp) \Lambda(p_\perp)} - \sigma_p \frac{M_3(k_\perp) M_3(q_\perp)} {\Lambda(k_\perp) \Lambda(q_\perp)} \right ) 
\left ( \sigma_k k_\perp^2 \frac{\Lambda_(k_\perp) }{M_3(k_\perp)}
+  \{ k \leftrightarrow p  \} +  \{ p \leftrightarrow q \}
\right )\nonumber \\
&& + \frac{M_3(k_\perp)}{\Lambda_k} \left ( \frac{p_\perp^2 \Lambda(p_\perp) M_3(q_\perp)}{\Lambda(q_\perp) M_3(p_\perp)}
- \{ p \leftrightarrow q  \}\right ),
\end{eqnarray}
where $\{p \leftrightarrow q\}$  refers to the same expression where $(\sigma_p, p_\perp)$ and $(\sigma_q, q_\perp)$ have been exchanged. 
Introducing the functions $\xi(k_\perp) =  \Lambda(k_\perp) / M_3(k_\perp)=s/v_{ph}(k_\perp)$ and $R(k_\perp) ={(1-M_1(k_\perp))}/{\Lambda(k_\perp)}$,
one gets
\begin{eqnarray}
&&V_{{\boldsymbol k}{\boldsymbol p}{\boldsymbol q}}^{\sigma_k \sigma_p \sigma_q}= \frac{1}{8} \frac{{\widehat {\boldsymbol z}}\bcdot({\boldsymbol p}\times {\boldsymbol q})}{k_\perp p_\perp q_\perp} \Big \{ \frac{\sigma_k}{\xi(k_\perp)}  \left ( \frac{\sigma_p}{\xi(p_\perp)} - 
\frac{\sigma_q}{\xi(q_\perp)}\right ) \sigma_k\sigma_p\sigma_q
\left ( \sigma_k k_\perp^2 \xi(k_\perp) 
+ \{ k \leftrightarrow p  \} +  \{ p \leftrightarrow q \}\right )
\nonumber \\
&& + \frac{1}{\xi(k_\perp) \xi(p_\perp) \xi(q_\perp)}\left ( p_\perp^2\xi^2(p_\perp) - q^2_\perp \xi^2(q_\perp) +
R(p_\perp)\xi(p_\perp)-R(q_\perp)\xi(q_\perp)
\right ) \Big \}.
\end{eqnarray}
One easily checks that 
\begin{equation}
p_\perp^2\xi^2(p_\perp) - q^2_\perp \xi^2(q_\perp) +
R(p_\perp)\xi(p_\perp)-R(q_\perp)\xi(q_\perp)=0.
\end{equation}
It is furthermore convenient to define the  quantity 
\begin{equation}
S_{{\boldsymbol k}{\boldsymbol p}{\boldsymbol q}}^{\sigma_k \sigma_p \sigma_q}= \frac{1}{8} {\widehat {\boldsymbol z}}\bcdot({\boldsymbol p}\times {\boldsymbol q}) \frac{\sigma_k\sigma_p\sigma_q}{k_\perp p_\perp q_\perp}
\left ( \sigma_k k_\perp^2 \xi(k_\perp) 
+ \{ k \leftrightarrow p  \} +  \{ p \leftrightarrow q \}\right )
\end{equation}
and to rewrite
\begin{equation}
V_{{\boldsymbol k}{\boldsymbol p}{\boldsymbol q}}^{\sigma_k \sigma_p \sigma_q}= \frac{\sigma_k}{\xi(k_\perp)}  \left ( \frac{\sigma_p}{\xi(p_\perp)} - 
\frac{\sigma_q}{\xi(q_\perp)}\right )S_{{\boldsymbol k}{\boldsymbol p}{\boldsymbol q}}^{\sigma_k \sigma_p \sigma_q}.\label{vertex}
\end{equation}
The resulting vertex for the coupling of the $\mu_{\boldsymbol k}$-modes is consistent with Eq. (6.4) of \citet{Voitenko98a}, directly derived from the Vlasov-Maxwell equations.

When $k_\perp$, $p_\perp$ and $q_\perp$ are the sides of a triangle, one has the property that, relatively to a permutation
of these variables, $S_{{\boldsymbol k}{\boldsymbol p}{\boldsymbol q}}^{\sigma_k \sigma_p \sigma_q}$
is invariant or change sign, depending on the positive or negative signature of the permutation.

The next step is to substitute the vertex given by Eq. (\ref{vertex}) into the kinetic equation (\ref{kinetic}).
One first easily checks that 
\begin{eqnarray}
&&V_{{\boldsymbol k}{\boldsymbol p}{\boldsymbol q}}^{\sigma_k \sigma_p \sigma_q} V_{{\boldsymbol p}{\boldsymbol q}{\boldsymbol k}}^{\sigma_p \sigma_q \sigma_k} = 
\frac{\sigma_k \sigma_p}{\xi(k_\perp)\xi(p_\perp)} 
 \left ( \frac{\sigma_p}{\xi(p_\perp)} - 
 \frac{\sigma_q}{\xi(q_\perp)}\right )
 \left ( \frac{\sigma_q}{\xi(q_\perp)} - 
 \frac{\sigma_k}{\xi(k_\perp)}\right )(S_{{\boldsymbol k}{\boldsymbol p}{\boldsymbol q}}^{\sigma_k \sigma_p \sigma_q})^2 \nonumber \\
&&V_{{\boldsymbol k}{\boldsymbol p}{\boldsymbol q}}^{\sigma_k \sigma_p \sigma_q} V_{{\boldsymbol q}{\boldsymbol p}{\boldsymbol k}}^{\sigma_q \sigma_p \sigma_k} = 
-\frac{\sigma_k \sigma_q}{\xi(k_\perp)\xi(q_\perp)} 
\left (\frac{\sigma_p}{\xi(p_\perp)} - 
\frac{\sigma_q}{\xi(q_\perp)}\right )
\left ( \frac{\sigma_p}{\xi(p_\perp)} - 
\frac{\sigma_k}{\xi(k_\perp)}\right )(S_{{\boldsymbol k}{\boldsymbol p}{\boldsymbol q}}^{\sigma_k \sigma_p \sigma_q})^2 \nonumber \\
&&(V_{{\boldsymbol k}{\boldsymbol p}{\boldsymbol q}}^{\sigma_k \sigma_p \sigma_q})^2 =  \frac{1}{\xi^2(k_\perp)} 
\left (\frac{\sigma_p}{\xi(p_\perp)} - 
\frac{\sigma_q}{\xi(q_\perp)}\right )^2 (S_{{\boldsymbol k}{\boldsymbol p}{\boldsymbol q}}^{\sigma_k \sigma_p \sigma_q})^2.
 \end{eqnarray}
Note that exchanging  $p_\perp$ and  $q_\perp$ together with $\sigma_p$ and $\sigma_q$ in $V_{{\boldsymbol k}{\boldsymbol p}{\boldsymbol q}}^{\sigma_k \sigma_p \sigma_q} V_{{\boldsymbol q}{\boldsymbol p}{\boldsymbol k}}^{\sigma_q \sigma_p \sigma_k}$ gives $V_{{\boldsymbol k}{\boldsymbol p}{\boldsymbol q}}^{\sigma_k \sigma_p \sigma_q} V_{{\boldsymbol p}{\boldsymbol q}{\boldsymbol k}}^{\sigma_p \sigma_q \sigma_k}$. Under this transformation, $\delta({\boldsymbol k}+ {\boldsymbol p}+{\boldsymbol q})$ and  
$\delta (\Omega_{{\boldsymbol k}{\boldsymbol p} {\boldsymbol q}}^{\sigma \sigma_p \sigma_q} )$ remain unchanged. It follows that
\begin{eqnarray}
&&\int \sum_{\sigma_{p},\sigma_{q}}\delta({\boldsymbol k}+ {\boldsymbol p}+{\boldsymbol q}) 
\delta (\Omega_{{\boldsymbol k}{\boldsymbol p} {\boldsymbol q}}^{\sigma \sigma_p \sigma_q}) V_{{\boldsymbol k}{\boldsymbol p}{\boldsymbol q}}^{\sigma \sigma_p \sigma_q} V_{{\boldsymbol q}{\boldsymbol p} {\boldsymbol k}}^{\sigma_q \sigma_p \sigma} Q^{\sigma_{p}}_{\boldsymbol p}  Q^{\sigma}_{\boldsymbol k} d{\boldsymbol p} d{\boldsymbol q} \nonumber \\
&& =\int \sum_{\sigma_{p},\sigma_{q}}\delta({\boldsymbol k}+ {\boldsymbol p}+{\boldsymbol q}) 
\delta (\Omega_{{\boldsymbol k}{\boldsymbol p} {\boldsymbol q}}^{\sigma \sigma_p \sigma_q} ) 
V_{{\boldsymbol k}{\boldsymbol p}{\boldsymbol q}}^{\sigma \sigma_p \sigma_q} V_{{\boldsymbol p}{\boldsymbol q} {\boldsymbol k}}^{\sigma_p \sigma_{q} \sigma} Q^{\sigma_{q}}_{\boldsymbol q} 
 Q^{\sigma}_{\boldsymbol k} 
d{\boldsymbol p} d{\boldsymbol q}.
\end{eqnarray}
The kinetic equation thus rewrites
\begin{eqnarray}
&&\partial_t Q^{\sigma}_{\boldsymbol k} = 4\pi  \sum_{\sigma_{p},\sigma_{q}}\int 
\delta({\boldsymbol k}+ {\boldsymbol p}+{\boldsymbol q}) \delta (\Omega_{{\boldsymbol k}{\boldsymbol p} {\boldsymbol q}}^{\sigma \sigma_p \sigma_q} )Q^{\sigma_{p}}_{\boldsymbol p}\Big\{\Big[
\frac{1}{\xi^2(k_\perp)} 
\left (\frac{\sigma_p}{\xi(p_\perp)} - 
\frac{\sigma_q}{\xi(q_\perp)}\right )^2 (S_{{\boldsymbol k}{\boldsymbol p}{\boldsymbol q}}^{\sigma \sigma_p \sigma_q})^2\Big ]Q^{\sigma_q}_{\boldsymbol q} \nonumber \\
&&- 2 \Big [\frac{\sigma \sigma_q}{\xi(k_\perp)\xi(q_\perp)} 
\Big (\frac{\sigma_p}{\xi(p_\perp)} - 
\frac{\sigma_q}{\xi(q_\perp)}\Big )
\Big ( \frac{\sigma_p}{\xi(p_\perp)} - 
\frac{\sigma}{\xi(k_\perp)}\Big )(S_{{\boldsymbol k}{\boldsymbol p}{\boldsymbol q}}^{\sigma\sigma_p \sigma_q})^2 
\Big]  Q^{\sigma}_{\boldsymbol k}  \Big\} d{\boldsymbol p} d{\boldsymbol q}. 
\end{eqnarray}
Note that, as
$\xi = s/v_{ph}$ and $\sigma/\xi = \sigma v_{ph}/s$, the kinetic effects enter the above weak turbulence equations only through the definition of the phase velocity $v_{ph}$.

\subsection{Detailed conservation laws}

Detailed-conservation of the invariants is easily checked by 
rewriting the kinetic equation in the more symmetric form
\begin{eqnarray}
&&\partial_t Q^{\sigma}_{\boldsymbol k} = 4\pi  \sum_{\sigma_{p},\sigma_{q}}\int 
\delta({\boldsymbol k}+ {\boldsymbol p}+{\boldsymbol q}) \delta (\Omega_{{\boldsymbol k}{\boldsymbol p} {\boldsymbol q}}^{\sigma \sigma_p \sigma_q} )\Big\{
\frac{1}{\xi^2(k_\perp)} 
\left (\frac{\sigma_p}{\xi(p_\perp)} - 
\frac{\sigma_q}{\xi(q_\perp)}\right )^2 (S_{{\boldsymbol k}{\boldsymbol p}{\boldsymbol q}}^{\sigma \sigma_p \sigma_q})^2
Q^{\sigma_{p}}_{\boldsymbol p} Q^{\sigma_{q}}_{\boldsymbol q} \nonumber \\
&&-\Big [\frac{\sigma \sigma_q}{\xi(k_\perp)\xi(q_\perp)} 
\Big (\frac{\sigma_p}{\xi(p_\perp)} - 
\frac{\sigma_q}{\xi(q_\perp)}\Big )
\Big ( \frac{\sigma_p}{\xi(p_\perp)} - 
\frac{\sigma}{\xi(k_\perp)}\Big )(S_{{\boldsymbol k}{\boldsymbol p}{\boldsymbol q}}^{\sigma\sigma_p \sigma_q})^2 Q^{\sigma_{p}}_{\boldsymbol p} Q^{\sigma}_{\boldsymbol k}\nonumber \\
&&-\Big [\frac{\sigma \sigma_p}{\xi(k_\perp)\xi(p_\perp)} 
\Big (\frac{\sigma_q}{\xi(q_\perp)} - 
\frac{\sigma_p}{\xi(p_\perp)}\Big )
\Big ( \frac{\sigma_q}{\xi(q_\perp)} - 
\frac{\sigma}{\xi(k_\perp)}\Big )(S_{{\boldsymbol k}{\boldsymbol p}{\boldsymbol q}}^{\sigma\sigma_p \sigma_q})^2 Q^{\sigma_{q}}_{\boldsymbol q} Q^{\sigma}_{\boldsymbol k}
 \Big\} d{\boldsymbol p} d{\boldsymbol q}.
\end{eqnarray}
The resonance conditions
\begin{eqnarray}
&&k_\| + p_\| + q_\|=0\\
&&\sigma v_{ph}(k_\perp) k_\| + \sigma_p v_{ph}(p_\perp) p_\|
+ \sigma_q v_{ph}(q_\perp) q_\| =0
\end{eqnarray}
lead to 
\begin{equation}
\frac{1}{k_\|} \left (\frac{\sigma_p}{\xi(p_\perp)}- \frac{\sigma_q}{\xi(q_\perp)} \right) = 
\frac{1}{p_\|} \left (\frac{\sigma_q}{\xi(q_\perp)}- \frac{\sigma_k}{\xi(k_\perp)} \right)=
\frac{1}{q_\|} \left (\frac{\sigma_k}{\xi(k_\perp)}- \frac{\sigma_p}{\xi(p_\perp)} \right).
\end{equation}
It follows that 
\begin{eqnarray}
	&&\partial_t Q^{\sigma}_{\boldsymbol k} = 4\pi  \sum_{\sigma_{p},\sigma_{q}}\int 
	\delta({\boldsymbol k}+ {\boldsymbol p}+{\boldsymbol q}) \delta (\Omega_{{\boldsymbol k}{\boldsymbol p} {\boldsymbol q}}^{\sigma \sigma_p \sigma_q} )
	\omega_{\boldsymbol k}^\sigma\Big [\frac{1}{k_\|} 
	\left (\frac{\sigma_p}{\xi(p_\perp)} - 
	\frac{\sigma_q}{\xi(q_\perp)}\right )\Big ]^2 (S_{{\boldsymbol k}{\boldsymbol p}{\boldsymbol q}}^{\sigma \sigma_p \sigma_q})^2
	\nonumber \\
	&&\left ( \omega_{\boldsymbol k}^\sigma Q^{\sigma_p}_{\boldsymbol p}
	Q^{\sigma_{q}}_{\boldsymbol q}
	+ \omega_{\boldsymbol p}^{\sigma_p} Q^\sigma_{\boldsymbol k}
	Q^{\sigma_q}_{\boldsymbol q}
	+\omega_{\boldsymbol q}^{\sigma_q} Q^\sigma_{\boldsymbol k}
	Q^{\sigma_p}_{\boldsymbol p} \right) 
	d{\boldsymbol p} d{\boldsymbol q}.
\end{eqnarray}

Proof of the detailed conservation of energy and generalized cross helicity at the level of the kinetic equations is based on the observation that the integrand in the rhs appears as the product of $\omega_{\boldsymbol k}^\sigma$ and of a quantity $\Psi_{\omega_{\boldsymbol k}\omega_{\boldsymbol p}\omega_{\boldsymbol q}}^{\sigma_k \sigma_p\sigma_q}$ which is invariant under circular perturbation of the wavevectors and of the corresponding sigma's. When summing on $\sigma_k$ and integrating on ${\boldsymbol k}$ to get the energy, the resulting integrand can be rewritten  $(1/3) \left (\omega_{\boldsymbol  k}+\omega_{\boldsymbol p}+\omega_{\boldsymbol q}\right) \Psi_{\omega_{\boldsymbol k}\omega_{\boldsymbol p}\omega_{\boldsymbol q}}^{\sigma_k \sigma_p\sigma_q}$, 
which vanishes because of the resonance condition.
In the case of the generalized cross helicity where, before summation and integration, the equation is multiplied by $\sigma_k/v_{ph}(k_\perp)$, the factor $\omega_{\boldsymbol k}$ is replaced by $\sigma_k \omega_{\boldsymbol k}/v_{ph}(k_\perp)= k_\|$ and the same argument as before can be  used, based on the resonance condition for the parallel wave numbers.

\subsection{Transversally isotropic turbulence}

Assuming isotropy in the transverse plane,  $\int d{\boldsymbol p} d{\boldsymbol q}\delta({\boldsymbol k}+ {\boldsymbol p}+{\boldsymbol q})	$ can be replaced by $2\pi \int dp_\| dq_\| \delta(k_\|+ p_\| +  q_\|) \int_{\Delta_{k_\perp}} (1/\sin \alpha)dp_\perp dq_\perp $ where $\Delta_{k_\perp}$ is the domain of the $(p_\perp, q_\perp)$-
plane limited by the triangular inequalities $|p_\perp -q_\perp|\le k_\perp\le p_\perp + q_\perp$. It is also convenient 
to introduce the energy and generalized cross helicity spectra 
\begin{eqnarray}
&&E(k_\perp, k_\|)= \frac{\pi}{2} k_\perp\sum_\sigma  Q^\sigma_{\boldsymbol k}\\
&&E_C(k_\perp, k_\|)= \frac{\pi}{2}\frac{k_\perp}{v_{ph}(k_\perp)}\sum_\sigma  \sigma Q^\sigma_{\boldsymbol k},
\end{eqnarray}
so that
\begin{equation}
Q^{\sigma}_{\boldsymbol k} =\frac{1}{\pi k_\perp} 
\left (E(k_\perp, k_\|) + \sigma v_{ph}(k_\perp)E_C(k_\perp, k_\|)\right ).
\end{equation}
Noting that for a function $f(\sigma_k, \sigma_p, \sigma_q)$
such that $f(\sigma_k, \sigma_p, \sigma_q)= f(-\sigma_k, -\sigma_p, -\sigma_q)$, 
\begin{equation}
\sum_{\sigma_k, \sigma_p,\sigma_q} \sigma_p f(\sigma_k, \sigma_p, \sigma_q) =0,
\end{equation}
it turns out that only nonlinear couplings of the form $E E$ and $E_C E_C$ arise in the equation for $E$, while in the equation for $E_C$, only terms of the form $E E_C$ are present. The
resonance condition $\omega^{\sigma_k}_{\boldsymbol k} = -\omega^{\sigma_p}_{\boldsymbol p} - \omega^{\sigma_q}_{\boldsymbol q}$ leads to a fully symmetric equation in ${\boldsymbol p}$ and ${\boldsymbol q}$, which permits 
transformation of the coupling between modes ${\boldsymbol p}$ and ${\boldsymbol k}$  into coupling between modes ${\boldsymbol q}$ and ${\boldsymbol k}$. One finally gets
\begin{eqnarray}
&&\partial_t E(k_\perp, k_\|) = \frac{1}{16} \sum_{\sigma_k, \sigma_p,\sigma_q} \int_{p_\|,q_\|}\int_{\Delta_k}  \delta(k_\|+ p_\| + q_\|)  \delta (\Omega_{{\boldsymbol k}{\boldsymbol p} {\boldsymbol q}}^{\sigma_k \sigma_p \sigma_q} )\frac{1}{k_\perp p_\perp q_\perp}\left(\frac{\sin \gamma}{q_\perp}\right)\nonumber \\
&&\qquad \left [\frac{1}{k_\|}\left (\frac{\sigma_p}{\xi(p_\perp)} -\frac{\sigma_q}{\xi (q_\perp)}\right) \right]^2 (\sigma_k k_\perp^2 \xi(k_\perp) + \sigma_p p_\perp^2 \xi(p_\perp) + \sigma_q q_\perp^2 \xi(q_\perp))^2 \nonumber \\ 
&& \qquad\omega^{\sigma_k}_{\boldsymbol k}\omega^{\sigma_p}_{\boldsymbol p} \{ E(q_\perp, q_\|) [p_\perp E(k_\perp, k_\|) - k_\perp E(p_\perp, p_\|)] + \sigma_q v_{ph}(q_\perp) E_C(q_\perp, q_\|)
\nonumber \\
&&\qquad[\sigma_k p_\perp v_{ph}(k_\perp) E_C(k_\perp, k_\|)
-\sigma_p k_\perp v_{ph}(p_\perp) E_C(p_\perp, p
_\|)]  \}
dp_\perp dq_\perp dp_\| dq_\| \label{energy-spec}\\
&&\partial_t E_C(k_\perp, k_\|) =\frac{1}{16} \sum_{\sigma_k, \sigma_p,\sigma_q} \int_{p_\|,q_\|}\int_{\Delta_k}  \delta(k_\|+ p_\| + q_\|)  \delta (\Omega_{{\boldsymbol k}{\boldsymbol p} {\boldsymbol q}}^{\sigma_k \sigma_p \sigma_q} )\frac{1}{k_\perp p_\perp q_\perp}\left(\frac{\sin \gamma}{q_\perp}\right)\nonumber \\
&&\qquad \left [\frac{1}{k_\|}\left (\frac{\sigma_p}{\xi(p_\perp)} -\frac{\sigma_q}{\xi (q_\perp)}\right) \right]^2 (\sigma_k k_\perp^2 \xi(k_\perp) + \sigma_p p_\perp^2 \xi(p_\perp) + \sigma_q q_\perp^2 \xi(q_\perp))^2  \nonumber \\ 
&& \qquad\frac{\sigma_k\omega^{\sigma_k}_{\boldsymbol k}\omega^{\sigma_p}_{\boldsymbol p}}{v_{ph}(k_\perp)}\{ E(q_\perp, q_\|) [\sigma_k v_{ph}(k_\perp)p_\perp E_C(k_\perp, k_\|) - \sigma_p v_{ph}(p_\perp)k_\perp E_C(p_\perp, p_\|)] \nonumber \\
&&\qquad + \sigma_q v_{ph}(q_\perp) E_C(q_\perp, q_\|)
[ p_\perp E(k_\perp, k_\|)-k_\perp E(p_\perp, p_\|)]  \}
dp_\perp dq_\perp dp_\| dq_\|.\label{helicity-spec}
\end{eqnarray}
As previously defined, $\omega_{\boldsymbol k}^{\sigma_k} = \sigma_kv_{ph}(k_\perp)k_\|$ and $\xi(k_\perp) = (2/\beta_e)^{1/2} / v_{ph}(k_\perp)$. 

In the sub-ion range where $v_{ph} \sim k_\perp$, 
the above system formally identifies with the kinetic equations given by \citet{Galtier03} in the framework of EMHD for  whistler waves when the dynamics is quasi-transverse. 

{\bf Remark:}
At the level of the above kinetic equations for an isotropic turbulence in the transverse plane, the energy conservation is immediately recovered by noticing that the integrand changes sign when exchanging $(k_\perp, k_\|, \sigma_k)$ and  $(p_\perp, p_\|, \sigma_p)$. The conservation of the generalized cross helicity is less straightforward. The integrand appears in this case as the product of a fully invariant quantity when $(k_\perp, k_\|, \sigma_k)$, $(p_\perp, p_\|, \sigma_p)$ and $(q_\perp, q_\|, \sigma_q)$ are exchanged, multiplied by the sum of four terms
 \begin{eqnarray}
&& A= k_\| \omega_{\boldsymbol p}^{\sigma_p} \sigma_k v_{ph}(k_\perp) p_\perp E(q_\perp, q_\|)E_C(k_\perp, k_\|)\\
&&B= k_\| \omega_{\boldsymbol p}^{\sigma_p} \sigma_q  v_{ph}(q_\perp) p_\perp E(k_\perp, k_\|) E_C(q_\perp, q_\|)\\
&&C= -k_\| \omega_{\boldsymbol p}^{\sigma_p} \sigma_p v_{ph}(p_\perp) k_\perp E(q_\perp, q_\|)E_C(p_\perp, p_\|)\\
&&D= -k_\| \omega_{\boldsymbol p}^{\sigma_p} \sigma_q v_{ph}(q_\perp) k_\perp E(p_\perp, p_\|)E_C(q_\perp, q_\|)
 \end{eqnarray}
Here, $B$ can be replaced by $q_\| \omega_{\boldsymbol p}^{\sigma_p} \sigma_k v_{ph}(k_\perp) p_\perp E(q_\perp, q_\|) E_C(k_\perp, k_\|)$. Using both the resonance conditions, $A+ B$ can be replaced by $p_\|(\omega_{\boldsymbol k}^{\sigma_k} + \omega_{\boldsymbol q}^{\sigma_q)})v_{ph}(k_\perp) p_\perp E(q_\perp, q_\|) E_C(k_\perp, k_\|)$. Furthermore, $C$ can be changed into 
$-p_\| \omega_{\boldsymbol k}^{\sigma_k} \sigma_k v_{ph}(k_\perp) p_\perp E(q_\perp, q_\|)E_C(k_\perp, k_\|)$ and $D$, through a circular permutation, into 
$-p_\| \omega_{\boldsymbol q}^{\sigma_q} \sigma_k v_{ph}(k_\perp) p_\perp E(q_\perp, q_\|)E_C(k_\perp, k_\|)$. After these substitutions, the sum of the four contributions to the integrand vanishes, thus ensuring detailed conservation. 

\section{A strongly-local model in the weak turbulence regime}
\label{local_model}

Still considering scales large compared to $d_e$, we concentrate on a simplified model  which only retains interacting
triads such that $k_\perp \approx p_\perp \approx q_\perp$ and 
$k_\|\approx p_\| \approx q_\|$ (supposed to be small and roughly constant). In this limit, the weak-turbulence kinetic equations reduce to partial differential equations (see e.g. \citet{Dyachenko92} in the case of four-wave coupling).
\textcolor{black}{Note however  that nonlocal interactions can be significant in the presence of dissipative processes.  Their relevance in Alfvenic turbulence with Landau damping is considered theoretically in \cite{Howes11} and numerically in \citet{Told15} where it is observed that roughly $30\%$ of the energy transfer to small scales in the kinetic range of KAW turbulence is mediated by interaction with modes at much larger scale. Nevertheless, as dissipative effects are not retained in the present paper, we choose to keep the model fully local.}

Let us first formally rewrite  the kinetic equations in the form
\begin{eqnarray}
&&\partial_t E(k_\perp, k_\|) = \sum_{\sigma_k, \sigma_p, \sigma_q}
\int_{p_\|,q_\|}\int_{\Delta_k}  
T^{\sigma_k \sigma_p \sigma_q}_{{\boldsymbol k}{\boldsymbol p}{\boldsymbol q}}
dp_\perp dq_\perp dp_\| dq_\| \\
&&\partial_t E_C(k_\perp, k_\|) = \sum_{\sigma_k, \sigma_p, \sigma_q}
\int_{p_\|,q_\|}\int_{\Delta_k}  
T^{\sigma_k \sigma_p \sigma_q}_{C{\boldsymbol k}{\boldsymbol p}{\boldsymbol q}}dp_\perp dq_\perp dp_\| dq_\|, 
\end{eqnarray}
where in the equation for the energy spectrum
\begin{equation}
T^{\sigma_k \sigma_p \sigma_q}_{{\boldsymbol k}{\boldsymbol p}{\boldsymbol q}}= - T^{\sigma_p \sigma_k \sigma_q}_{{\boldsymbol p}{\boldsymbol k}{\boldsymbol q}} .
\end{equation}
Following \citet{GBuch10} who performed a similar analysis in the case of MHD turbulence, one multiplies the equation for the energy spectrum by an arbitrary test function $f(k_\perp, k_\|)$ and integrate on $0<k_\perp < +\infty$
and $-\infty<k_\|< + \infty$. Exchanging $k_\perp$ into $p_\perp$, $k_\|$ into
$p_\|$ and $\sigma_k$ into $\sigma_p$, one gets
\begin{eqnarray}
&&\partial_t \int_{k_\perp k_\|} E(k_\perp, k_\|) f(k_\perp, k_\|)dk_\perp dk_\|= \nonumber \\
&&\int_{k_\perp}\int_{k_\|}\frac{1}{2}\sum_{\sigma_k, \sigma_p, \sigma_q}
\int_{p_\|,q_\|}\int_{\Delta_k}  
T^{\sigma_k \sigma_p \sigma_q}_{{\boldsymbol k}{\boldsymbol p}{\boldsymbol q}}
[f(k_\perp, k_\|)-f(p_\perp, p_\|)] dk_\perp dp_\perp dq_\perp dk_\|dp_\| dq_\|.
\end{eqnarray}
For local interactions, 
\begin{equation}
f(k_\perp, k_\|)-f(p_\perp, p_\|) \approx (k_\perp -p_\perp) \frac{\partial f} {\partial {k_\perp}} 
+ (k_\| - p_\|) \frac{\partial f} {\partial {k_\|}},
\end{equation}
where in the rhs the second term is negligible compared to the first one.
Writing $p_\perp = k_\perp(1+ \epsilon_{p_\perp})$ and $q_\perp = k_\perp(1+ \epsilon_{q_\perp})$, 
assuming that only transverse wavenumbers $p_\perp$ and $q_\perp$ such that
$|k_\perp -p_\perp|/k_\perp < \epsilon$ and 
$|k_\perp -q_\perp|/k_\perp < \epsilon$ can significantly interact with $k_\perp$, one obtains, by integrating by part in $k_\perp$ and noting that the function $f$ is arbitrary,
\begin{equation}
\partial_t E(k_\perp, k_\|) = -\frac{1}{2} \frac{\partial}{\partial k_\perp} \left (\sum_{\sigma_k, \sigma_p, \sigma_q}
\int_{p_\|,q_\|} dp_\| dq_\|\int_{-\varepsilon}^{+\varepsilon} \int_{-\varepsilon}^{+\varepsilon} d\epsilon_{p_\perp} d\epsilon_{q_\perp}  \varepsilon_{p_\perp} k_\perp^3 T^{\sigma_k \sigma_p \sigma_q}_{{\boldsymbol k}{\boldsymbol p}{\boldsymbol q}}\right ).
\end{equation}

In order to get the asymptotic form of $T^{\sigma_k \sigma_p \sigma_q}_{{\boldsymbol k}{\boldsymbol p}{\boldsymbol q}}$, one first uses locality in the transverse directions. The factor 
$\displaystyle {\left [\frac{1}{k_\|}\left (\frac{\sigma_p}{\xi(p_\perp)} -\frac{\sigma_q}{\xi (q_\perp)}\right) \right]^2}$ implies that 
non-negligible couplings require $\sigma_p = -\sigma_q$, indicating that, in this asymptotics, only counter-propagative waves can interact, even in the dispersive range.
Furthermore, defining $\Omega_{{\boldsymbol k}{\boldsymbol p} {\boldsymbol q}}^{\sigma \sigma_p \sigma_q} =\Omega_{-{\boldsymbol k}{\boldsymbol p} {\boldsymbol q}}^{-\sigma \sigma_p \sigma_q} $
with $\Omega_{{\boldsymbol k}{\boldsymbol p} {\boldsymbol q}}^{\sigma \sigma_p \sigma_q} $ given by Eq. (\ref{Omega}), one has
\begin{equation}
\delta (\Omega_{{\boldsymbol k}{\boldsymbol p} {\boldsymbol q}}^{\sigma_k \sigma_p \sigma_q} )= \frac{1}{v_{ph}(k_\perp)} \delta(\sigma_k k_\| + 	\sigma _p p_\| + \sigma_q q_\|).
\label{delta(Omega)}
\end{equation}
The resonance conditions, which reduce to 
\begin{eqnarray}
&& k_\|+ p_\| + q_\| = 0\\
&& \sigma_k k_\| + 	\sigma _p p_\| + \sigma_q q_\| = 0, 
\end{eqnarray}
imply
\begin{equation}
(\sigma_k - \sigma_p) p_\| + (\sigma_k -\sigma_q) q_\| = 0,
\end{equation}
and thus 
\begin{equation}
(\sigma_k -\sigma_p) p_\| + (\sigma_k + \sigma_p) q_\| = 0.\label{RC}
\end{equation}
It turns out that  only the case $\sigma_p = \sigma_k$ is relevant. Otherwise, the resonance frequency condition reduces to $\delta(p_\|) =0$ and with the factor $p_\|$ originating from  $\omega_{\boldsymbol p}^{\sigma_p}$, the integral vanishes. It follows that  $q_\|=0$ and thus $k_\| = -p_\|$\footnote{\textcolor{black}{Note that this condition leads to the same difficulty as in the MHD range when assuming strong transverse locality, even in the presence of dispersion. The interaction with the $k_\|=0$ mode being still in need of a more complete and rigorous understanding (see  e.g. the different interpretations in \citet{Lithwick03}  and \citet{Scheko12}), we resorted to use the classical formulation where the $k_\|=0$ mode is treated on equal footing as the others.}}. 
As a consequence, $ \delta (k_\|+ p_\| + q_\|)  \delta (\Omega_{{\boldsymbol k}{\boldsymbol p} {\boldsymbol q}}^{\sigma \sigma_p \sigma_q} )$ can be replaced by $(1/v_{ph})\delta(q_\|) \delta(k_\| + p_\|)$. At the level of the discrete summation, as $\sigma_q= - \sigma_p = -\sigma_k$, only the summation on $\sigma_k$ is to be carried out. Within the local approximation,
\begin{eqnarray}
&&\frac{1}{k_\perp p_\perp q_\perp}\left(\frac{\sin \gamma}{q_\perp}\right)
\left [\frac{1}{k_\|}\left (\frac{\sigma_p}{\xi(p_\perp)} -\frac{\sigma_q}{\xi (q_\perp)}\right) \right]^2 (\sigma_k k_\perp^2 \xi(k_\perp) + \sigma_p p_\perp^2 \xi(p_\perp) + \sigma_q q_\perp^2 \xi(q_\perp))^2 
\omega^{\sigma_k}_{\boldsymbol k}\omega^{\sigma_p}_{\boldsymbol p}\nonumber \\
&& \approx -2 \sqrt{3} v_{ph}^2(k_\perp)  \label{factor1}
\end{eqnarray}

Furthermore, neglecting again the contribution of parallel variations,
\begin{equation}
p_\perp E(k_\perp, p_\|)- k_\perp E(p_\perp, k_\|) \approx \varepsilon_{p_\perp} k_\perp^3  \frac{\partial}{\partial k_\perp}\left(  \frac{E(k_\perp, k_\|)}{k_\perp}\right)
\end{equation}
and $E(q_\perp, 0)$ is also approximated by $E(k_\perp, k_\|)$.

As a consequence (after taking into account the sum on $\sigma _k$), the contribution of the non-helical term in the equation for the energy spectrum reduces to 
\begin{equation}
C \frac{\partial}{\partial k_\perp} \left ( k_\perp^6 v_{ph}(k_\perp) E(k_\perp, k_\|) \frac{\partial}{\partial k_\perp} \left ( \frac{E(k_\perp, k_\|)}{k_\perp} \right) \right)
\end{equation}

with 
\begin{equation}
C = \frac{\sqrt{3}}{16} \int_{-\epsilon}^{+\epsilon} \int_{-\epsilon}^{+\epsilon} d\epsilon_{p_\perp} d\epsilon_{q_\perp} \epsilon_{p_\perp}^2 = \frac{\epsilon^4} {4\sqrt{3}}.
\end{equation}
Similarly, the helical contribution to the same equation is given by
\begin{equation}
-C \frac{\partial}{\partial k_\perp} \left ( k_\perp^6 v_{ph}^2(k_\perp) E_C(k_\perp, k_\|) \frac{\partial}{\partial k_\perp} \left ( \frac{v_{ph}(k_\perp)E_C(k_\perp, k_\|)}{k_\perp} \right) \right).
\end{equation}

The same procedure is implemented for the generalized cross-helicity spectrum. In fact, $T^{\sigma_k \sigma_p \sigma_q}_{C{\boldsymbol k}{\boldsymbol p}{\boldsymbol q}}$
does not display exact antisymmetry properties as it is the
case for $T^{\sigma_k \sigma_p \sigma_q}_{{\boldsymbol k}{\boldsymbol p}{\boldsymbol q}}$, but this property is 
asymptotically recovered in the limit of local interactions
where the dominant interactions correspond to $\sigma_k=\sigma_p= -\sigma_q$ and where  $v_{ph}$ in the denominator of the third line of Eq.(\ref{helicity-spec})
is eliminated by using Eqs. (\ref{delta(Omega)}) and (\ref{factor1}).

One finally gets the local model
\begin{eqnarray}
&&\partial_t E(k_\perp) =
C \frac{\partial}{\partial k_\perp} \Big \{ k_\perp^6 v_{ph}(k_\perp) \Big [E(k_\perp) \frac{\partial}{\partial k_\perp} \Big ( \frac{E(k_\perp)}{k_\perp} \Big) \nonumber \\
&& \qquad - v_{ph}(k_\perp)E_C(k_\perp) \frac{\partial}{\partial k_\perp} \Big ( \frac{v_{ph}(k_\perp)E_C(k_\perp)}{k_\perp} \Big)\Big ]\Big \}\label{E-loc}\\
&&\partial_t E_C(k_\perp) =
C \frac{\partial}{\partial k_\perp} \Big \{ k_\perp^6 \Big [E(k_\perp) \frac{\partial}{\partial k_\perp} \Big ( \frac{v_{ph}(k_\perp) E_C(k_\perp)}{k_\perp} \Big) \nonumber \\
&&\qquad - v_{ph}(k_\perp) E_C(k_\perp) \frac{\partial}{\partial k_\perp} \Big ( \frac{E(k_\perp)}{k_\perp} \Big)\Big ]\Big \},\label{Ec-loc}
\end{eqnarray}
where, to simplify the writing, we dropped the $k_\|$ argument of the spectra, the longitudinal transfer being assumed to be negligible in the weak-turbulence regime.

It is of interest to rewrite the above system in terms of  $E^\pm (k_\perp) = (E(k_\perp) \pm v_{ph}(k_\perp) E_C(k_\perp))/2$ which, in the MHD range, identify with  the spectral densities of the usual  Elsasser variables, while in the sub-ion range, they involve the electron velocity.
They obey
\begin{eqnarray}
\partial_t  E^\pm(k_\perp)&=& 2 C \frac{\partial}{\partial k_\perp} \left [ k_\perp^6 v_{ph}(k_\perp)  E^\mp(k_\perp) \frac{\partial }{\partial k_\perp} \left ( \frac{E^\pm(k_\perp)}{k_\perp}\right) \right]\nonumber \\
&& \mp  C \frac{\partial v_{ph}}{\partial k_\perp} k^5_\perp (E^-(k_\perp))^{2} \frac{\partial}{\partial k_\perp} \left ( \frac{E^+(k_\perp)}{E^-(k_\perp)}\right) \label{eqpm}
\end{eqnarray}
which, in contrast with the MHD limit considered by \citet{GBuch10}, is non conservative.

\section{Heuristic extension to strong turbulence}
\label{strong_turbulence}

\textcolor{black}{\subsection{Estimates of the characteristic times}}

The aim of this Section is to construct a  model for strong Alfv\'en-wave turbulence, by phenomenologically adapting the above weak turbulence model. 
A main difference between weak and strong turbulence regimes  concerns the transfer time. In  weak turbulence, it is given by \textcolor{black}{$\tau_{tr,w} = \tau^2_{NL,w} \omega_L$ where $\tau_{NL,w}$ is the nonlinear time and $\omega_L$ the frequency of the linear waves at the corresponding wave vector.} The latter is given by $\omega_L= v_{ph}k_\|$ where, in this regime, $k_\|$ can be viewed as essentially constant and taken equal to the injection wavenumber $k_f$, as longitudinal transfer is negligible. 
Inspection of Eq. (\ref{eqpm}) suggests estimating the transfer time of $E^\pm$ as $\tau^\pm_{tr,w}= (k^3_\perp v_{ph} E^\mp)^{-1}$ and thus the corresponding nonlinear time by $\tau^\pm_{NL,w}= (k^3_\perp v^2_{ph} k_\|E^\mp)^{-1/2}$.

In the strong turbulence regime, parallel transfer, while remaining small, is no longer negligible, so that $k_\|$ cannot be assumed constant anymore. \textcolor{black}{Neglecting the effect of dynamical alignment and the formation of sheet-like structures,} it seems reasonable in this case to  replace  the nonlinear time $\tau^\pm_{NL,w}$ of the weak turbulence model by $\tau^{\pm}_{NL,st}=(k^3_\perp v^2_{ph}{\widebar E}^\mp)^{-1/2}$, where
 ${\widebar E}^\mp = \int^{+\infty}_{-\infty}  E^{\mp}(k_\perp, k_\|) dk_\|$. 
 This time is characteristic of the stretching by the transverse electron velocity in  balanced strong turbulence  \citep{PS15, PST18,Cranmer03} and is also consistent with that given by \cite{Lithwick07} for  imbalanced  strong  MHD turbulence. \textcolor{black}{In all the regimes,}
the transfer time $\tau_{tr}^\pm$  can be written \citep{Matthaeus09,PS15}
\begin{equation}
(\tau_{tr}^\pm)^{-1} = \frac{(\tau_{NL,st}^\pm)^{-2}}{v_{ph}{\widetilde k}_\|^\mp },
\end{equation}
where ${\widetilde k}_\|^\pm$ is the typical inverse parallel correlation length of a $\pm$ magnetic eddy. 
If both waves undergo a strong cascade, ${\widetilde k}_\|^\pm \approx  k_\|^\pm \equiv (k_\perp^3 {\widebar E}^\pm)^{1/2}$, which reproduces the transfer time of \citet{Lithwick07}.
In fact, in a strongly imbalanced regime, the most energetic wave (hereafter the $+$ wave), will affect the parallel correlation length of the $-$ wave in such a way that ${\widetilde k}_\|^-$ becomes larger than $(k_\perp^3 {\widebar E}^-)^{1/2}$. We model this effect by changing the definition of ${\widetilde k}_\|^-$  into ${\widetilde k}_\|^-= (k_\perp^3 {\widebar E}^-)^{1/2}({\widebar E}^+/{\widebar E}^-)^{\nu/2}$ (with $0\le \nu \le 1$) in order to ensure a weaker cascade for the $+$ wave than for the $-$ wave. The case $\nu=1$ corresponds to the equality of the two parallel correlation lengths, as in the  model of \citet{Chandran08}. 
This value of $\nu$ does not seem to be supported by numerical simulations where arbitrarily large ratios $\varepsilon^+/\varepsilon^-$ of the $E^\pm$ injection and thus transfer rates can be prescribed by supplementing appropriate random drivings on the Elsasser variables \citep{Beresnyak09}. The model of \citet{Beresnyak08}, which consists in taking for ${\widetilde k}_\|^-$ the geometrical average $(k_\|^+ k_\|^-)^{1/2}$, corresponds to $\nu=1/4$ in our formulation.
As will be argued below, the case $\nu<1$ can provide a better modeling for the situations where transfer rates are imposed, but they turn out to lead to unphysical behavior in the presence of dispersion. This difficulty could result from the conflict between the prescription of a direct helicity flux with the tendency of the system to develop \textcolor{black}{ an inverse helicity cascade} in the dispersive range \citep{PST18}.   
Alternatively to the prescription of $\varepsilon^+$ and $\varepsilon^-$, one can consider a problem where 
the spectra $E^\pm(k_\perp)$ are fixed at the outer scale, the fluxes adjusting to accommodate the small-scale behavior. In this case, we show in the following that the model  with $\nu=1$ provides an adequate modeling. The observation that in the problem with prescribed  energy fluxes one has to take $\nu<1$, while $\nu=1$ is more suitable for situations where the energy ratio is fixed at the outer scale can be argued as follows. In the former case, the wave driving maintains the characteristics of each type of  waves, while in the latter case the frequency (or wavenumber) of the $-$ wave is affected by the interaction with the $+$ wave, with a much weaker influence of the external constraint. In the solar wind, the quantities $\varepsilon^+$ and $\varepsilon^-$ are in fact measured to be of the same order of magnitude \citep{Carbone09,Marino09}. Correlating the measure of   $\varepsilon^+/\varepsilon^-$ with that of $E^+/E^-$ at the outer scale would be of great interest.

In order to provide a formalism permitting the description of both weak and strong turbulence regimes, and in particular the transition from the former to the latter at small scales, we further extend the definition of the effective parallel wavenumber as 
\begin{equation}
{\widetilde k}_\|^{(r)}=k_f + (k_\perp^3 {\widebar E}^{(r)})^{1/2}({\widebar E}^+/{\widebar E}^-)^{(1-r)\nu/4},
\end{equation} where $k_f$ is a typical parallel wavenumber at the outer scales (or at scales where energy is injected). Here $(r)=\pm 1$ and the weak turbulence limit is recovered when ${\widebar E}^\pm$ is small enough so that ${\widetilde k}_\|^\pm\approx k_f$. 
\textcolor{black}{Extension of the transfer time formula for retaining nonlocal interactions can easily be done
	using integral formulations as in \citet{PS15} (see also \citet{Pouquet76}). As previously mentioned, such nonlocal interactions are important in the presence of dissipation. They are required in order to obtain a correct dissipation range when extending the Leith equation for hydrodynamic turbulence to scales where viscosity is relevant \citep{Clark09}}.

\subsection{\textcolor{black}{Fully three-dimensional dynamics}}  
In order to account for the nonlinear transfer along the direction of the  ambient magnetic field, a parallel diffusion is also added to Eq. (\ref{eqpm}). Following \cite{Cranmer03} and \cite{Chandran08}, we introduce a term of the form $C_\| (\tau^{\pm}_{tr})^{-1} {\widetilde k_\|}^{2} \left ( \partial^2 E^\pm(k_\perp, k_\|)/\partial k_\|^2\right )$ where $C_\|$ is a constant of the same order of magnitude as $C$ and ${\widetilde k_\|}=\max({{\widetilde k_\|}^+},{\widetilde k_\|}^-)$. In the strong turbulence regime where $k_f$ is negligible with respect to $(k_\perp^3 {\widebar E}^\mp)^{1/2}$, the ratio  ${\widetilde k_\|}^\pm/k_\|$ reduces to the nonlinearity parameter of the $\pm$ wave for $\nu=0$ and of the $+$ wave for $\nu=1$.
This leads to the system
\begin{eqnarray}
\partial_t  E^\pm(k_\perp, k_\|) &=& 2 C' \frac{\partial}{\partial k_\perp} \left [ \frac{k^{6}_\perp v_{ph} {\widebar E}^\mp}{{\widetilde k}_\|^\mp} \frac{\partial }{\partial k_\perp} \left ( \frac{E^\pm(k_\perp, k_\|)}{k_\perp}\right) \right] \nonumber \\
&&\mp  C' \frac{\partial v_{ph}}{\partial k_\perp}k_\perp^{6} \left[\frac{{\widebar E}^-}{{\widetilde k}_\|^-} \frac{\partial }{\partial k_\perp} \left ( \frac{E^+(k_\perp, k_\|)}{k_\perp}\right)-
\frac{{\widebar E}^+}{{\widetilde k}_\|^+}  \frac{\partial }{\partial k_\perp} \left ( \frac{E^-(k_\perp, k_\|)}{k_\perp}\right) \right] \nonumber \\
&&+ C_\|  v_{ph}  \frac{k^{3}_\perp  {\widebar E}^\mp} {{\widetilde k}_\|^\mp} {\widetilde k}_\|^2 \frac{\partial^2}{\partial k^2_\|}E^\pm(k_\perp, k_\|) .          \label{strongpm}
\end{eqnarray}
The constant $C'$ is a priori different from the constant $C$ that appears in Eqs. (\ref{E-loc})-(\ref{Ec-loc}). Continuity between the two approaches is ensured when  $C'=C k_f$.
In the case of strong turbulence (with $k_f=0$), it may be of interest to look for a special class of solutions of Eq. (\ref{strongpm}), of the form
\begin{equation}
E^\pm (k_\perp,k_\|)=\frac{{\widebar E}^\pm(k_\perp)}{\widetilde k_\|}f^\pm (\zeta) 
\end{equation}
where $\zeta=k_\|/{\widetilde k_\|}$ and the functions $f^\pm$ can be chosen such that $\int_{-\infty}^{+\infty} f^\pm (\zeta)d\zeta=1$.
Assuming power laws, it is easily seen that, when $\nu\ne 1$,  the existence of separable solutions requires equal exponents for the two spectra, both in the MHD and the dispersive regimes. In contrast, no condition holds when $\nu=1$. Such self-similar solutions were analyzed in the MHD case by \citet{Cranmer03} and by \citet{Chandran08} in a model similar  to the present one when $\nu=1$.

\subsection{\textcolor{black}{Transverse dynamics}}
In the following, we concentrate on the transverse dynamics and thus integrate Eq. (\ref{strongpm})
over $k_\|$. Dropping the  overbar, we get, for the spectra integrated over $k_\|$,
\begin{eqnarray}
\partial_t  {E}^\pm(k_\perp) &=& 2 C' \frac{\partial}{\partial k_\perp} \left [ \frac{k^{6}_\perp v_{ph} {E}^\mp(k_\perp)}{{\widetilde k}_\|^\mp} \frac{\partial }{\partial k_\perp} \left ( \frac{E^\pm(k_\perp)}{k_\perp}\right) \right] \nonumber \\
&&\mp  C' \frac{\partial v_{ph}}{\partial k_\perp}k_\perp^{6} \left[\frac{{E}^-(k_\perp)}{{\widetilde k}_\|^-} \frac{\partial }{\partial k_\perp} \left ( \frac{E^+(k_\perp)}{k_\perp}\right)-
\frac{{E}^+(k_\perp)}{{\widetilde k}_\|^+}  \frac{\partial }{\partial k_\perp} \left ( \frac{E^-(k_\perp)}{k_\perp}\right) \right]. \label{eq:modelEpm}
\end{eqnarray}

These equations can be rewritten in a conservative form by using  the energy and generalized cross helicity spectra $E(k_\perp) = E^+ (k_\perp)+ E^- (k_\perp)$ and $E_C(k_\perp) = (E^+ (k_\perp)- E^- (k_\perp))/v_{ph}$.
One gets
\begin{eqnarray}
\partial_t E(k_\perp) &=& \frac{C'}{2} \frac{\partial}{\partial k_\perp}
\Big \{ k^{6}_\perp v_{ph}\Big [\frac{E(k_\perp)-v_{ph}E_C(k_\perp)}{{\widetilde k}^-_\|} \frac{\partial}{\partial k_\perp}  \Big(\frac{E(k_\perp)+v_{ph}E_C(k_\perp)}{k_\perp}\Big)\nonumber \\
&&+\frac{E(k_\perp)+v_{ph}E_C(k_\perp)}{{\widetilde k}^+_\|} \frac{\partial}{\partial k_\perp}  \Big (\frac{E(k_\perp)-v_{ph}E_C(k_\perp)}{k_\perp}\Big)\Big]\Big\}\equiv - \frac{\partial \varepsilon}{\partial k_\perp}  \label{strong-energy}\\
\partial_t E_C(k_\perp) &=& \frac{C'}{2} \frac{\partial}{\partial k_\perp}
\Big \{ k^{6}_\perp \Big [\frac{E(k_\perp)-v_{ph}E_C(k_\perp)}{{\widetilde k}^-_\|} \frac{\partial}{\partial k_\perp}  \Big(\frac{E(k_\perp)+v_{ph}E_C(k_\perp)}{k_\perp}\Big)\nonumber \\
&&-\frac{E(k_\perp)+v_{ph}E_C(k_\perp)}{{\widetilde k}^+_\|} \frac{\partial}{\partial k_\perp}  \Big (\frac{E(k_\perp)-v_{ph}E_C(k_\perp)}{k_\perp}\Big)\Big]\Big\} \equiv - \frac{\partial \eta}{\partial k_\perp} 
 \label{strong-helicity}
\end{eqnarray}
\textcolor{black}{where $\varepsilon$ and $\eta$  refer to the energy and the general cross-helicity fluxes.} 
These equations provide a model for imbalanced turbulence, either weak or strong, depending on the definition of ${\widetilde k}_\|^\pm$. Equation (\ref{eq:modelEpm}) shows that this model only retains  interactions between counter-propagating waves, not only in the MHD range but also at the dispersive scales. A discussion on the possibility of including co-propagating wave interactions is given in Section \ref{sec:coprop}. 

\section{Transverse cascades}\label{transverse-spectra}

In this Section, finite flux stationary solutions of Eqs. (\ref{strong-energy})-(\ref{strong-helicity}) are discussed, first in the weak turbulence limit where analytical solutions can be found, and then in the case of strong turbulence where a numerical integration of the governing  ordinary differential equations is needed. 
They correspond to energy and/or generalized cross-helicity cascades where  $\varepsilon$ and $\eta$ are constant and satisfy
\begin{eqnarray}
&& -\frac{\varepsilon}{2C'} =k^6_\perp v_{ph}\left ( \frac{E^-(k_\perp)}{{\widetilde k}^-_\|} 
\frac{\partial}{\partial k_\perp}\frac{E^+(k_\perp)}{k_\perp}  +\frac{E^+(k_\perp)}{{\widetilde k}^+_\|} \frac{\partial}{\partial k_\perp}\frac{E^-(k_\perp)}{k_\perp} \right)\\
&&-\frac{\eta}{2C'} =k^6_\perp \left ( \frac{E^-(k_\perp)}{{\widetilde k}^-_\|} 
\frac{\partial}{\partial k_\perp}\frac{E^+(k_\perp)}{k_\perp}  -\frac{E^+(k_\perp)}{{\widetilde k}^+_\|} \frac{\partial}{\partial k_\perp}\frac{E^-(k_\perp)}{k_\perp} \right),
\end{eqnarray}
or, defining $u^\pm(k_\perp)=E^\pm(k_\perp)/k_\perp$,
\begin{equation}
\frac{d}{d k_\perp} u^\pm(k_\perp)=-\frac{\varepsilon\pm\eta v_{ph}}{4C'k^7_\perp v_{ph}}\frac{{\widetilde k}^\mp_\|}{u^\mp(k_\perp)}. \label{eq:upm}
\end{equation}
When $\varepsilon=\eta=0$, these equations  have solutions in the form of thermodynamic spectra $E^\pm(k_\perp)\sim k_\perp$ and thus $E(k_\perp)\sim k_\perp$ and $E_C(k_\perp)\sim k_\perp v^{-1}_{ph}(k_\perp)$.

Equation (\ref{eq:upm}) can be considered either as an initial value problem when the energy and generalized cross-helicity fluxes $\varepsilon$ and $\eta$  are given and the spectra $E^\pm$  prescribed at a wavenumber $k_\perp=k_0$, or alternatively as
a nonlinear eigenvalue problem for $\varepsilon$ and $\eta$ where the spectra are specified at  $k_0$ and prescribed to decrease to zero at infinity. In the simple case of the original Leith equation for three-dimensional hydrodynamic turbulence which involves only the energy flux (Appendix \ref{app1}), the two problems are equivalent, but this is not necessarily the case in the present context.
It is indeed immediately seen that Eq. (\ref{eq:upm}) does not have a satisfactory solution when $\eta\ne 0$. The linear growth of the phase velocity $v_{ph}$ at small scales implies that for $k_\perp$ large enough, depending on the sign of $\eta$, one of the spectra $E^+$ or $E^-$ increases with $k_\perp$. The saturation of $v_{ph}$ due to electron inertia is not sufficient, as an acceptable solution would strongly constrain the helicity flux $\eta$. Thus, both in weak and strong turbulence, simultaneous cascades of energy and generalized cross-helicity cannot extend to arbitrary small scales. This observation is not really surprising since, as mentioned in \cite{PST18}, an inverse generalized cross-helicity cascade is expected in the dispersive range. Such a cascade can possibly be captured in the framework of the time evolution problem with an injection at small scale, as the system could adjust to ensure a  zero helicity flux  towards large wavenumbers. This issue is beyond the scope of the present paper.

\subsection{Weak turbulence regime} \label{cascades}

In the weak turbulence case,  it is convenient to  turn back to 
Eqs. (\ref{E-loc})-(\ref{Ec-loc}). 
In the  energy cascade, one has 
\begin{equation}
-\varepsilon = \frac{C}{2}  k^7_\perp v_{ph}(k_\perp) \frac{\partial}{\partial k_\perp} \left [
\left( \frac{E(k_\perp)}{k_\perp}\right )^2 - \left( \frac{v_{ph}(k_\perp) E_C(k_\perp)}{k_\perp})\right )^2 \right ],\label{eq:energycascade}
\end{equation} 
while in the generalized cross-helicity cascade, 
\begin{equation}
-\eta= C k^7_\perp \left[ \frac{E(k_\perp)}{k_\perp}\frac{\partial}{\partial k_\perp}
\left (\frac{v_{ph}(k_\perp)E_C(k_\perp)}{k_\perp}\right)-
\frac{v_{ph}(k_\perp)E_C(k_\perp)}{k_\perp}\frac{\partial}{\partial k_\perp}
\left (\frac{E(k_\perp)}{k_\perp}\right)\right].\label{eq:helicitycascade}
\end{equation}

We first consider these equations independently as governing energy and generalized cross helicity cascades, 
like in \citet{Galtier15} who focus on the inverse cascade of helicity with no constraint on the energy flux (the  problem being in this case only locally stationary).
Away from the transition zone between the MHD and the sub-ion ranges, the phase velocity is a power law $v_{ph}(k_\perp) \sim k_\perp^{m_v}$, so that we can  consider power law solutions for the energy and generalized cross-helicity spectra in the form $E(k_\perp)\sim k^{-m}$ and $E_C(k_\perp)\sim k^{-m_c}$.   
One immediately gets $m=2+m_v/2$ and $m_c=2+3m_v/2$ in the energy cascade (where we find $\eta=0$), consistently with the predictions of \citet{Galtier03}. Differently, in the helicity cascade, we only get the entanglement relation $m+m_c=4+m_v$, which is consistent with a power counting argument performed on Eqs. (B7) and (B8) of \citet{Galtier03}
\footnote{\textcolor{black}{Using Eq. (C3) of \citet{Galtier15} (obtained by applying  the Zakharov transformation to the weak-turbulence kinetic equation), together with a corrected form of Eq. (C2) involving  a cartesian form of the divergence operator (due to  the fact that $H_k$ is the helicity spectral density per perpendicular and parallel wavenumbers), one finds that the helicity flux  associated to this entanglement solution is logarithmic, while it is constant for the diffusion model. This indicates that in the helicity cascade spectra deviate from pure power laws.  Similar self-similarity breaking arises in the enstrophy cascade of two-dimensional Navier-Stokes turbulence where nonlocal interactions lead to a logarithmic correction in the energy spectrum \citep{Kraichnan71}. Specifying the correcting factors in the present problem requires further analysis of the kinetic equations.}}
		
We now turn to the case of simultaneous cascades in order to study the possible continuation in the dispersive range of the direct energy and cross-helicity cascades that are expected  in the MHD range.
Excluding the case where $E(k_\perp)=|v_{ph}(k_\perp)E_C(k_\perp)|$ (i.e. either $E^+(k_ \perp)=0$ or $E^-(k_\perp)=0$), a situation where $\varepsilon=\eta=0$ (due to the assumption of ultra-locality which only selects interactions between counter-propagating waves), we can write
\begin{eqnarray}
&&\frac{E(k_\perp)}{k_\perp}=\rho(k_\perp)\cosh\phi(k_\perp)\\
&&\frac{v_{ph}(k_\perp)E_C(k_\perp)}{k_\perp}=\rho(k_\perp)\sinh\phi(k_\perp),
\end{eqnarray}
and thus
\begin{equation}
E^\pm(k_\perp)=\frac{1}{2}k_\perp \rho(k_\perp)e^{\pm\phi(k_\perp)}.
\end{equation}
Note that the condition
\begin{equation}
|E_C(k_\perp)| \le \frac{E(k_\perp)}{v_{ph}(k_\perp)}, \label{ineqEcE}
\end{equation} 
prescribed by the definition of these spectra, is automatically satisfied.

The system rewrites
\begin{eqnarray}
&& \frac{\partial}{\partial k_\perp}\rho^2(k_\perp)=-\frac{2\varepsilon}{Ck^7_\perp v_{ph}(k_\perp)}\label{rho}\\
&&\rho^2(k_\perp)\frac{\partial}{\partial k_\perp}\phi(k_\perp)=-\frac{\eta}{C k^7_\perp}. \label{phi}
\end{eqnarray}
Several regimes are then to be distinguished.

\subsubsection{Zero generalized cross-helicity flux}
When $\eta=0$, we can  prescribe vanishing spectra at infinity, and solve as
\begin{equation}
\rho^2(k_\perp)=\frac{2\varepsilon}{C}\int_{k_\perp}^\infty\frac{1}{k'^7_\perp v_{ph}(k'_\perp)} dk'_\perp , \label{eq:rho2}
\end{equation}
with $\phi=\phi_0$ given by $\tanh(\phi_0)=v_{ph}(k_0) E_c(k_0)/E(k_0)$.
The spectra thus also behave like power laws with, for  spectra $E^\pm(k_\perp)\sim k_\perp^{-m^\pm}$, the relations $m^+=m^-=2+m_v/2$ and $m_c=2+3m_v/2$. The ratio $E^+_\perp(k)/E^-_\perp(k)=e^{2\phi_0}=(1+v_{ph}(k_0)E_C(k_0)/E(k_0))/(1-v_{ph}(k_0)E_C(k_0)/E(k_0))$ is independent of the wavenumber. 

\subsubsection{Finite generalized cross helicity flux}
Let us first address the case of pure MHD (i.e. with a constant $v_{ph}$ equal to $s$), for which solutions can be continued to infinity, making Eq. (\ref{eq:rho2}) valid.  The energy flux $\varepsilon$ is  necessarily non zero and we find $\rho^2(k_\perp)= \varepsilon/(3Cs) k^{-6}_\perp$. 
One then obtains, when integrating from a wavenumber $k_0$  to $k_\perp$,
\begin{equation}
\phi=\phi_0-\int_{k_0}^{k_\perp} \frac{\eta}{C k'^7_\perp \rho^2(k'_\perp)}dk'_\perp=\phi_0-\frac{3\eta s}{\varepsilon}\ln \frac{k_\perp}{k_0},
\end{equation}
and thus
\begin{equation}
E^\pm(k_\perp)=\frac{1}{2}\left (\frac{\varepsilon}{3Cs} \right )^{1/2}e^{\pm\phi_0}k_\perp^{-2} \left (\frac{k_\perp}{k_0}\right )^{\mp {3\eta s}/{\varepsilon}}. \label{eq:sppm-weak}
\end{equation}
Writing  $\varepsilon^\pm=(\varepsilon\pm\eta s )/2$, 
one gets
\begin{eqnarray}
m^+ &=& \frac{5\varepsilon^+ -\varepsilon^-}{\varepsilon^+ + \varepsilon^-}\label{eq:mp}\\
m^- &=& \frac{5\varepsilon^- -\varepsilon^+}{\varepsilon^+ + \varepsilon^-},\label{eq:mm}
\end{eqnarray}
which satisfy the entanglement condition $m^+ + m^- =4$. In this case, the nonlinear eigenvalue problem does not have a unique solution. The spectra can a priori intersect at a wavenumber which can only be determined by additional physical effects. For example, in the presence of viscosity, the pinning condition that both spectra become equal at the dissipation scale (taken to be the same for the two waves), and remain so in the whole dissipation range, selects a unique solution,  as  discussed in \citet{Lithwick03} and \citet{Chandran08}.
Note also that there is a constraint on the fluxes for the spectral exponents to remain in a range ensuring locality of the interactions \citep{Galtier15}.

Let us now turn to the more delicate situation where both helicity flux and dispersion are present.
At small scales, where $v_{ph}(k_\perp)\sim k_\perp $, assuming again that $\rho$ vanishes at infinity, $\rho^2(k_\perp)\sim \varepsilon k^{-7}_\perp$, and thus $\phi(k_\perp)\sim a+b k_\perp$ where $a$ and $b$ are constants, the latter being negative and proportional to $\eta/\varepsilon$. It results that $E^\pm (k_\perp)\sim k^{-5/2}_\perp e^{\pm b k_\perp}$. These spectra are not physically relevant in the entire spectral range, as one of the solutions diverges as $k_\perp\rightarrow +\infty$.  In fact, the phase velocity should saturate at $k_\perp d_e\ge 1$ due to electron inertia, so that the exponential behavior of the spectra is only a transient that can be acceptable if $\eta/\varepsilon$ and the saturated value of $v_{ph}$ are  sufficiently small.
The above behavior nevertheless indicates an exponential decrease of the imbalance at scales where the system starts  becoming dispersive.
To support this point we calculate from Eqs. (\ref{eq:upm}) the local slopes of the energy spectra defined as $m^\pm(k_\perp)=-d\ln(E^\pm(k_\perp))/d\ln(k_\perp)$ and find
\begin{equation}
m^\pm(k_\perp)=\frac{\varepsilon\pm \eta v_{ph}}{4C'k^6_\perp v_{ph}}\frac{{\widetilde k}^\mp_\|}{u^+(k_\perp)u^-(k_\perp)}-1.
\end{equation}
We thus have 
\begin{equation}
m^+(k_\perp)-m^-(k_\perp)=\frac{2\eta v_{ph} k_f}{4C'k^6_\perp v_{ph}u^+(k_\perp)u^-(k_\perp) },
\end{equation}
which is positive (zero  when $\eta=0$ and growing with $k_\perp$ otherwise), showing that the spectrum of the dominant wave is steeper than that of the smaller-amplitude one and, as a consequence, that the two spectra will cross each other.

\subsubsection{Numerical solutions}
In order to illustrate the previous discussion, we  numerically integrated Eqs. (\ref{eq:upm}). Since the original nonlinear eigenvalue problem does not have solutions in the general case, we resorted to prescribe both the transfer rates $\varepsilon$ and $\eta$ and the  values of $u^\pm$ at a finite wavenumber $k_d$. 
Inspection of Eqs. (\ref{rho})-(\ref{phi}) shows that the value of $\rho$ at $k_d$ should be chosen large enough to prevent the occurrence of singularities. In this case, for $k_\perp> k_d$, $\phi$ becomes constant, leading to absolute equilibrium spectra. The obtained solution  is  similar to a warm cascade \citep{CN04}. Nevertheless, while the range of absolute equilibrium can be suppressed by an accurate prescription of the energy flux in the Kolmogorov cascade (see Appendix \ref{app1}), in the present situation the ultraviolet divergence cannot be prevented.
From a physical point of view,  $k_d$ can be viewed as mimicking the  wavenumber where dissipation starts acting, the model being unable 
to describe the regime of arbitrarily small dissipation. 
At this dissipation scale, pinning of the two spectra is often reported in numerical simulation of viscous- diffusive MHD \citep{Perez12}.

\textcolor{black}{It turns out that both in the weak and strong turbulence regimes, there exists a critical value $u_c=r(\varepsilon/k_d^7)^{1/2}$ of $u^\pm(k_d)$ (with a coefficient $r$  of order unity that we empirically estimate by a dichotomy process) such that the small-scale spectrum displays  a thermodynamic behavior or a singularity depending on whether $u^\pm(k_d)$ is chosen above or below $u_c$.}
	
In all  \textcolor{black}{the simulations discussed below}, we prescribe $\beta_e=2$\footnote{In the absence of electron inertia, $\beta_e$ is not constrained to be small.}, $\tau=1$ and zero electron inertia. It turns out that using Bessel functions in the definition of  $v_{ph}$ leads to a strong increase in the computer time. We thus replaced the function $\Gamma_0(x)$ by its usual Pad\'e form $1/(1+x)$ and use for $\Gamma_1(x)$ the function 
$(x/2) (1+0.8 x^2)^{-1}$. Both functions provide the correct asymptotic behavior at $x=0$ and  decay for large $x$. The coefficient $0.8$ was adjusted to provide a good global fit at the level of the function $v_{ph}$, with a maximal error less than $3\%$, localized near $x=1$.

\begin{figure}
	\centerline{
		\includegraphics[width=0.32\textwidth]{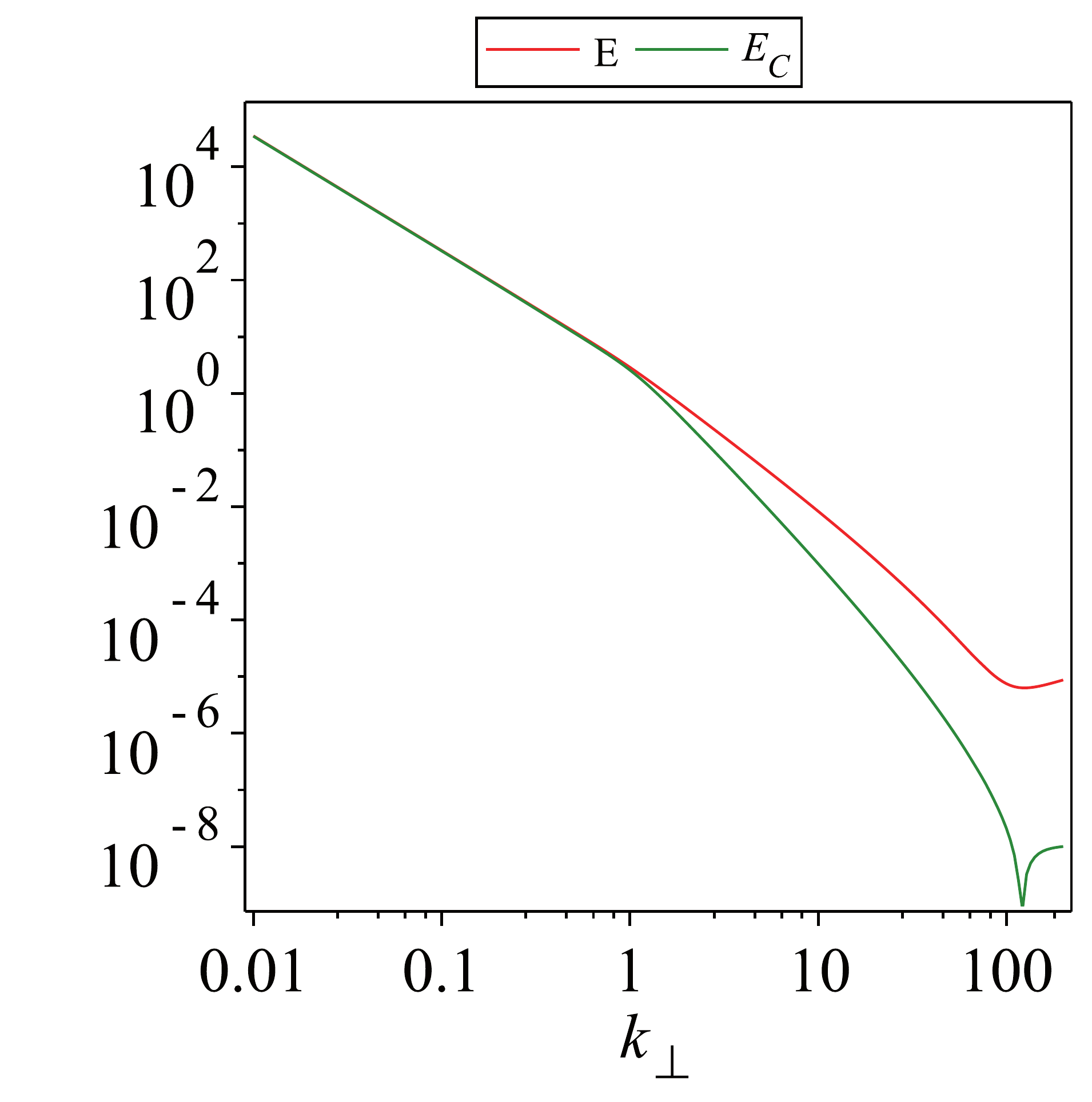}
		\includegraphics[width=0.32\textwidth]{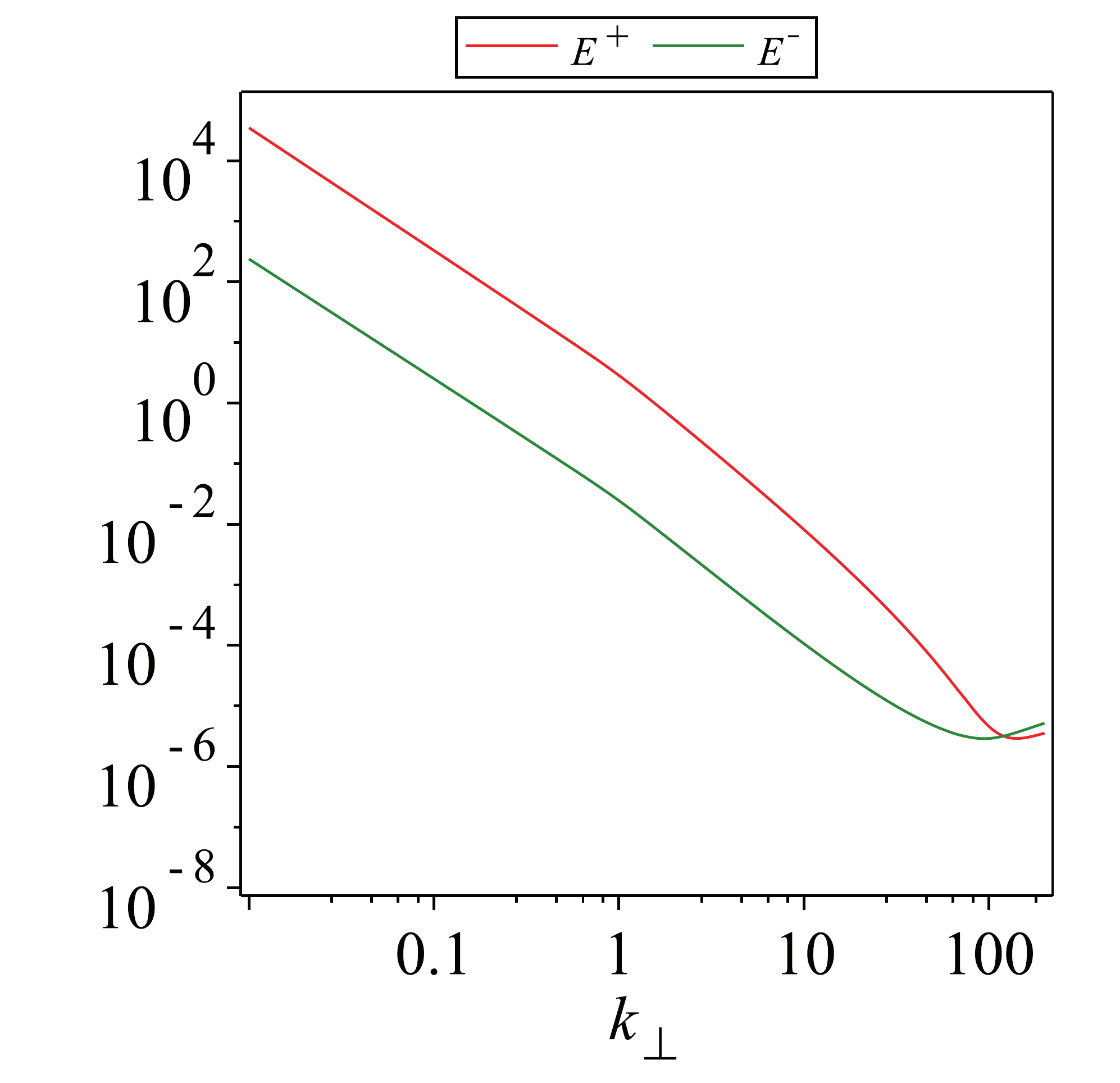}
	    \includegraphics[width=0.32\textwidth]{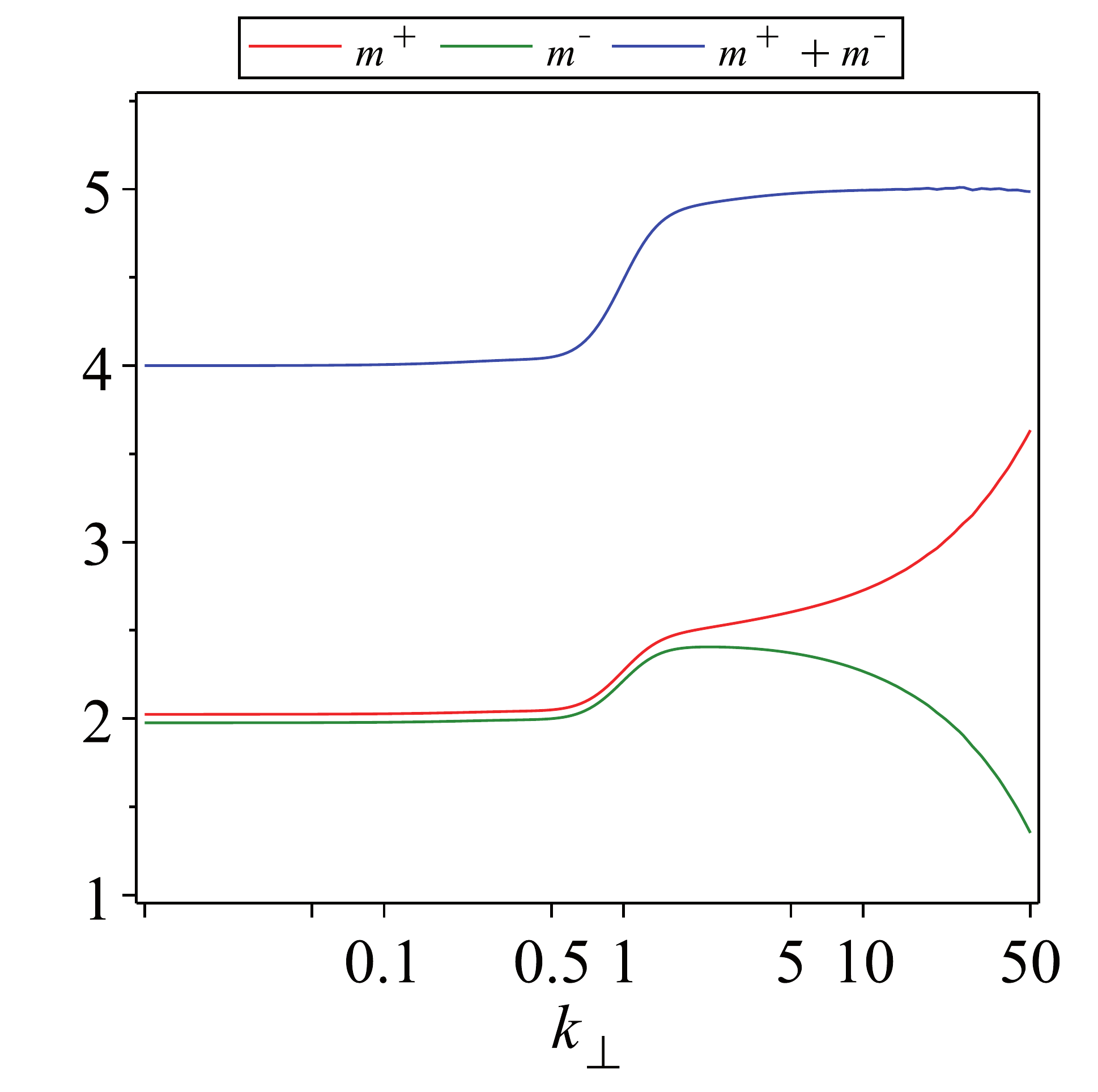}}
	\caption{Left: Spectra $E(k_\perp)$ (red) and $|E_C(k_\perp)|$ (green); Middle:  spectra $E^+(k_\perp)$ (red) and $E^-(k_\perp)$ (green); 
	Right: 	Slopes $m^+(k_\perp)$ (red), $m^-(k_\perp)$ (green) and $m^++m^-$ (blue). The simulation was performed with  $\beta_e=2$, $\tau=1$ in the weak turbulence regime in the case of a weak transfer rate of  generalized cross helicity  $\eta/\varepsilon = 0.008$, but a relatively large value of $k_d=120$. Other parameters are  ${\widetilde k^\pm_\|=1}$, $C'=1$, $\varepsilon=1$,  $u^+(k_d)=u^-(k_d)=(1/2)(\varepsilon/k_d^7)^{1/2}$.  }
	\label{fig:Weak1}
\end{figure}

We show,  in Fig. \ref{fig:Weak1},  $E(k_\perp)$ (red solid line) and $|E_C(k_\perp)|$ (green solid line) (left panel), as well as $E^+(k_\perp)$ (red) and $E^-(k_\perp)$ (green) (middle panel) for a case of weak  generalized cross-helicity transfer rate  ($\eta/\varepsilon = 0.008$), but  relatively large  $k_d=120$. Additional parameters are ${\widetilde k^\pm_\|=1}$, $C'=1$, $\varepsilon=1$ and  $u^+(k_d)=u^-(k_d)= (1/2)(\varepsilon/k_d^7)^{1/2}$. 
\textcolor{black}{The value $1/2$ of the coefficient slightly exceeds the critical value $r$.}     
The right panel displays the local slopes $m^+(k_\perp)$ and $m^-(k_\perp)$, together with their sum, showing in particular that, beyond the transition range, where the slopes reach 2.5,  the sub-ion range is dominated by an exponential zone. The sum of the slopes is $4$ in the MHD range and $5$ in the sub-ion one.  Close to $k_d$,  the cross-helicity spectrum  becomes negative. Note that changing the sign of $\eta$ leads to exchange $E^+(k_\perp)$ and $E^-(k_\perp)$ and thus the sign of $E_C(k_\perp)$.

\begin{figure}
	\centerline{
		\includegraphics[width=0.32\textwidth]{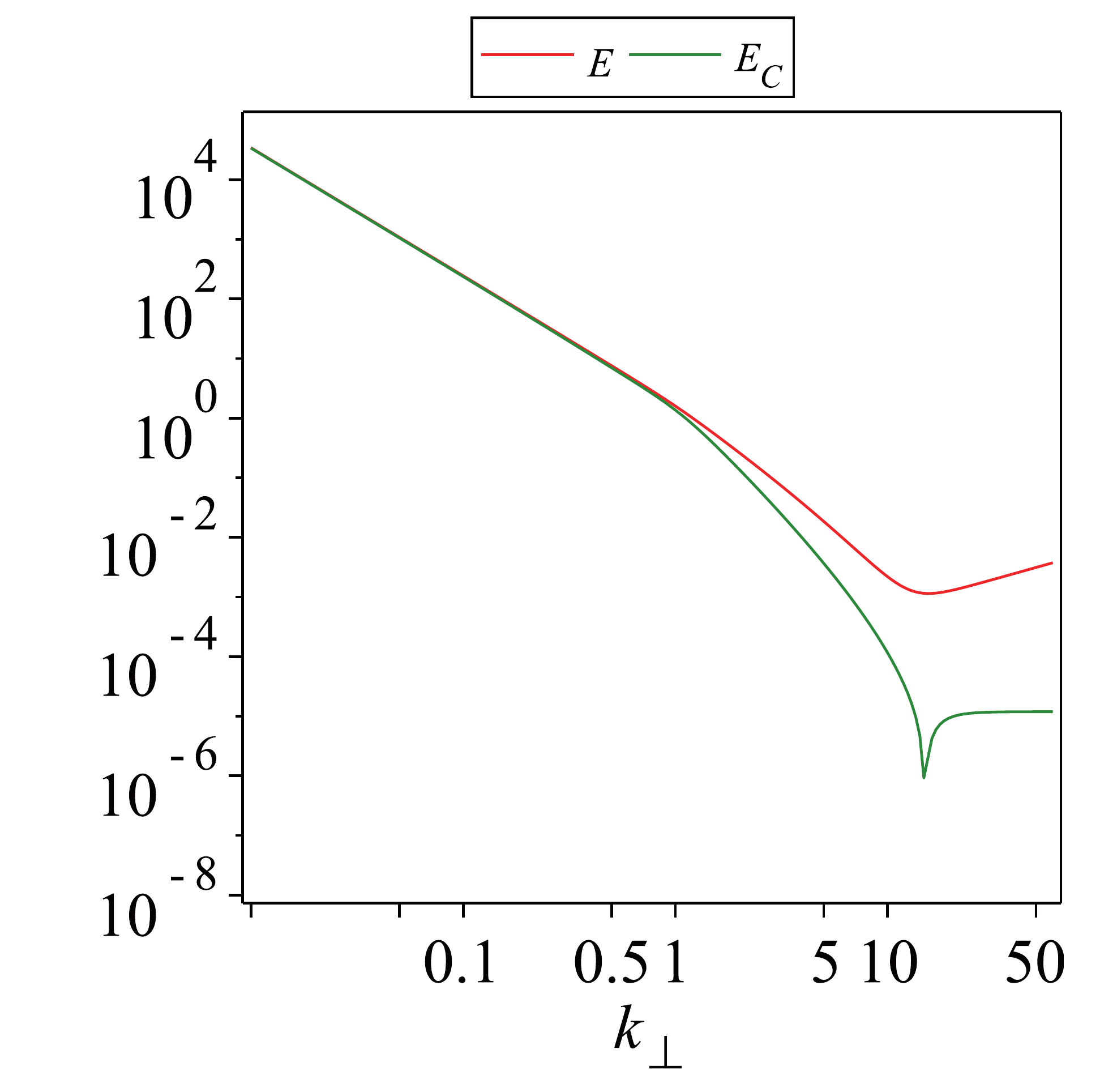}
		\includegraphics[width=0.32\textwidth]{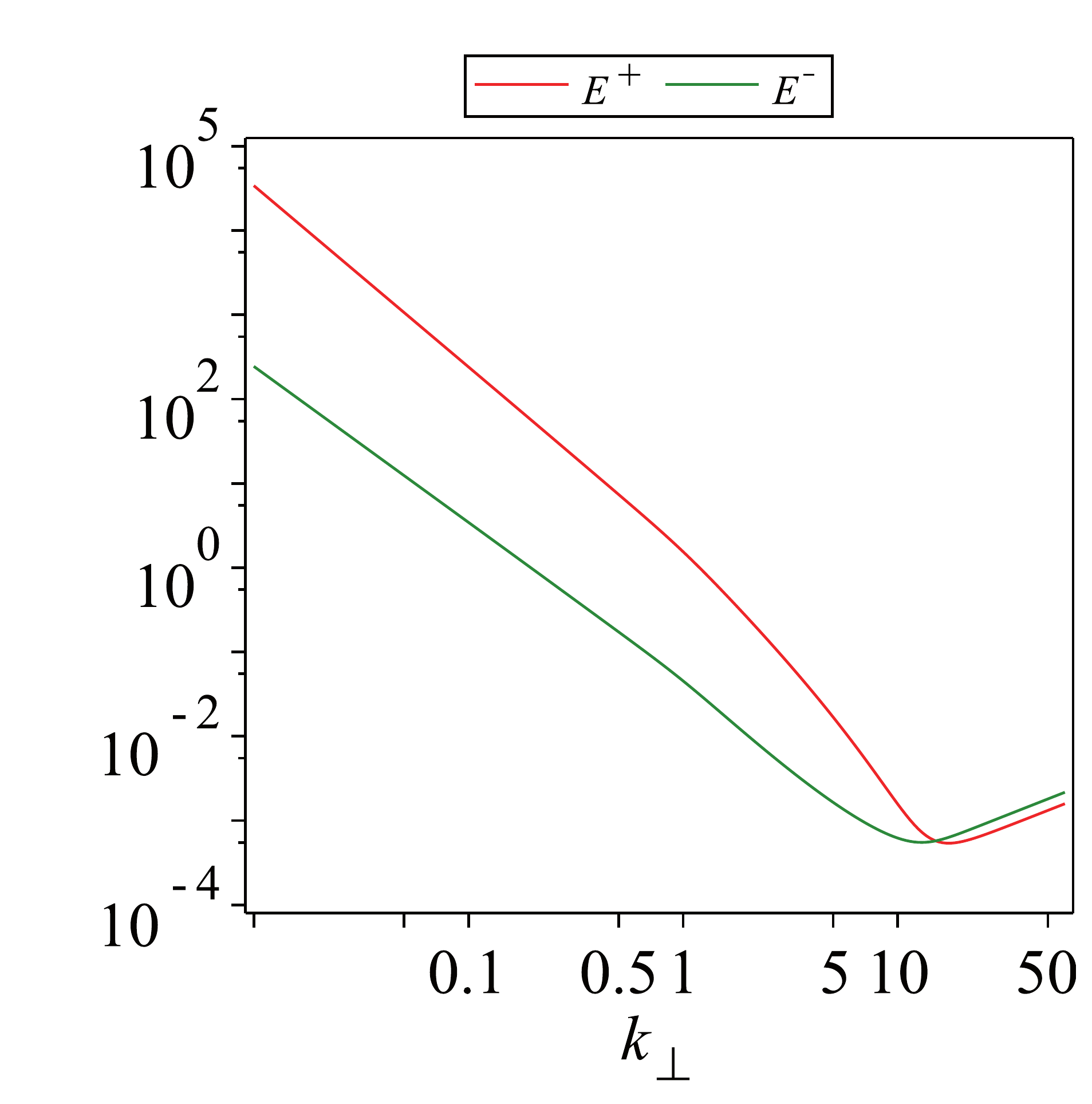}
		\includegraphics[width=0.32\textwidth]{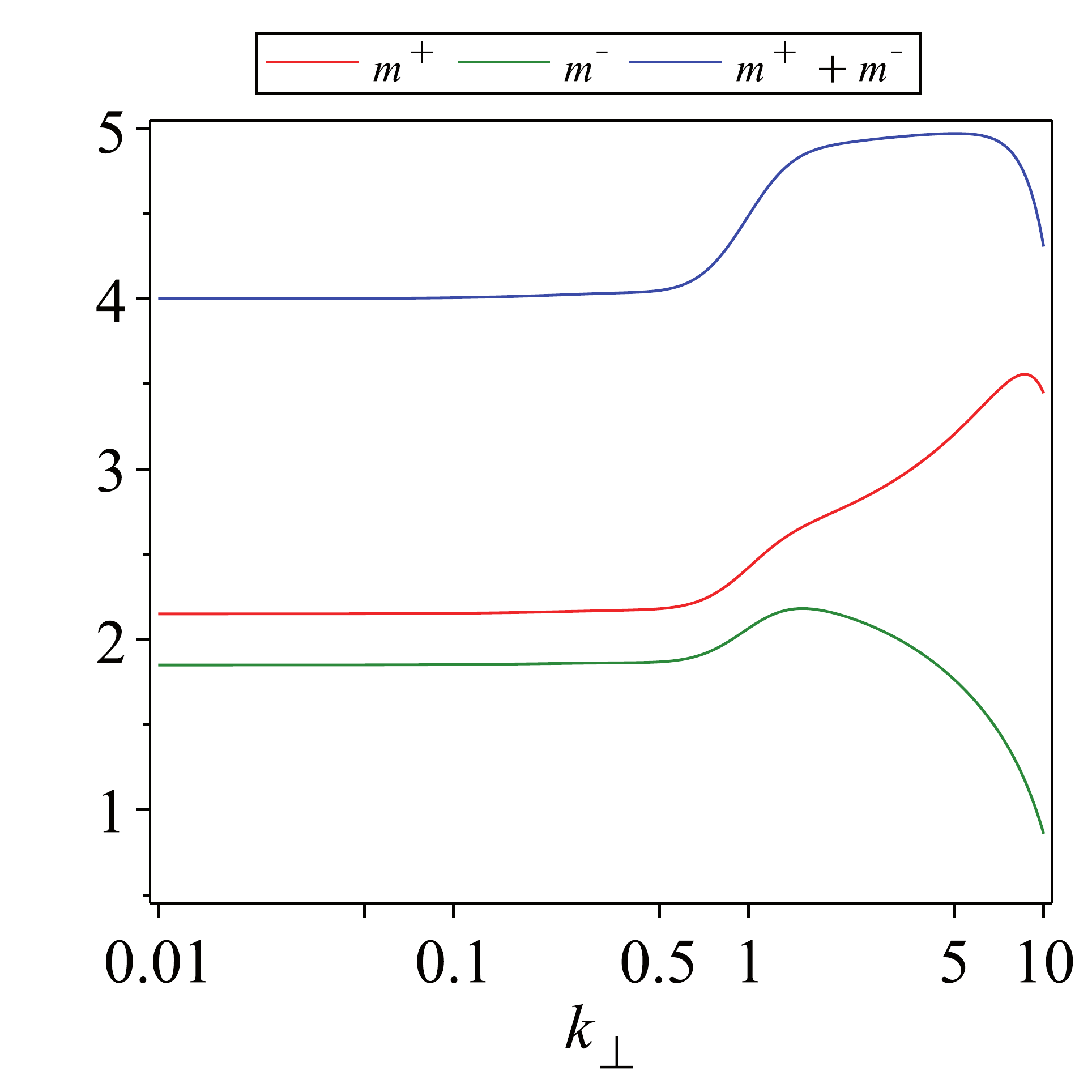}}
	\caption{Same as Fig. \ref{fig:Weak1}, but for $\eta/\varepsilon = 0.05$ and $k_d=15$.}
	\label{fig:Weak2}
\end{figure}

As mentioned in \citet{Chandran08} for  imbalanced MHD, as a consequence of pinning, the larger the dissipation wavenumber, the smaller $\varepsilon^+/\varepsilon^-$ should be for a given value of $E^+/E^-$ at the outer scale. The situation is here similar, pinning being replaced by the boundary condition at $k_d$. This point is illustrated in Fig. \ref{fig:Weak2} which displays the same graphs as  Fig. \ref{fig:Weak1}, but for $\eta/\varepsilon = 0.05$ and $k_d=15$. 

\subsection{Strong turbulence regime}
It is of interest to analyze, in the various regimes, the predictions that can be made from Eqs. (\ref{eq:upm}) in cases where the parallel wavenumber is no longer constant.
\subsubsection{MHD regime}
When concentrating on  scales large enough for dispersion to be negligible, it is possible to look for cascades associated to power-law transverse spectra $E^\pm(k_\perp)\sim k_\perp^{-m^\pm}$ and constant fluxes $\varepsilon^\pm$. Several regimes can be distinguished, depending on the amplitude of the fluctuations. 

\begin{figure}
	\centerline{
		\includegraphics[width=0.32\textwidth]{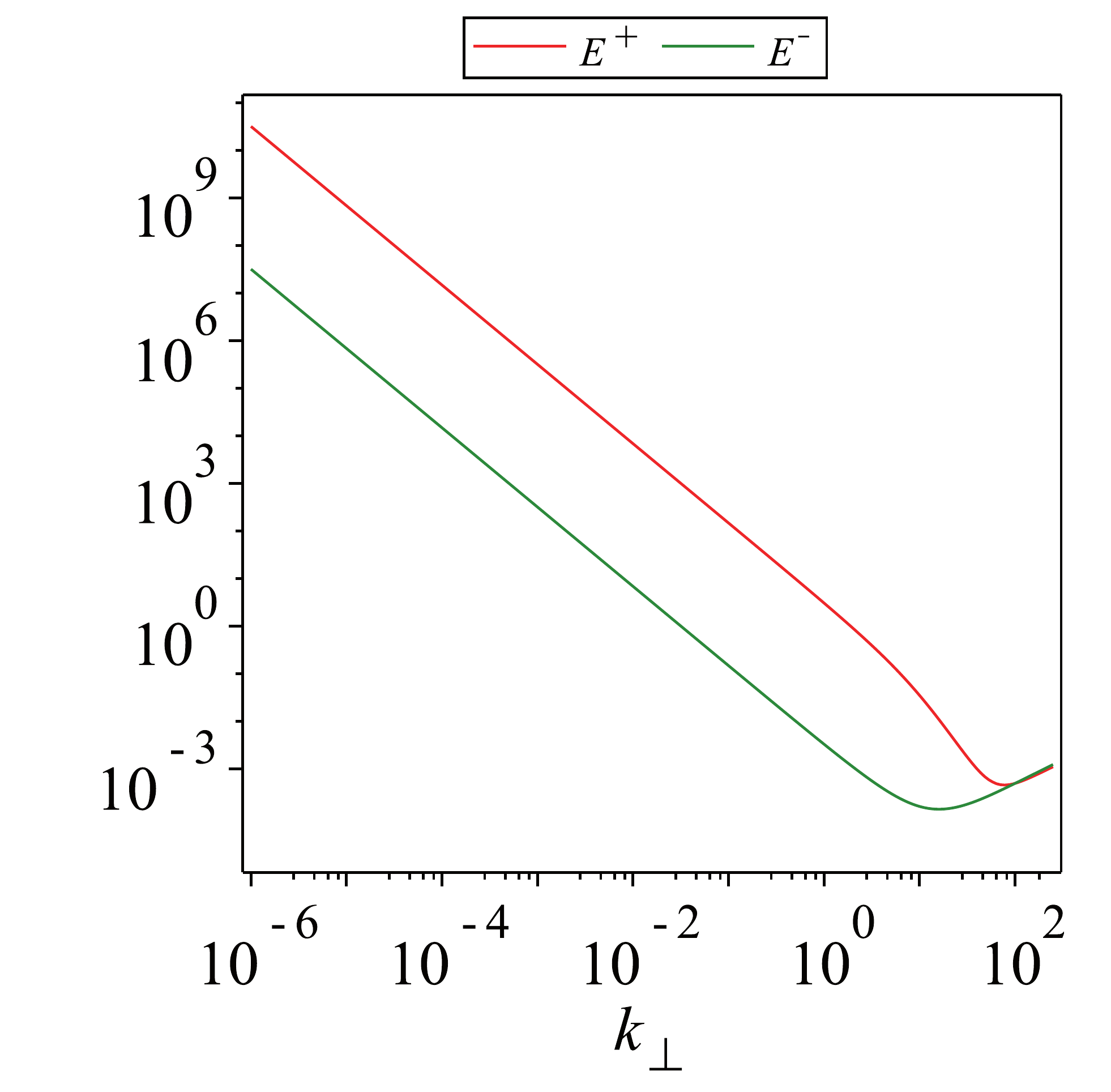}
		\includegraphics[width=0.32\textwidth]{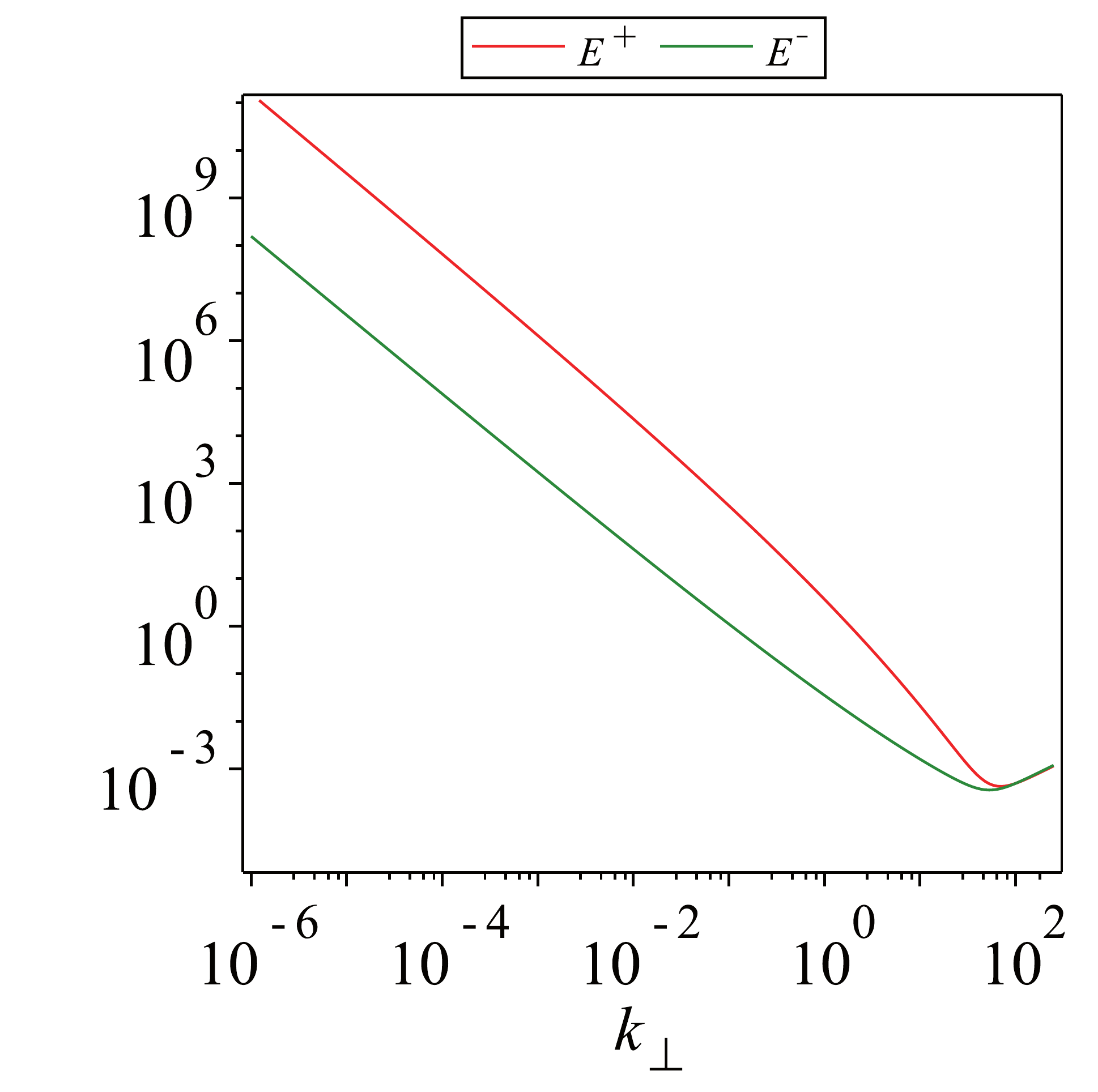}
		\includegraphics[width=0.32\textwidth]{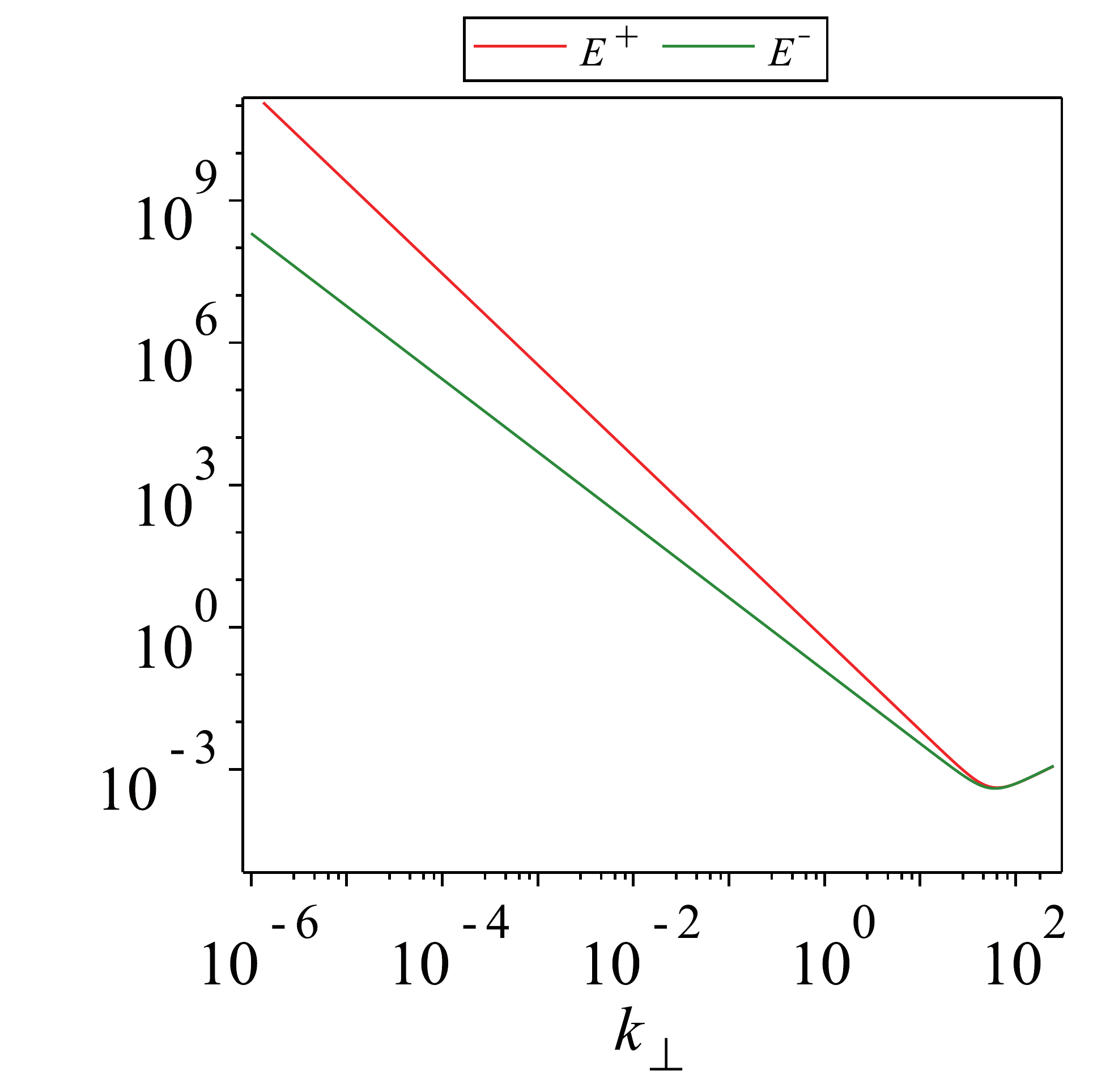}
	}
	\centerline{
	\includegraphics[width=0.32\textwidth]{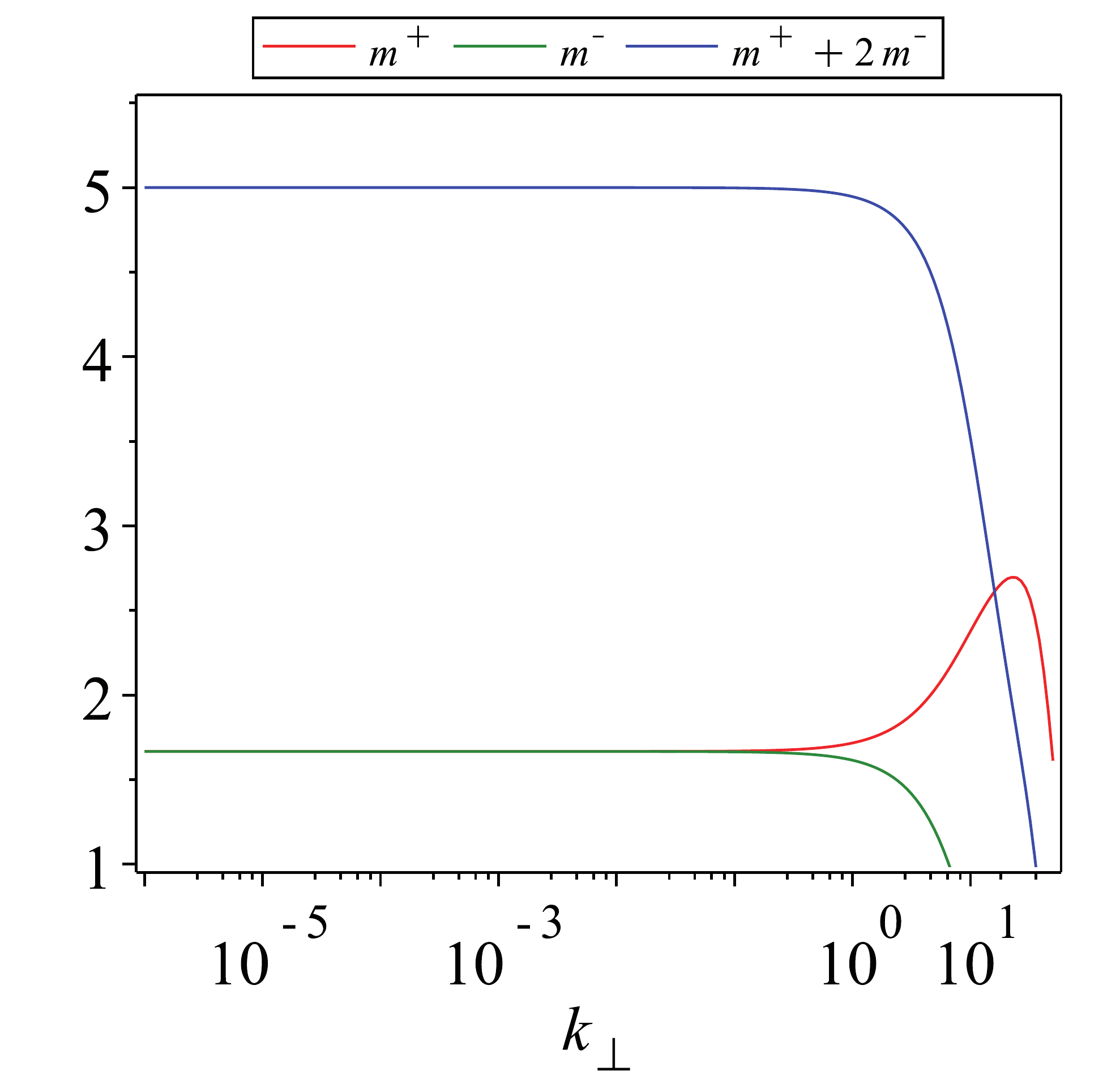}
	\includegraphics[width=0.32\textwidth]{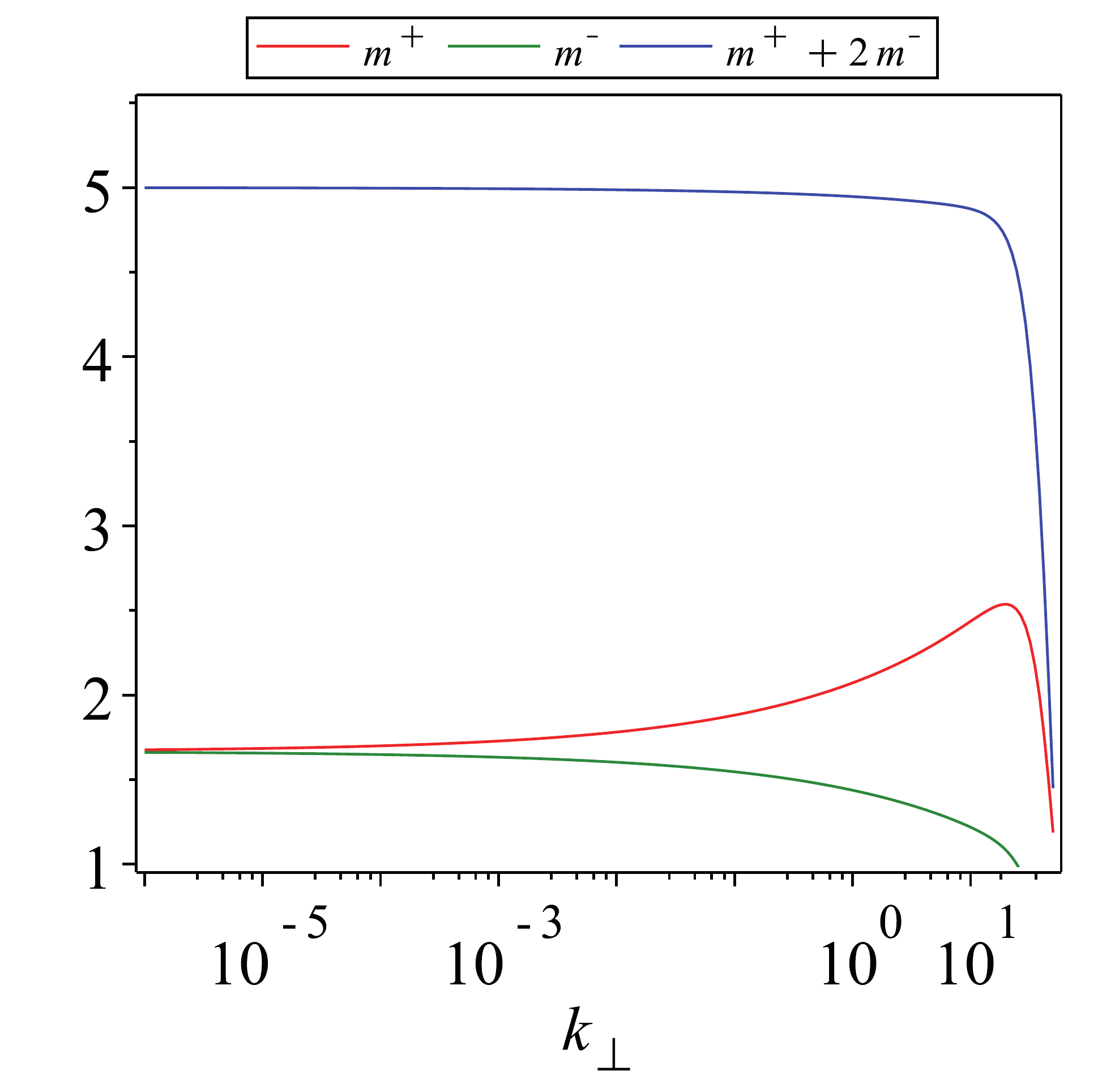}
	\includegraphics[width=0.32\textwidth]{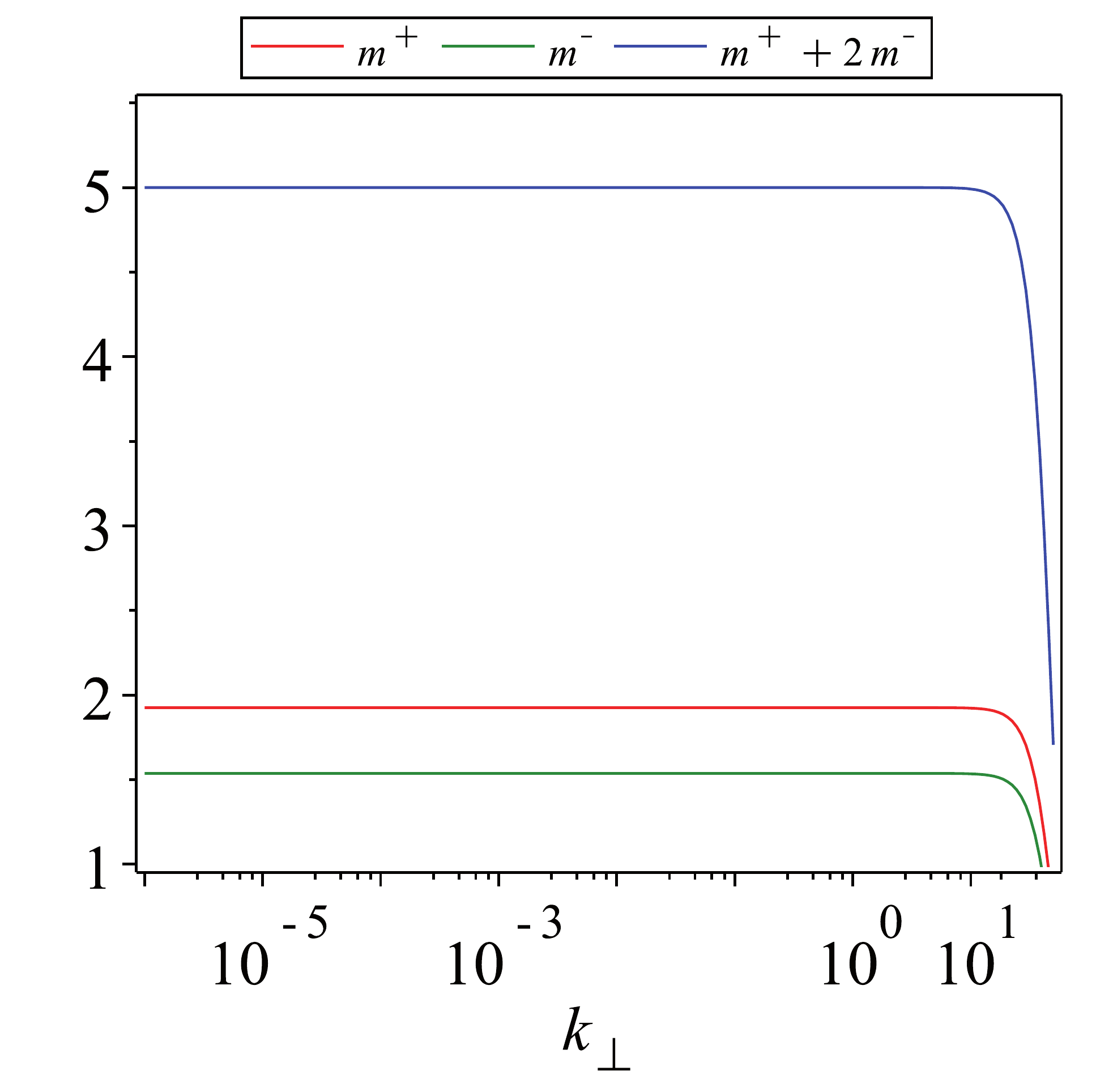}
}
	\caption{$E^+$ (red) and $E^-$ (green) spectra (top) together with the local slopes (bottom) $m^+$ (red), $m^-$ (green) and $m^++2m^-$ (blue) in the strong turbulence regime, for $\nu=0$ (left), $\nu=0.8$ (middle) and  $\nu=1$ (right). The other parameters are $\varepsilon=1$,  $k_f=0$, $k_d=100$ with  $u^+(k_d)=u^-(k_d)=50(\varepsilon/k_d^7)^{1/2}$. The values $\eta=0.93869$ for $\nu=0$, $\eta=0.33477$ for $\nu=0.8$ and $\eta=0.07115$ for $\nu=1$ are chosen such that $E^+(10^{-6})/E^-(10^{-6})\approx 1000$ in all the cases. }
	\label{fig:MHD-strong-nu}
\end{figure}

\begin{itemize}[label=$\bullet$] 
	\item \hskip 0.3cm
	When both the parallel and anti-parallel propagating Alfv\'en modes have a strong amplitude, it is legitimate to take $k_f=0$. One has

\begin{equation}
-\frac{\varepsilon^{(r)}}{4C' s} = k_\perp^{9/2} \sqrt{E^{(-r)}}(\frac{E^+}{E^-})^{-\nu(r+1)/4}\frac{d}{dk_\perp} \left(\frac{E^{(r)}}{k_\perp}\right),\label{eq:epsr}
\end{equation}
which, by simple power counting, implies for the spectral indices
\begin{align}
(1-\frac{\nu}{2})m^+ + \frac{1+\nu}{2} m^-=\frac{5}{2}\\
\frac{1}{2}m^+ + m^- =\frac{5}{2},
\end{align}
leading to  $m^\pm=5/3$ when $\nu\ne1$ \citep{Lithwick07} and to the sole entanglement relation
$m^+ +2m^-=5$, as in \citet{Chandran08}, for $\nu=1$.  
\begin{itemize}[label=$\star$] 
\item \hskip 0.3cm
For $\nu\ne 1$, the nonlinear eigenvalue problem (fixing $E^\pm(k_0)= E^\pm_0$ at a
finite transverse wavenumber $k_0$ together with $E^\pm=0$ at infinity) has a well-defined solution given by
\begin{eqnarray}
&& \varepsilon^+ =\frac{32C' s}{3}k^{5/2}_0 (E^+_0)^{1-\nu/2} (E^-_0)^{(1+\nu)/2} \label{eq:epsp} \\
&& \varepsilon^- =\frac{32C' }{3}k^{5/2}_0 E^-_0 \sqrt{E^+_0}, \label{eq:epsm}
\end{eqnarray}
which leads to
\begin{equation}
r_{\varepsilon}\equiv\frac{\varepsilon^+}{\varepsilon^-}=\left (\frac{E^+_0}{E^-_0}\right )^{(1-\nu)/2}.
\end{equation}
The relation $E^+_0/E^-_0=r_{\varepsilon}^2$ predicted by the model of \citet{Lithwick07} is recovered for $\nu=0$. Direct numerical MHD simulations by \citet{Beresnyak09} where the Elsasser fields are
randomly driven, seem to indicate that a similar relation holds but with a slightly larger power of the flux-rate ratio, that can easily be fitted with an appropriate choice of $\nu$. From Fig. 13 of \citet{Beresnyak09}, one can estimate $\nu\approx 0.2$, comparable to the value $\nu=1/4$ corresponding to the model of \citet{Beresnyak08}. Nevertheless, the numerical simulations display spectral indices that differ from $-5/3$, an effect that can be related to finite Reynolds number or hyperviscosity effects.

\item \hskip 0.3cm
For $\nu=1$, the problem is in contrast under-determined and we find, as in \citet{Chandran08}, that $\varepsilon^+$ and $\varepsilon^-$ should be equal for infinitely extended power-law spectra to exist. Other values of $r_{\varepsilon}$ are possible if different boundary conditions are prescribed,
although the model cannot accommodate for a ratio $r_{\varepsilon}$ larger than $2$. Indeed, Eqs. (\ref{eq:epsr}) imply $r_{\varepsilon}=(m^++1)/(m^-+1)$ which, supplemented by the entanglement relation, gives, for $r_{\varepsilon}=2$, $m^+=3$ and $m^-=1$, which are the limiting values of the spectral exponents ensuring the convergence of $\int E^-(k_\perp)dk_\perp$. 
For given fluxes such that $r_{\varepsilon}<2$, the
spectral exponents are uniquely defined as 
\begin{align}
m^+=(7r_{\varepsilon}-2)/(r_{\varepsilon}+2) \\
m^-=(6-r_{\varepsilon})/(r_{\varepsilon}+2).
\end{align}

To specify the amplitudes and thus obtain a complete solution of the problem, pinning at the dissipation scale is requested, as in the weak turbulence case. 
Equations (\ref{eq:epsp})-(\ref{eq:epsm}) determines $\sqrt{E^+_0}E^-_0$ and the equality of the spectra at the  wavenumber $k_d$ provides the extra condition to fix the amplitudes. Differently, if the amplitudes are given, the extra pinning condition permits the determination of the fluxes, and thus of the spectral slopes.
\end{itemize}

\item \hskip 0.3cm For a strongly imbalanced regime where  the amplitude of one type of waves is large and that of the other sufficiently small for ${\widetilde k}^-_\|$ to be dominated by $k_f$,
the spectral exponents obey the conditions $m^+ + m^- =4$ and $m^+ + 2 m^- = 5$, which implies $m^+ = 3$ and $m^-= 1$. Interestingly, the constraint on the fluxes mentioned for the weak turbulence regime is no longer necessary.

The behavior of the spectra, together with the local slopes, in the case where they are taken equal at the wavenumber $k_d$ with a  ratio $E^+/E^-=1000$ prescribed at $k_\perp=10^{-6}$ is displayed in Fig. \ref{fig:MHD-strong-nu} for $\nu=0$, $\nu=0.8$ and $\nu=1$. For any value of $\nu\ne 1$, both spectral exponents become equal to $5/3$ at scales that are larger and larger when $\nu$ approaches 1. When $\nu=1$, they stay different at all the scales. In all the cases, the entanglement relation $m^++2m^-=5$ is satisfied, except for $k_\perp$ very close to $k_d$.
The case $\nu=0.25$ is almost undistinguishable  from the one with $\nu=0$.
\end{itemize}

\subsubsection{Dispersive regime with $\eta=0$}
In the case $\eta=0$, we have
\begin{equation}
m^+(k_\perp)-m^-(k_\perp)=\frac{\varepsilon \sqrt{u^-(k_\perp)}}{4C'k^4_\perp v_{ph}u^+(k_\perp)u^+(k_\perp) }\left ( \left (\frac{u^+(k_\perp)}{u^-(k_\perp)}\right )^{\frac{\nu-1}{2}}-1\right).\label{eq:mpmm-eta0}
\end{equation}
When $\nu=1$, we find that $m^+(k_\perp)=m^-(k_\perp)$ (equal to $5/3$ or $7/3$ in the MHD or sub-ion ranges respectively), whatever the ratio $u^+(k_\perp)/u^-(k_\perp)$.
In contrast, when $\nu<1$, if we assume $u^+(k_\perp)>u^-(k_\perp)$, we find that $m^+(k_\perp)<m^-(k_\perp)$, indicating that the spectrum of the more energetic wave is shallower than that of the lower amplitude one. Under this hypothesis, the two spectra diverge from each other as $k_\perp$ increases, which is unphysical. The physically relevant solution is thus $u^+(k_\perp)=u^-(k_\perp)$ for all $k_\perp$ corresponding to $m^+=m^-$ (equal to $5/3$ or $7/3$ depending on the wavenumber range).
We can then conclude that in the case $\nu=1$, an imbalanced regime can be obtained with a zero helicity flux, as in the weak turbulence regime (see discussion at the end of Sec. \ref{cascades}) or, in the presence of a finite $k_d$, with a value of $\eta/\varepsilon$ that is  smaller as $k_d$ is increased). Such a regime is impossible for $\nu<1$.

\subsubsection{The dispersive regime with $\eta\ne 0$}

\begin{figure}
	\centerline{
		\includegraphics[width=0.32\textwidth]{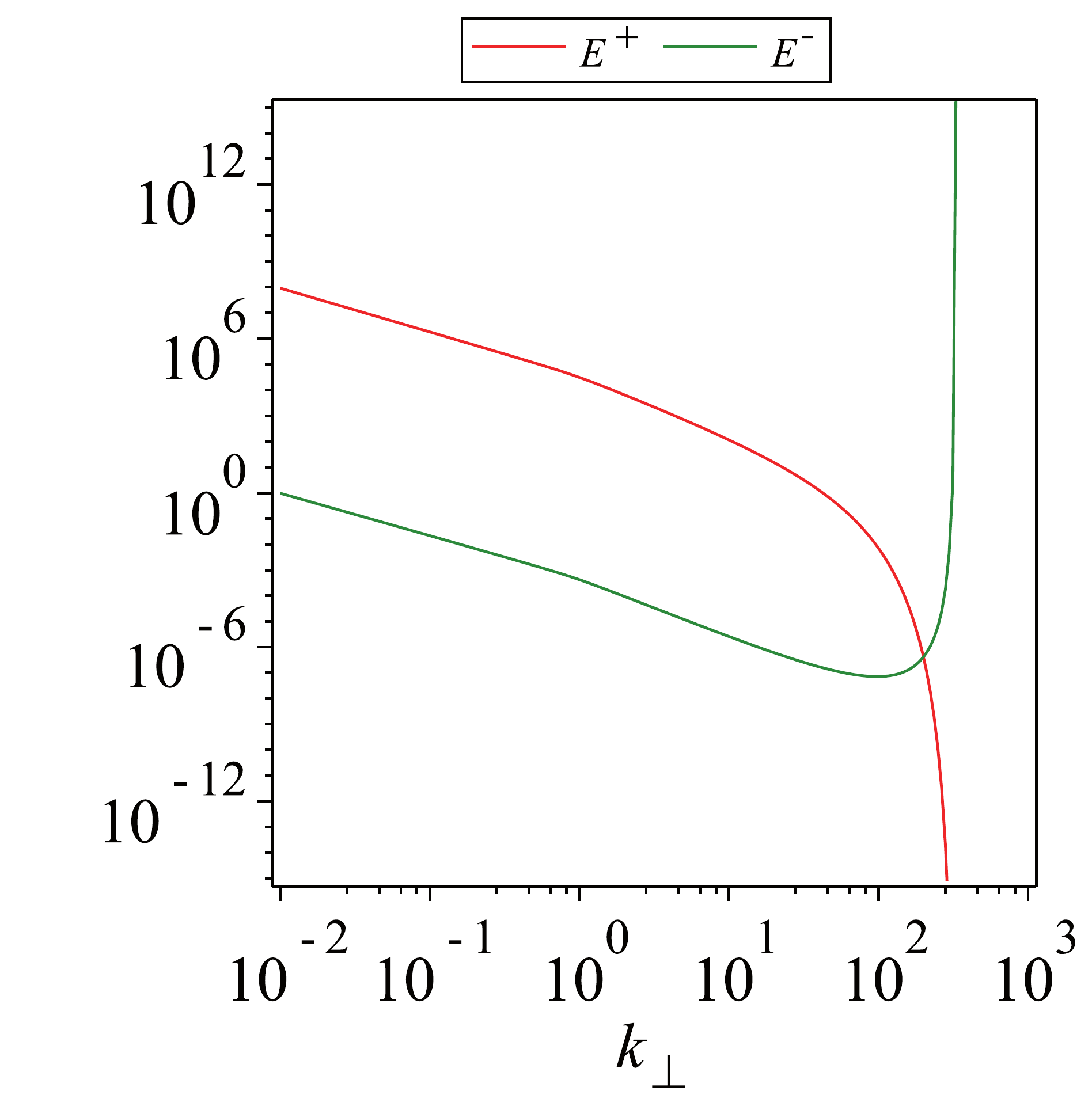}
		\includegraphics[width=0.32\textwidth]{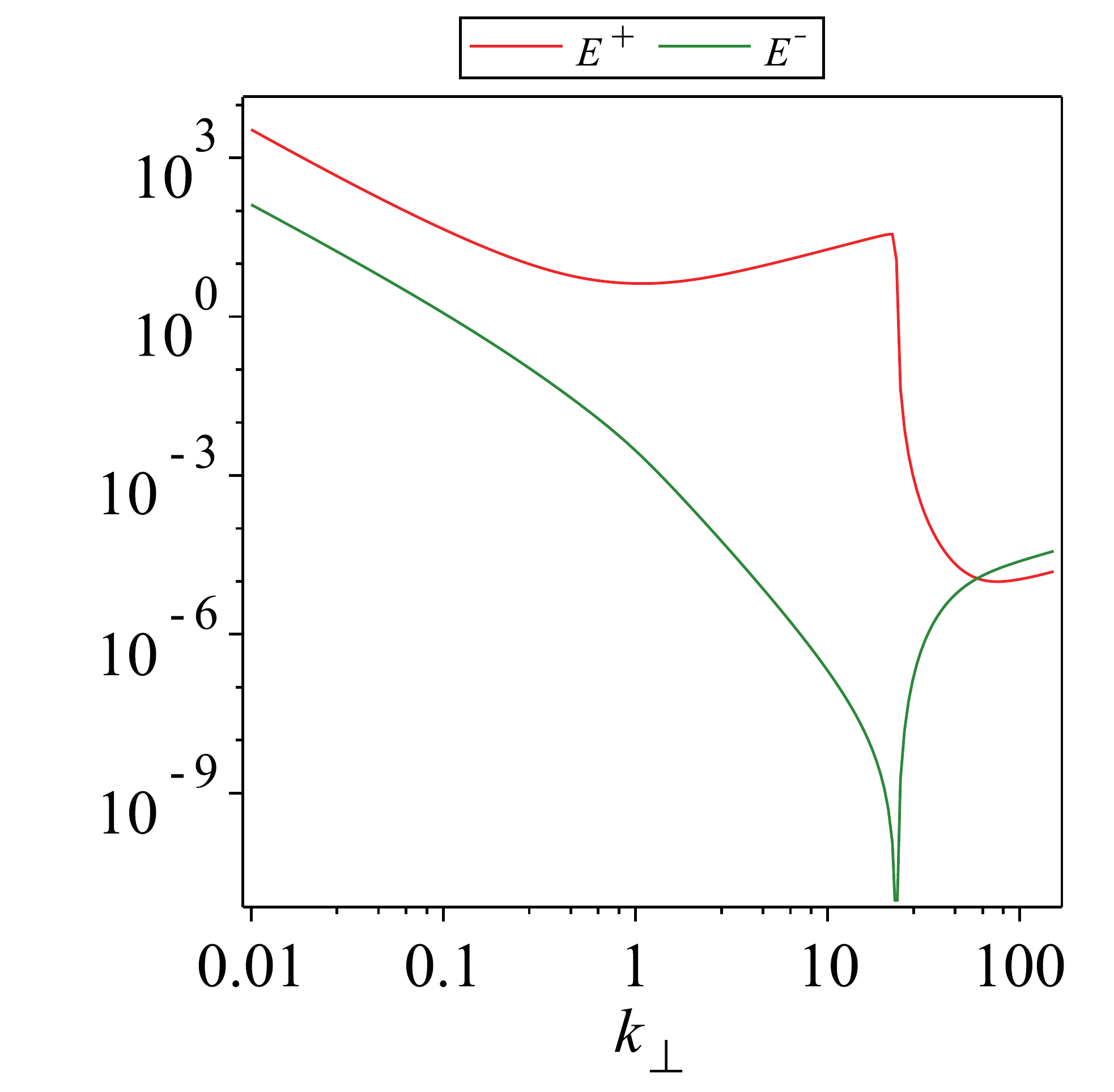}
		\includegraphics[width=0.32\textwidth]{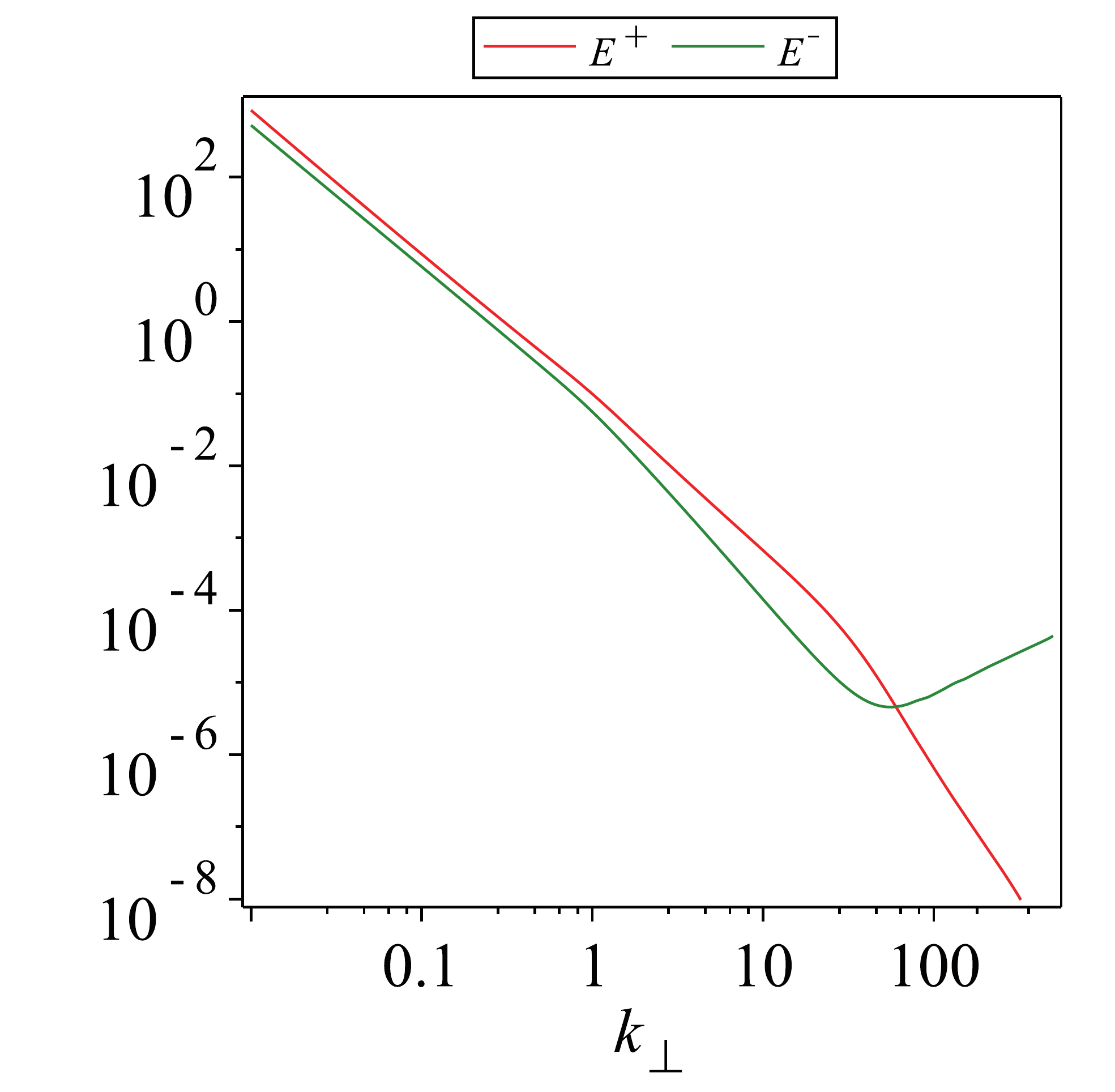}
	}
	\caption{$E^+$ (red) and $E^-$ (green) spectra in the strong turbulence regime. Left: for $\nu=1$,  $\varepsilon=1$, $\eta=0.01$,  $k_d=200$, $k_f=0$ with  $u^+(k_d)=u^-(k_d)=0.23552(\varepsilon/k_d^7)^{1/2}$; Middle: for $\nu=0$, $\varepsilon=0.1$, $k_f=0.5$, $k_d=60$ with  $u^+(k_d)=u^-(k_d)=(\varepsilon/k_d^7)^{1/2}$, $\eta=0.053752 \varepsilon$ (singular point $E^-=0$ at $k_*\approx 23$); Right:  case $k_*>k_d$ for $\nu=0$,    $\varepsilon=0.1$, $\eta=0.015$ (giving $k_*\approx 82$) $k_f=0.5$,  $k_d=60$ when choosing the critical value $u^+(k_d)=u^-(k_d) =u_c\equiv 0.405176(\varepsilon/k_d^7)^{1/2}$ for which $E^+$  decays but  $E^-$ tends to the absolute equilibrium. }
	\label{fig:Strong1}
\end{figure}

The typical behavior of Eq. (\ref{eq:upm}) at  small scales is easily obtained when assuming $\eta v_{ph}\gg \varepsilon$, a situation where one can reasonably assume that $k_f$ is negligible. The equations then reduce to
\begin{align}
\sqrt{u^+(k_\perp)}\frac{d}{d k_\perp} u^-(k_\perp)=\frac{\eta }{4C'k^5_\perp}\\
\sqrt{u^-(k_\perp)}\frac{d}{d k_\perp} u^+(k_\perp)= -\frac{\eta }{4C'k^5_\perp} \left ( \frac{u^+(k_\perp)} {u^-(k_\perp)}  \right)^{\nu/2},\label{eq:upmss}
\end{align}
from which it follows that
\begin{equation}
\left (E^+(k_\perp)\right )^{\frac{1-\nu}{2}}=\lambda k^{\frac{1-\nu}{2}}_\perp - \left (E^-(k_\perp)\right )^{\frac{1-\nu}{2}},\label{eq:EpEm-disp-eta-ne-0}
\end{equation}
where $\lambda$ is a positive constant.
 \begin{figure}
	\centerline{\includegraphics[width=0.48\textwidth]{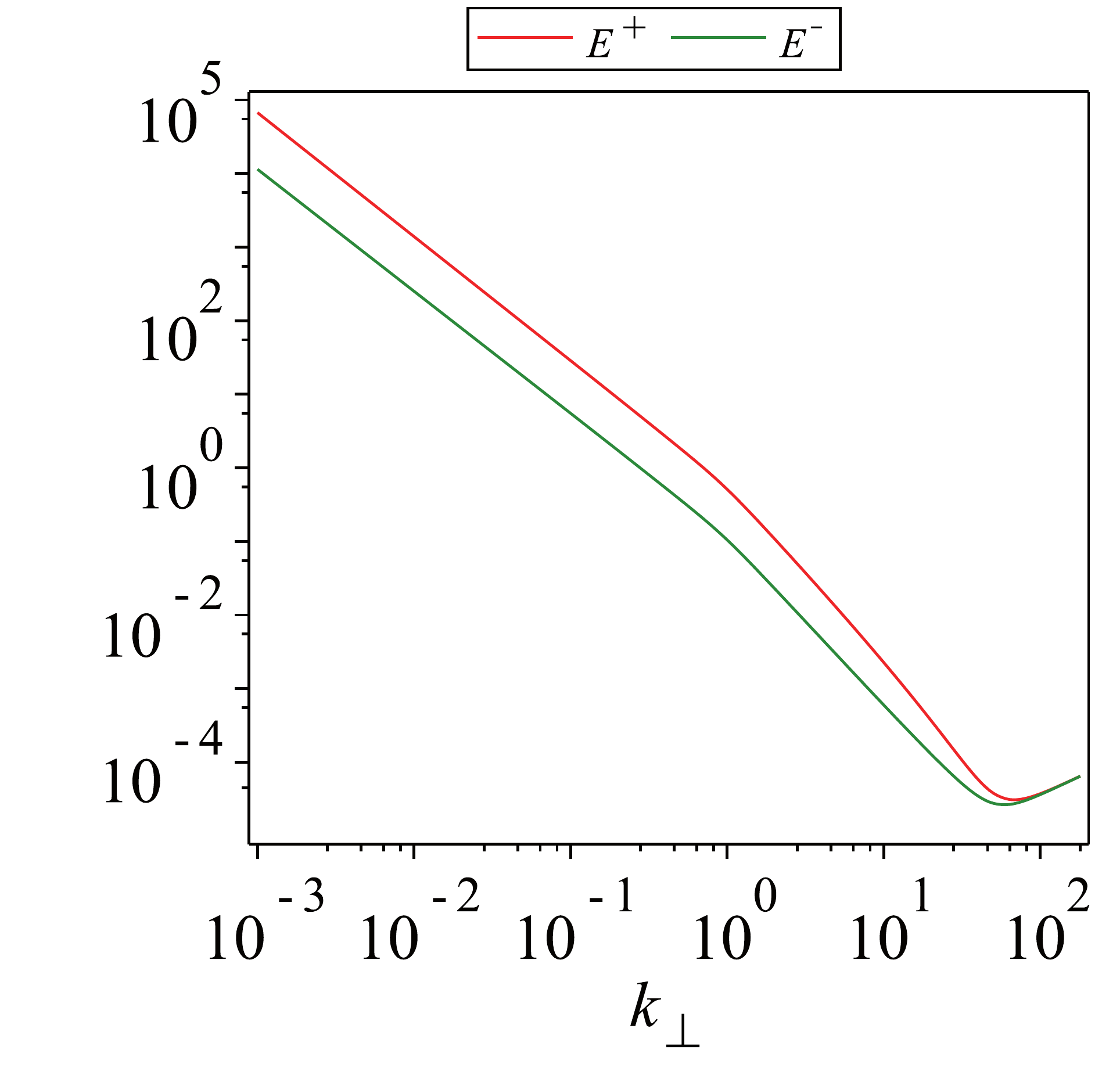}
		\includegraphics[width=0.48\textwidth]{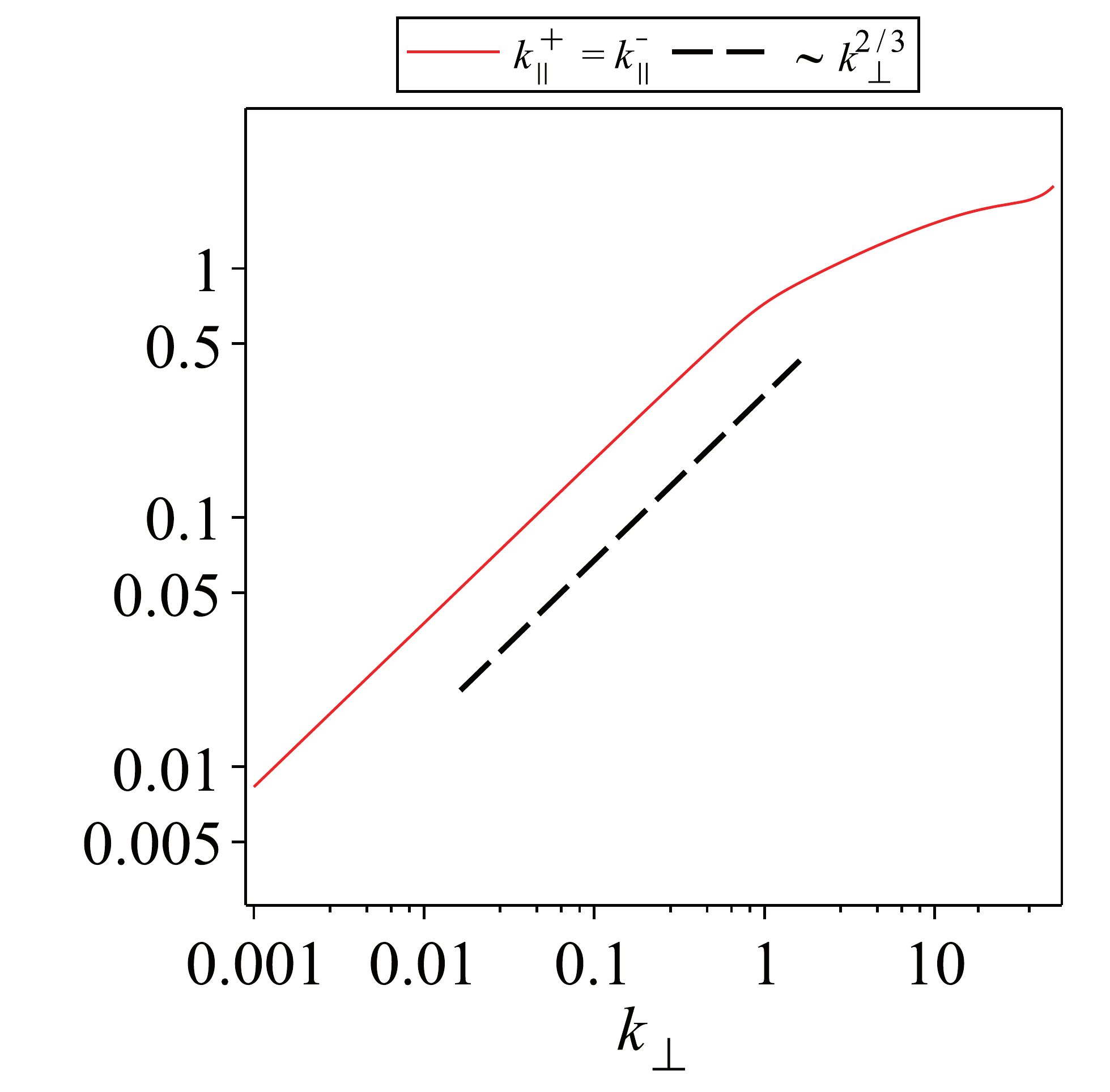}}
	\centerline{\includegraphics[width=0.48\textwidth]{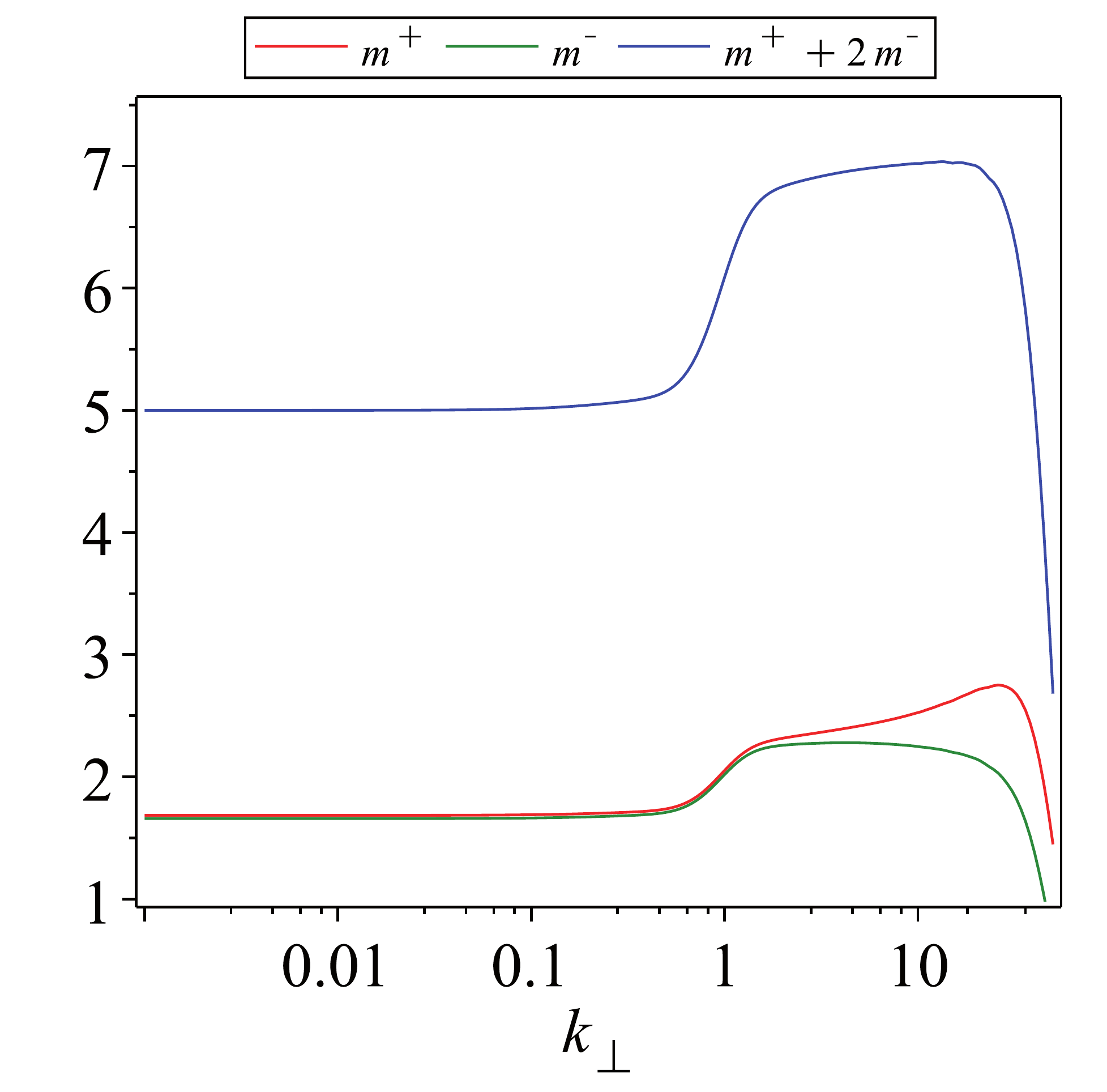}
		\includegraphics[width=0.48\textwidth]{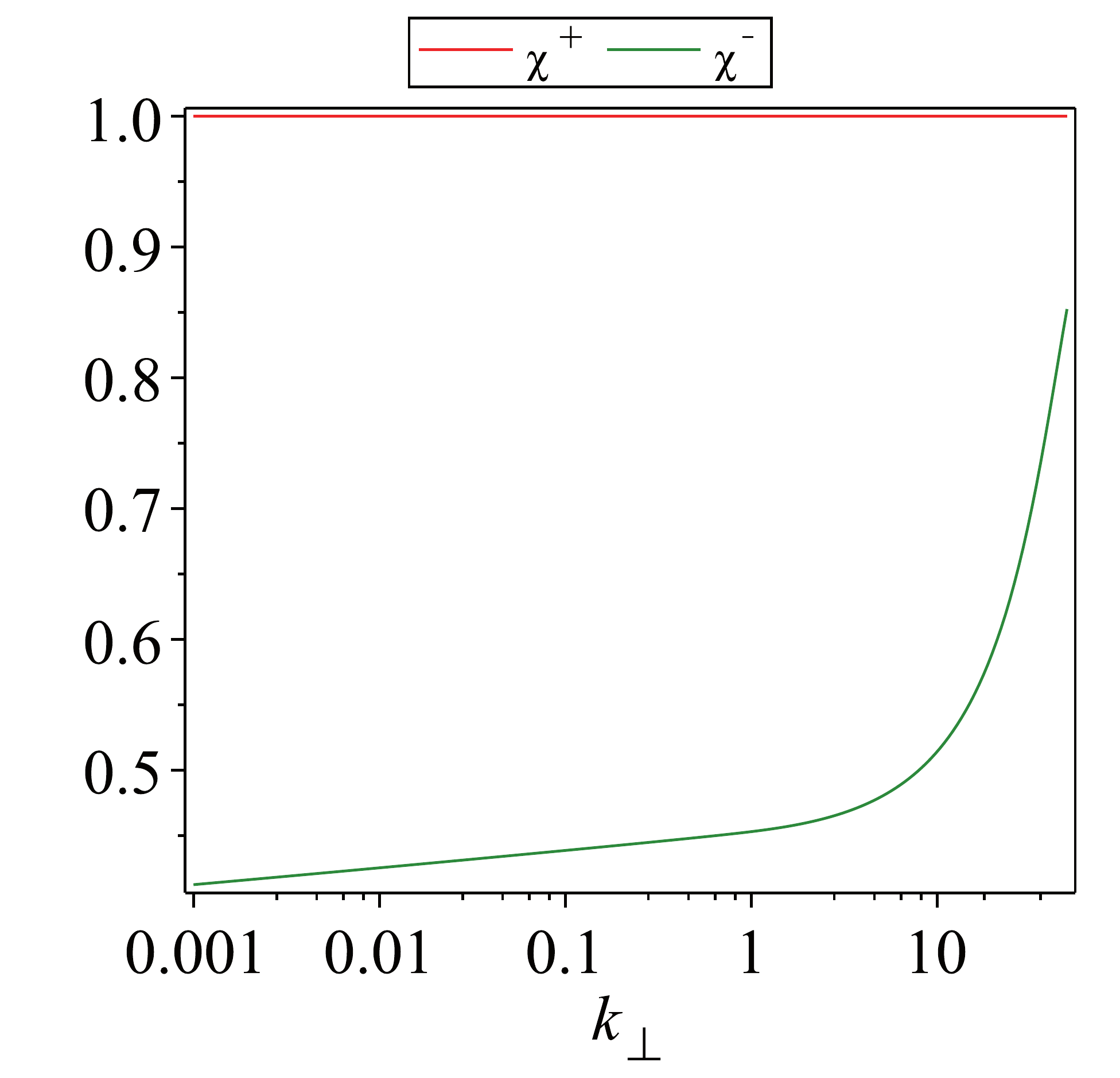}
	}
	\caption{A case with $\nu=1$, $\varepsilon=1$, $\eta/\varepsilon=0.005$,  $k_d=500$, $u^+(k_d)=u^-(k_d)= 10^3(\varepsilon/k_d^7)^{1/2}$, $k_f=0$.
		In the top panels are shown the $E^\pm$ spectra (left), and ${\widetilde k}^{\pm}_{\|}$ (right), while in the bottom panels are displayed the local slopes $m^\pm(k_\perp)$ together with $m^++2m^-$ (in blue) (left) and the nonlinearity parameters $\chi^\pm=(k^3_\perp E^\pm (k_\perp))^{1/2}/{\widetilde k}^\pm_\| $ (right). In all these graphes, red (green) color refer to $+$ ( $-$) waves.}
	\label{fig:WS1}
\end{figure}

 \begin{figure}
\centerline{\includegraphics[width=0.48\textwidth]{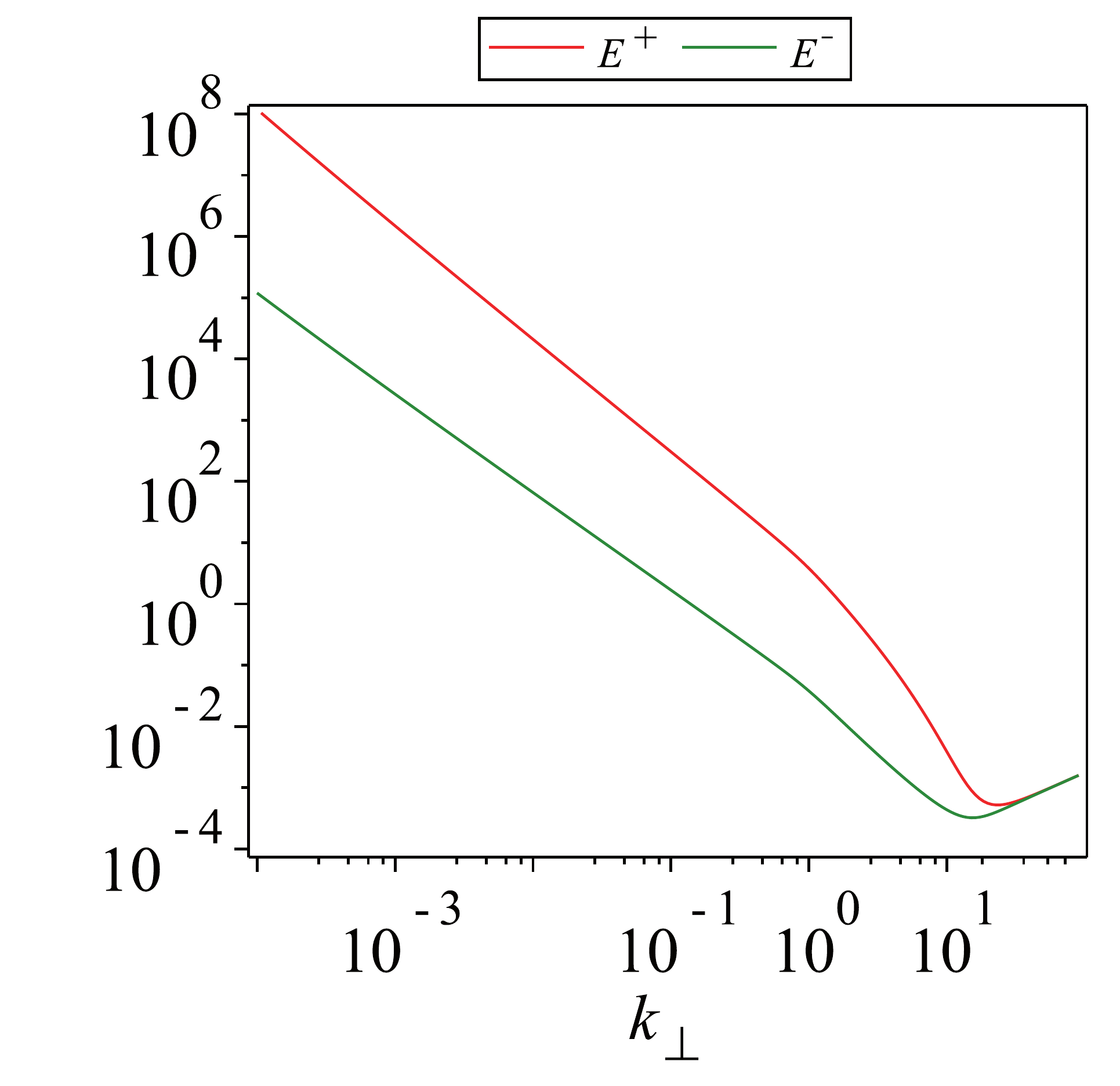}
\includegraphics[width=0.48\textwidth]{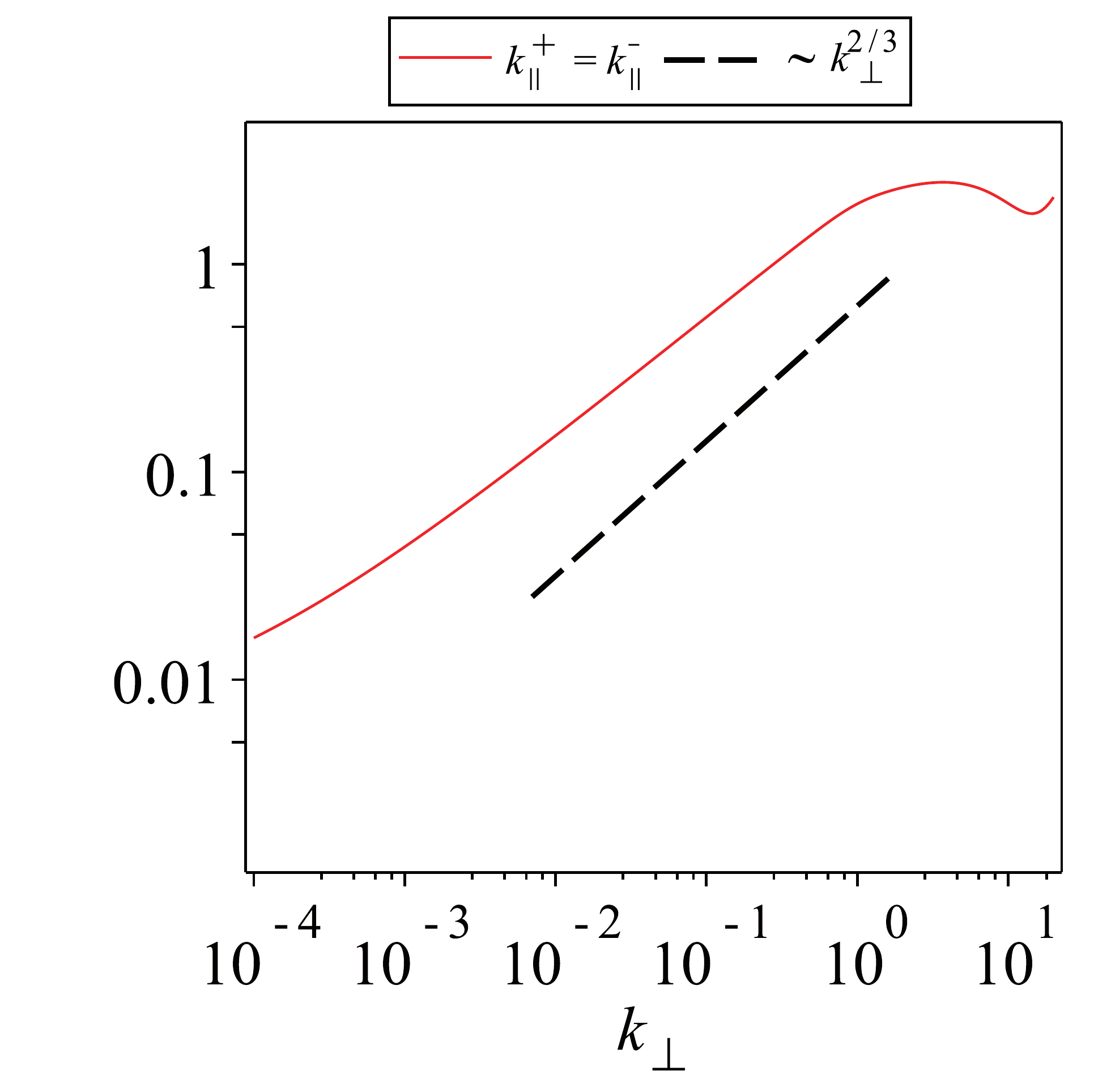}}
\centerline{\includegraphics[width=0.48\textwidth]{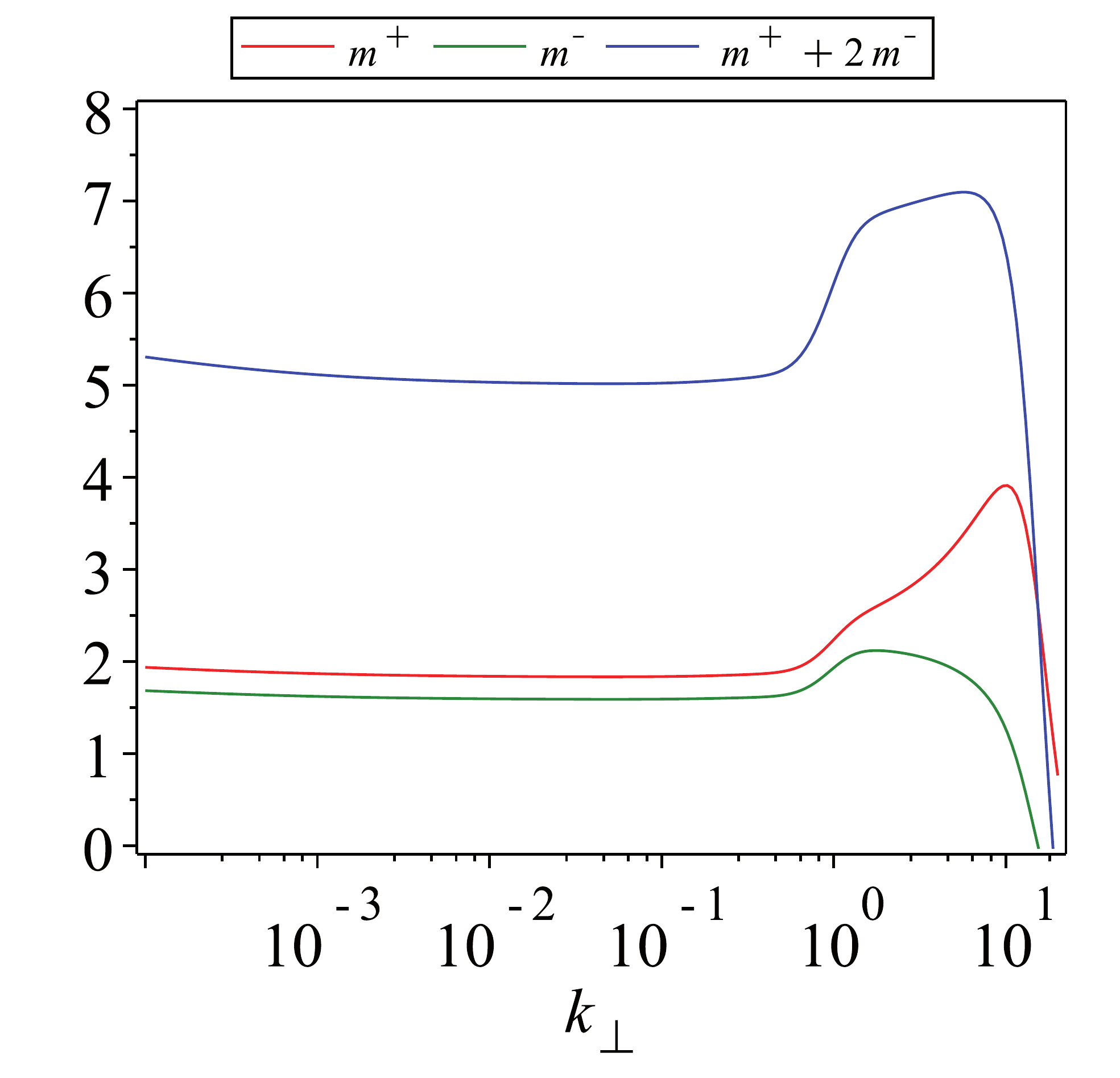}
\includegraphics[width=0.48\textwidth]{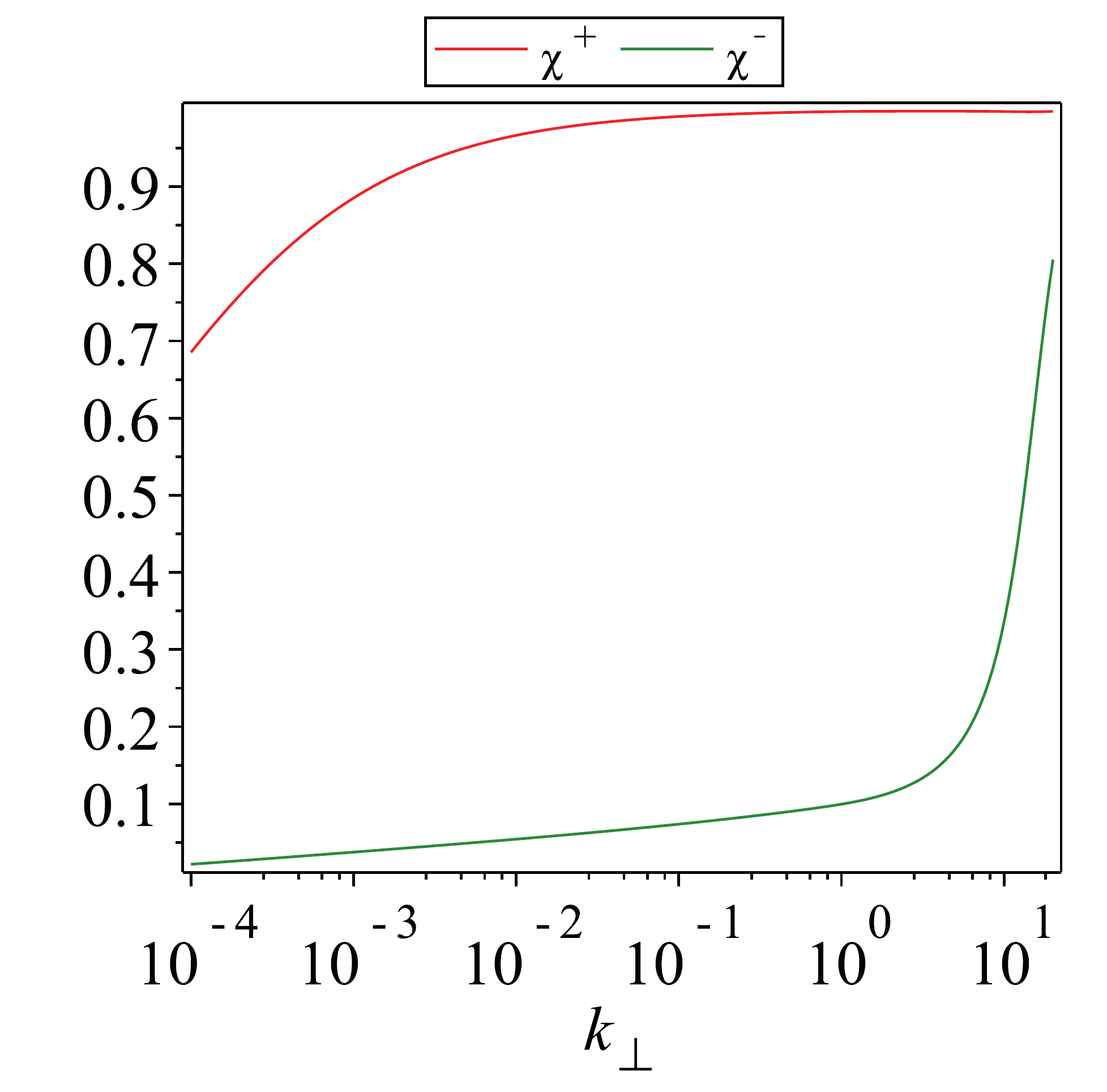}
}
\caption{A case with $\nu=1$, $\varepsilon=1$, $\eta/\varepsilon=0.045$,  $k_d=200$, $u^+(k_d)=u^-(k_d)=2\times 10^3(\varepsilon/k_d^7)^{1/2}$, $k_f=5\times10^{-3}$ where the $-$ component  remains in the weak turbulence regime, while the $+$ one undergoes a transition to strong turbulence.
In the top panels are shown the $E^\pm$ spectra (left), and ${\widetilde k}^{\pm}_{\|}$ (right), while in the bottom panels are displayed the local slopes $m^\pm(k_\perp)$ together with $m^++2m^-$ (in blue) (left) and the nonlinearity parameters $\chi^\pm=(k^3_\perp E^\pm (k_\perp))^{1/2}/{\widetilde k}^\pm_\| $ (right). In all these graphs, red (green) color refer to $+$ ( $-$) waves.}
\label{fig:WS2}
\end{figure}

For $\nu=1$,  we find
\begin{equation}
E^+(k_\perp)E^-(k_\perp)=\lambda k^2,
\end{equation}
compatible with small-scale absolute equilibria and showing that both spectra cannot tend to zero at infinity. A nearly singular behavior is depicted in Fig. \ref{fig:Strong1} (left) for $\varepsilon=1$, $\eta=0.01$,  $k_d=200$, $k_f=0$, when choosing  $u^+(k_d)=u^-(k_d)=r(\varepsilon/k_d^7)^{1/2}$ with $r=0.23552$. For a slightly smaller value of $r$, $E^-$ (respectively $E^+$) tends to infinity (respectively zero) at a finite wavenumber, while for larger values of $r$, they both tend to absolute equilibria.

For $\nu\ne 1$,  Eq. (\ref{eq:EpEm-disp-eta-ne-0}) implies that, for $\eta\ne 0$, it is impossible to prescribe  that both spectra vanish at infinity (and in fact one of them will diverge), as was already mentioned at the beginning of this Section.
\textcolor{black}{It also indicates that $E^-(k_\perp)$ vanishes at a wavenumber $k_*$ such that $\eta v_{ph}(k_*)\approx \varepsilon$, close to which the above approximation is no longer valid. This regime is exemplified in Fig \ref{fig:Strong1} (middle) which displays the $E^\pm$ spectra for $\nu=0$, $\varepsilon=0.1$, $k_f=0.5$, starting the integration at $k_d=60$ with  $u^+(k_d)=u^-(k_d)=(\varepsilon/k_d^7)^{1/2}$ and choosing (by a dichotomy process)  $\eta=0.053752 \varepsilon$ such that the singular wavenumber $k_*$ is  slightly smaller than $23$, satisfying $\eta v_{ph}(k_*)= \varepsilon$. This result indicates that it is more physically appropriate to choose a value of $\eta$ such that $k_*>k_d$.}
 
 \begin{figure}
	\centerline{\includegraphics[width=0.48\textwidth]{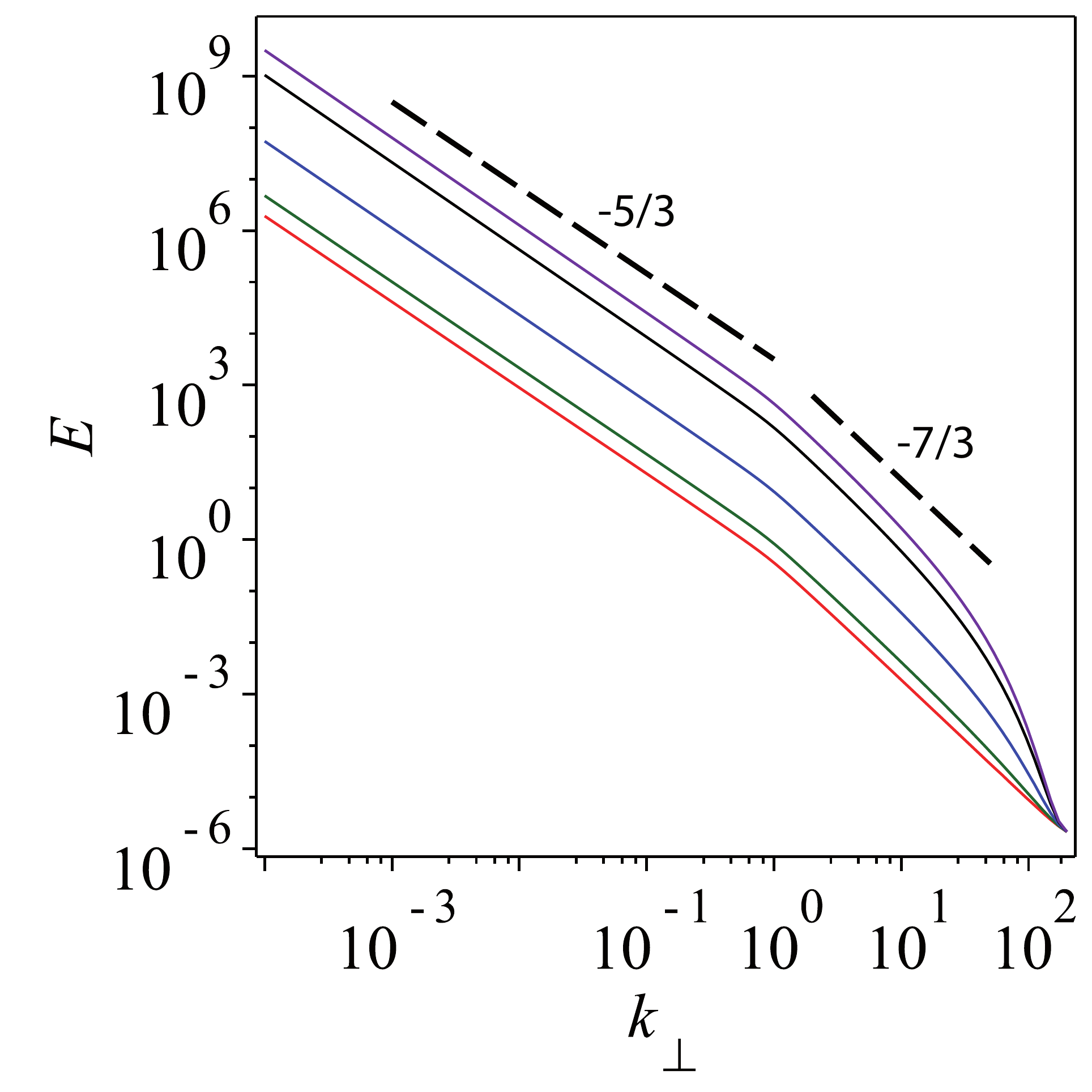}
		\includegraphics[width=0.48\textwidth]{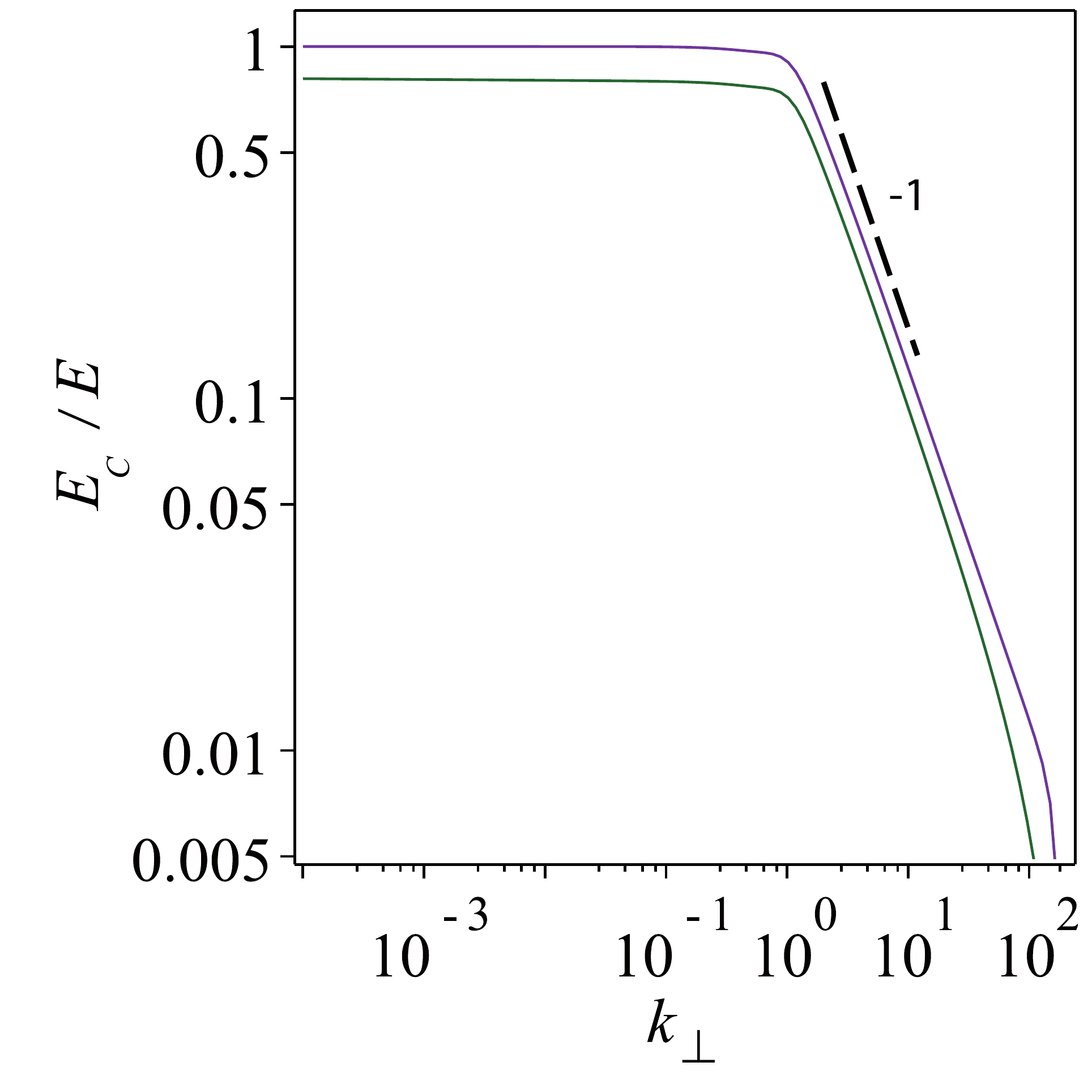}
	}
	\caption{\textcolor{black}{Left: total energy spectra obtained for $\eta=0$ (red), $\eta=2\times10^{-3}$ (green), $\eta=5\times10^{-3}$ (blue), $\eta=8\times10^{-3}$ (black) and $\eta=9\times10^{-3}$ (violet), with $\varepsilon =1$, $k_d=200$, $k_f=0$, $u^+(k_d)=u^-(k_d)=0.6 (\varepsilon/k_d^7)^{1/2}$; Right: normalized cross-helicity spectrum $E_C(k_\perp)/E(k_\perp)$ for  $\eta=2\times10^{-3}$  and $\eta=9\times10^{-3}$, the other parameters being the same as in the left panel.}}
	\label{fig:E-sigma}
\end{figure}

 
The case $k_*>k_d$ is illustrated in Fig. \ref{fig:Strong1} (right) in a situation where   $\nu=0$, $\varepsilon=0.1$, $\eta=0.015$ (giving $k_*\approx 82$) $k_f=0.5$, $k_d=60$ with  $u^+(k_d)=u^-(k_d)$.
We find (again by dichotomy) that there exists a value $u_c=0.405176(\varepsilon/k_d^7)^{1/2}$ of $u^+$, such that if $u^+(k_d)>u_c$, absolute equilibrium spectra establish at small scale and  if $u^+(k_d)<u_c$, a singularity is present. When choosing $u^+(k_d)=u_c$, one can ensure that the $E^+$ spectrum decays. The $E^-$ spectrum nevertheless tends to the absolute equilibrium solution. The smaller $\eta/\varepsilon$ is, the larger is the wavenumber $k_*$  and the smaller is the imbalance at large scale. Note that retaining electron inertia in the function $v_{ph}$ does not change qualitatively the above conclusion, and in particular the fact that one of the spectra displays an absolute equilibrium range at small scales.
Figure \ref{fig:Strong1} also shows that the ratio $E^+/E^-$ is small in the MHD range and tends to increase at dispersive  scales before decreasing to unity at the pinning scale. As announced, this observation suggests that in the presence of dispersion, the case $\nu=1$ provides a better model. Indeed, Fig. \ref{fig:WS1} (left) shows in this case (for $\varepsilon=1$, $\eta/\varepsilon=0.005$,  with $k_d=500$, $u^+(k_d)=u^-(k_d)= 10^3(\varepsilon/k_d^7)^{1/2}$ and  $k_f=0$) that, in the MHD range, the $E^\pm$ spectra display a significant imbalance (in spite of a ratio $\varepsilon^+/\varepsilon^-= 1.01$) with similar slopes close to $-5/3$ (bottom left). \textcolor{black}{Note the large values of the coefficient entering the expression of $u^\pm(k_d)$. This choice is necessary, when $k_d$ is taken large, to ensure  moderate values of $E^+(k_\perp)/E^-(k_\perp)$ in the MHD range.}

In the sub-ion range, the low-amplitude wave displays a clear transition towards a $k_\perp^{-7/3}$ energy spectrum, while the spectrum of the more energetic one is not 
a power law, the local slope increasing up to the pinning wavenumber. This phenomenon is also visible on the parallel wavenumber (identical for both waves) displayed in the top right panel, which scales like $k_\perp^{2/3}$ at large scales and undergoes a transition with local slopes smaller than $1/3$. As expected, the nonlinearity, parameters $\chi^\pm=(k^3_\perp E^\pm (k_\perp))^{1/2}/{\widetilde k}^\pm_\| $ (bottom right) are such that $\chi^+$ is equal to unity and $\chi^-\ll 1$, with a tendency to increase at small scales.

A situation displaying a transition from weak to strong turbulence within the framework of the model with $\nu=1$ is displayed in Fig. \ref{fig:WS2}, where $\varepsilon=1$, $\eta/\varepsilon=0.045$,  with $k_d=200$, $u^+(k_d)=u^-(k_d)=2\times 10^3(\varepsilon/k_d^7)^{1/2}$ and  $k_f=5\times10^{-3}$.
In the top panels are shown the $E^\pm$ spectra (left) and ${\widetilde k}^{\pm}_{\|}$ (right), while in the bottom panels are displayed the local slopes $m^\pm=-({d}/{d\ln k})\ln E^\pm (k_\perp)$,
together with $m^++2m^-$ (in blue) (left) and the nonlinearity parameters $\chi^\pm $ (right). We clearly see that the $"+"$ component starts in a weak regime but rapidly  undergoes a  transition towards strong turbulence, while the $"-"$ component remains in the weak regime. 
\textcolor{black}{In the graph displaying the parallel wavenumber, we have  superimposed a $2/3$ power law showing that the critical balance is reached for $k_\perp >0.1$, as confirmed by the values of the nonlinearity parameters.}

\textcolor{black}{Another issue concerns the sensitivity of the energy spectrum to the degree of imbalance. We considered the strong turbulence model with $\nu=1$, as it does not lead to unphysical divergences. When prescribing  $\eta=0$ and varing the ratio $E^+/E^-$, the energy spectrum always reduces to  power laws with exponents  $-5/3$ at MHD scales and $-7/3$ in the sub-ion range. 	More interesting is the case of a non-zero generalized flux of cross-helicity. Figure 7 (left) displays the energy spectra obtained for $\eta=0$ (red), $\eta=2\times10^{-3}$ (green), $\eta=5\times10^{-3}$ (blue), $\eta=8\times10^{-3}$ (black) and $\eta=9\times10^{-3}$ (violet), keeping all the other parameters fixed ($\varepsilon =1$, $k_d=200$, $k_f=0$, $u^+(k_d)=u^-(k_d)=0.6 (\varepsilon/k_d^7)^{1/2}$). The imbalance, measured by the ratio $E^+/E^-$ at $k_0=10^{-4}$ increases with $\eta$, taking the respective values  $0$, $9.55$, $425.56$, $36175$ and $1.91\times 10^5$. As $\eta$ increases, the energy spectrum becomes slightly steeper than $k_\perp^{-5/3}$ in the MHD range while, in the  sub-ion range, the $k^{-7/3}$ domain that establishes in  the balanced regime reduces, being replaced by a 
decay faster than a power law. Such a steepening of the sub-ion KAW spectrum was also observed in fully kinetic particle-in-cell simulations in the imbalanced regime \citep{Groselj18}. We also show in Fig. 7b the normalized cross-helicity spectrum $E_C(k_\perp)/E(k_\perp)$ for  $\eta=2\times10^{-3}$  and $\eta=9\times10^{-3}$, the other parameters being the same as in the left panel. This quantity is constant in the MHD range,  as observed in the solar wind \citep{Podesta10b}. However, while the decay of $E_C/E$ observed at small scales is due, as mentioned by the authors, to numerical noise in the data processing, in the present model, the $k_\perp^{-1}$ range obtained at sub-ion scales is a physical effect.}

\subsubsection{Effect of co-propagating waves interactions}\label{sec:coprop}
It is remarkable that for strong turbulence with $\nu<1$, the imbalance measured as the ratio $E^+/E^-$ increases with $k_\perp$ in the dispersion range. In order to check that this behavior is not a consequence of the absence of the co-propagating waves couplings, following \citet{Voitenko16}, we  heuristically modify Eqs. (\ref{eq:upm}) to account for self-interactions between wavenumbers which are comparable but not necessarily asymptotically close.
Inspection of Eq. (\ref{eq:ak}) suggests to define two  nonlinear characteristic frequencies for the  mode ${\boldsymbol k}$ propagating in the $\pm$ direction,  $\gamma_{\boldsymbol k}^{\pm (\uparrow\uparrow)}= V_{\boldsymbol k}^{(\uparrow\uparrow)} |a_{\boldsymbol k}^\pm|$ 
and $\gamma_{\boldsymbol k}^{\pm (\uparrow\downarrow))}= V_{\boldsymbol k}^{(\uparrow\downarrow)} |a_{\boldsymbol k}^\mp|$, \textcolor{black}{associated with the}
interactions between co-propagating $(\uparrow\uparrow)$ or counter-propagating $(\uparrow\downarrow)$ waves respectively. These frequencies  are phenomenologically estimated by noting that $|a^\pm_{\boldsymbol k}| = k_\perp\sqrt {u^\pm(k_\perp)}$ and using for $V_{\boldsymbol k}^{(\uparrow\uparrow)}$ or $V_{\boldsymbol k}^{(\uparrow\downarrow)}$ a semi-local approximation of the 
vertex $V_{{\boldsymbol k}{\boldsymbol p}{\boldsymbol q}}^{\sigma_k \sigma_p \sigma_q}$ given by Eq. (\ref{vertex}). This vertex includes a factor $s^{-1}\left (\sigma_p v_{ph}(p_\perp) -\sigma_q v_{ph}(q_\perp)\right)$ 
which vanishes when the interactions between co-propagating waves ($\sigma_p= \sigma_q)$ are strongly local. In contrast, for moderately local interactions, this factor is not zero and, since $v_{ph}/s \approx \sqrt{1+[1-1/(1+s^2)]k_\perp^2}$ for $k_\perp \ll 1$ and $v_{ph} \sim k_\perp$ for $k_\perp\gg 1$, it behaves  at large scales like a constant or like $k_\perp^2$ depending on whether contra or co-propagating waves are interacting, while at small scales it scales like $k_\perp$ in both cases. Furthermore, the factor $s(\sigma_k k_\perp^2/v_{ph}(k_\perp) +  \sigma_p p_\perp^2/v_{ph}(p_\perp) + \sigma_q q_\perp^2/v_{ph}(q_\perp))$ in Eq. (\ref{vertex}) is proportional to  $k_\perp^2 /v_{ph}(k_\perp)$ with an undetermined  numerical factor which,  in the case of strongly-local interactions, is equal to 3  or 1 for co or counter-propagating waves.  The global inverse transfer time writes
\begin{equation}
(\tau_{tr, G}^\pm)^{-1} = \frac{(\gamma_{\boldsymbol k}^{\pm (\uparrow\uparrow)})^2}{v_{ph} {\widetilde k}^\pm_\|} +\frac{(\gamma_{\boldsymbol k}^{\pm(\uparrow\downarrow)})^2}{v_{ph}{\widetilde k}^\mp_\|} .
\end{equation}
Noting that ${(V_{\boldsymbol k}^{(\uparrow\downarrow)})^2 |a^\mp_{\boldsymbol k}|^2}/{v_{ph}{\widetilde k}^\mp_\|}=(\tau_{tr}^\pm)^{-1}= (k^3_\perp v_{ph}{\widebar E}^\mp)/{\widetilde k}^\mp_\|$, and defining $\alpha(k_\perp)=V_{\boldsymbol k}^{(\uparrow\uparrow)} / V_{\boldsymbol k}^{(\uparrow\downarrow)}$, which  is an increasing function of $k_\perp$ that is essentially zero in the MHD range and saturates to a finite value  in the dispersive range, the new inverse transfer time rewrites
\begin{equation}
(\tau_{tr, G}^\pm)^{-1} \approx \frac{(k^4_\perp v_{ph}u^\mp)}{{\widetilde k}^\mp_\|}
\left(1+ \alpha^2(k_\perp)\frac{u^\pm(k_\perp)}{u^\mp(k_\perp)} 
\frac{{\widetilde k}^\mp_\|}{{\widetilde k}^\pm_\|}
\right).
\end{equation}
Equation (\ref{eq:upm}) is thus replaced by 
\begin{equation}
\frac{d}{d k_\perp} u^\pm(k_\perp)=-\frac{\varepsilon\pm\eta v_{ph}}{4C'k^7_\perp v_{ph}}\frac{1}
 {\left (\frac{u^\mp(k_\perp)}{{\widetilde k}^\mp_\|}  + \alpha^2(k_\perp) \frac{u^\pm(k_\perp)}{{\widetilde k}^\pm_\|} \right )}.\label{eq:coprop-interaction} \\
\end{equation}
The function $\alpha$ is taken as $\alpha(k)=3((1+ck^2)^{1/2}-1)/((1+ck^2)^{1/2}+1)$ in \citet{Voitenko16}, where $c$ is a slowly varying function of $k_\perp$ that we can here  assume to be constant \footnote{Assuming power law spectra, Eq. (\ref{eq:coprop-interaction}) reproduces Eq. (6) of \cite{Voitenko16} for $\varepsilon^-/\varepsilon^+$ where $\varepsilon^\pm = \varepsilon\pm\eta v_{ph}$.}. Within  Eq. (\ref{eq:coprop-interaction}), the energy spectra $E^\pm$ still separate as $k_\perp$ increases in the dispersive range, thus not qualitatively changing the predictions of the original model. This feature also holds in the case $\nu=1$ which, in the absence of counter-propagating waves, was displaying a satisfactory behavior. On the other hand, if one makes the (unjustified) assumption that $\varepsilon\pm\eta v_{ph}$ is independent of $k_\perp$, the results of \citet{Voitenko16} are  qualitatively recovered by integrating the resulting system with $\nu=0$. 

\section{Conclusion}  \label{conclusion}

A two-fluid model is used to study KAW turbulence from the MHD to the electron scales when neglecting the coupling to other waves. In the weak turbulence regime, kinetic equations are systematically derived in the case of zero electron inertia, from which a system of two nonlinear diffusion equations somewhat similar to  Leith's model for hydrodynamic turbulence is obtained, under the sole assumption of strong spectral locality. An interesting property of both the kinetic equations and of the diffusion model is that all the kinetic effects only appear through the Alfv\'en-wave phase velocity. By adjusting the  transfer time, a phenomenological extension of this diffusion  model to strong turbulence is also presented.  This model involves the frequencies and thus the longitudinal correlation lengths of  wave packets of opposite polarities. While for the most energetic wave this length is uniquely defined by the critical balance condition, for the weaker amplitude one, it is affected by its interactions with the stronger wave in a way that depends on the process by which turbulence is maintained in a stationary state. It turns out that in the MHD regime, when the system is driven by prescribing the $\varepsilon^\pm$ transfer rates, the small amplitude wave correlation rate is reduced by a moderate amount characterized by a free exponent $\nu$ that can be determined by fitting the scaling between $E^+/E^-$ and $\varepsilon^+/\varepsilon^-$ against numerical simulations. This model provides a satisfactory description of the energy spectra for collisional MHD. Differently, if the dissipation scale is smaller than the ion transition scale where dispersion starts acting, the above model displays unphysical features. This 
results from the conflict between a prescribed cross-helicity flux in the forward direction and the tendency of the cross-helicity which, at the dispersive scales, identifies with the magnetic helicity, to undergo an inverse cascade \citep{Galtier15,PST18}, \textcolor{black}{as argued on the basis of a \citet{Fjortoft53} argument in HRMHD  \citep{Schekochihin09} and observed in the numerical simulations of EMHD \citep{Cho16}.}
In contrast, if the energy spectra are prescribed at large scales, the transfer rates can adjust freely. This regime being less constrained, the interaction between the waves lead to comparable correlation length for both waves as in \citet{Chandran08}, corresponding to $\nu=1$. This system can permit a significant imbalance even for a ratio $\varepsilon^+/\varepsilon^-$ close to unity, with a ratio $E^+(k_\perp)/E^-(k_\perp)$ that starts to decrease at the ion scale and reaches $1$ at the pinning wavenumber.

In the framework of the diffusion model, the nonlinear eigenvalue problem for the energy and generalized cross-helicity fluxes when the spectra are prescribed at a wavenumber $k_0$ (corresponding to the outer scales) and assumed to decay to zero at infinity, has a unique solution in the strong MHD turbulence regime (for $\nu <1$) and a manifold of solutions characterized by a spectral entanglement condition either in weak turbulence or in strong turbulence when $\nu=1$. In the presence of dispersive effects, this problem is only well posed when $\eta=0$. The case where both dispersion and helicity flux are present leads to at least one of the $E^\pm$ spectra diverging at infinity, whatever the choice of the fluxes. Technically, the difficulty originates from the fact that the effective transfer rates of $E^+$ (respectively $E^-$) increases (respectively decreases) as $k_\perp$ increases in the dispersive range  (where these quantities are not conserved).
In this situation, we resorted to study the problem in a finite range of wavenumbers, prescribing the energy spectra at a large but finite wavenumber $k_d$. Depending on their amplitudes, the solution can either be singular or develop an absolute equilibrium at smaller scales.
Physically relevant solutions can nevertheless be obtained in the latter case at wavenumbers smaller than $k_d$.
Analogous to what happens with  the pinning effect in visco-diffusive MHD, the ratio of the generalized cross-helicity to the energy flux rates must be small enough, the more so if $k_d /k_0$ is large. In other words, if $k_d$ is viewed as the wavenumber limiting the inertial range at small scales, the present model appears to only permit a finite "Reynolds number" depending on the prescribed fluxes.  Even though the MHD case is much simpler, one of the striking consequence of the pinning effect is the sensitivity of the "inertial" exponents to the magnitude of the dissipation scale and thus to the Reynolds number, in contrast with the universal behavior of  usual hydrodynamic turbulence.

\textcolor{black}{Another issue concerns the sensitivity of the total energy spectrum to the degree of imbalance. As the generalized cross-helicity increases, the spectrum becomes steeper, decaying faster than a power law in the sub-ion range. Observational results on this issue are expected from the Parker Solar Probe space mission \citep{Fox16} which will permit exploring solar-wind turbulence at heliocentric distances inside $0.3$ au where the cross-helicity is expected to be larger than at larger distances \citep{Stansby19}.}

Several options are open for further development. It is first natural to regularize the unphysical small-scale behavior by dissipative processes, such as magnetic diffusivity, Landau damping or other kinetic processes as discussed in \cite{Cranmer03}. Looking for stationary solutions in the dissipative case is  possible by supplementing the model with equations for  scale-dependent transfer rates, possibly including integral expressions for the nonlinear times \citep{Clark09,PS15}.  Another development concerns the time dependent problem where the transfer rates are not prescribed but adjust dynamically. Such simulations were performed in the case of imbalanced MHD by \citet{GBuch10} and \citet{Chandran08} and for balanced turbulence with kinetic effects by \citet{Cranmer03}. Extending such simulations to the  dispersive unbalanced regime is outside the scope of the present paper and will be addressed separately. It will in particular permit the  study of the inverse helicity cascade in the presence of a small-scale forcing.

Other issues concern the model itself. Spectral localization eliminates the interaction of waves with the same polarization, but this  does not seem to strongly affect the qualitative behavior.  Furthermore, the transfer times are taken independent from the parallel wavenumber, thus imposing a constant flux in the transverse direction. If this is not the case, a closed equation cannot be written for the spectrum integrated over $k_\|$, thus permitting to the energy and/or the helicity to also flow in the longitudinal direction. This effect could be consistent with a reduction of the anisotropy at small scales, reported both in simulations of a reduced model \citep{Boldyrev12}  and with kinetic numerical simulations \citep{Cerri17}.
In any case, direct numerical simulations  of the two-field gyrofluid equations  could help clarify these issues and will be performed in a near future.

\appendix
\section{Limiting forms of the interaction vertex $V_{{\boldsymbol k}{\boldsymbol p}{\boldsymbol q}}^{\sigma \sigma_p \sigma_q} $} \label{app:vertex}
\subsection{RMHD limit}
In this case, $M_1= 0$, $M_2=k_\perp^2$, $M_3=1$, $D_e =k$, and thus $\Lambda=1$. One has
\begin{equation}
L_{{\boldsymbol k}{\boldsymbol p}{\boldsymbol q}}^{\sigma \sigma_p \sigma_q} = \frac{1}{4} \frac{1}{k_\perp p_\perp q_\perp}
\left ( q_\perp^2 - \sigma_p \sigma_q q_\perp^2 - \sigma \sigma_p k_\perp^2 \right)
\end{equation}
and thus
\begin{equation}
V_{{\boldsymbol k}{\boldsymbol p}{\boldsymbol q}}^{\sigma \sigma_p \sigma_q}= \frac{1}{8} \frac{{\widehat {\boldsymbol z}}\bcdot({\boldsymbol p}\times {\boldsymbol q})}{k_\perp p_\perp q_\perp} \left \{ (q_\perp^2 -p_\perp^2) (1 -\sigma_p \sigma_q)-\sigma k_\perp^2 (\sigma_p - \sigma_q) \right\},
\end{equation}
where one can also write
\begin{equation}
(q_\perp^2 -p_\perp^2) (1 -\sigma_p \sigma_q)-\sigma k_\perp^2 (\sigma_p - \sigma_q) = (\sigma_q-\sigma_p) (\sigma k_\perp^2 + \sigma_p p_\perp^2 + \sigma_q q_\perp^2).
\end{equation}
As a consequence, $V_{{\boldsymbol k}{\boldsymbol p}{\boldsymbol q}}^{\sigma \sigma_p \sigma_q} \neq 0$ requires $\sigma_p
= -\sigma_q$, expressing that at large scales, only counter-propagating waves can interact. In this case,
\begin{equation}
V_{{\boldsymbol k}{\boldsymbol p}{\boldsymbol q}}^{\sigma \sigma_p -\sigma_p}= \frac{1}{4} \frac{{\widehat {\boldsymbol z}}\bcdot({\boldsymbol p}\times {\boldsymbol q})}{k_\perp p_\perp q_\perp} (q_\perp^2 -p_\perp^2-\sigma_p\sigma k_\perp^2  ).
\end{equation}
The only non zero elements of the vertex are thus \citep{Galtier02,Nazarenko11, Tronko13}
\begin{eqnarray}
V_{{\boldsymbol k}{\boldsymbol p}{\boldsymbol q}}^{\sigma \sigma -\sigma}&=& \frac{1}{4} \frac{{\widehat {\boldsymbol z}}\bcdot({\boldsymbol p}\times {\boldsymbol q})}{k_\perp p_\perp q_\perp} (q_\perp^2 -p_\perp^2 -k_\perp^2)
=-\frac{1}{2} \frac{{\widehat {\boldsymbol z}}\bcdot({\boldsymbol p}\times {\boldsymbol q})({\boldsymbol p}_\perp \bcdot {\boldsymbol k}_\perp)}{k_\perp p_\perp q_\perp} \\
V_{{\boldsymbol k}{\boldsymbol p}{\boldsymbol q}}^{\sigma -\sigma \sigma}&=& \frac{1}{4} \frac{{\widehat {\boldsymbol z}}\bcdot({\boldsymbol p}\times {\boldsymbol q})}{k_\perp p_\perp q_\perp} (q_\perp^2 -p_\perp^2 +k_\perp^2)=\frac{1}{2} \frac{{\widehat {\boldsymbol z}}\bcdot({\boldsymbol p}\times {\boldsymbol q})({\boldsymbol q}_\perp \bcdot {\boldsymbol k}_\perp)}{k_\perp p_\perp q_\perp}. 
\end{eqnarray}

\subsection{Sub-ion range $\rho^{-1}_i \ll k_\perp \ll d_e^{-1}$}
When taking the large-wavenumber limit while neglecting electron inertia, the two-field model identifies with the ERMHD equations.  One finds that in this case the operators 
$M_1$ and $M_2$ reduce to constants $m_1$ and $m_2$. Furthermore $D_e= k_\perp$ and $\Lambda= \lambda k_\perp^{-1}$ where $\lambda=(1+m_2-m_1)^{1/2} m_2^{1/2}$. In this regime
\begin{equation}
L_{k_ \perp p_\perp q_\perp}^{\sigma \sigma_p \sigma_q}=\frac{\lambda}{4m_2} \left\{ 1 -\sigma_p\sigma_q \frac{q_\perp}{p_\perp} -\sigma \sigma_p  \frac{k_\perp}{p_\perp} \right\}.
\end{equation}
and thus,
\begin{equation}
L_{k_ \perp p_\perp q_\perp}^{\sigma \sigma_p \sigma_q} - 
L_{k_ \perp q_\perp p_\perp}^{\sigma \sigma_q \sigma_p} =\frac{\lambda}{4m_2}\frac{1}{p_\perp q_\perp}\left \{
\sigma_p\sigma_q (p_\perp^2 -q_\perp^2) - k_\perp \sigma(q_\perp \sigma_p - p_\perp \sigma_q)\right\}.
\end{equation}
Writing
\begin{eqnarray}
\sigma_p\sigma_q (q_\perp^2 -p_\perp^2) &=& \sigma_p\sigma_q (\sigma_q q_\perp - \sigma_p p_\perp) (\sigma_q q_\perp + \sigma_p p_\perp) \nonumber \\
&=& (\sigma_p q_\perp - \sigma_q p_\perp) (\sigma_p p_\perp + \sigma_q q_\perp),
\end{eqnarray}
one has
\begin{eqnarray}
L_{k_ \perp p_\perp q_\perp}^{\sigma \sigma_p \sigma_q} - 
L_{k_ \perp q_\perp p_\perp}^{\sigma \sigma_q \sigma_p} &=&-\frac{\lambda}{4m_2}\frac{1}{p_\perp q_\perp}
(\sigma_p q_\perp - \sigma_q p_\perp) (\sigma k_\perp + \sigma_p p_\perp + \sigma_q q_\perp)\nonumber \\
&=& -\frac{\lambda}{4m_2}\frac{\sigma_p\sigma_q}{p_\perp q_\perp}(\sigma_q q_\perp - \sigma_p p_\perp)
(\sigma k_\perp + \sigma_p p_\perp + \sigma_q q_\perp),
\end{eqnarray}
leading to 
\begin{equation}
V_{{\boldsymbol k}{\boldsymbol p}{\boldsymbol q}}^{\sigma \sigma_p \sigma_q}= \frac{1}{8} \sqrt{\frac{1+m_2-m_1}{m_2}} \sigma k_\perp  (\sigma_p p_\perp - \sigma_q q_\perp) 
\left\{\sigma \sigma_p\sigma_q \frac{\sin\alpha}{k_\perp} (\sigma k_\perp + \sigma_p p_\perp + \sigma_q q_\perp)\right\},
\end{equation}
consistent with \citet{Galtier03}.

\section{Leith's equation for hydrodynamic turbulence}\label{app1}

Let us consider the spectral equation phenomenologically introduced by \citet{Leith67} to describe the direct energy cascade in three-dimensional hydrodynamic turbulence (see also \citet{CN04})
\begin{equation}
\partial_t E = \frac{1}{8} \partial_k\left (k^{11/2} E^{1/2}  
\partial_k \left ( \frac {E}{k^2}\right ) \right).
\end{equation}
Stationnary solutions associated with a constant energy flux $\varepsilon$ are given by 
\begin{equation}
-\varepsilon = \frac{1}{8}  k^{11/2} E^{1/2}  
\partial_k \left ( \frac {E}{k^2}\right ). 
\end{equation}
Defining $X= \sqrt{E/k^2}$, one easily solves as
\begin{equation}
X^3 = \alpha +\frac{24}{11}\varepsilon k^{-11/2},
\end{equation}
where $\alpha$ is a constant, and gets the general solution
\begin{equation}
E= k^2(\alpha + \frac{24}{11} \varepsilon k ^{-11/2})^{2/3}.
\end{equation}
Prescribing that the spectrum vanishes at infinity and is given by $E(k_0) = E_0$ at a finite wavenumber $k_0$ (which physically corresponds to the injection wavenumber), one is led to choose
an energy flux given by 
\begin{equation}
\varepsilon = \varepsilon_0\equiv \frac{11}{24} E_0^{3/2} k_0^{5/2},
\end{equation}
which just expresses the usual phenomenological estimate $\varepsilon_0=V^3_0/L_0$ where $V_0$ refers to the typical velocity at the outer scale $L_0$.
If this condition is not satisfied and 
$\varepsilon$ is given by $\varepsilon= \varepsilon_0 + \zeta$,
then $\alpha = -(24/11) \zeta k_0^{-11/2}$ and the  solution rewrites
\begin{equation}
E= (\frac{24}{11})^{2/3} k^2 \left [-\zeta k_0^{-11/2} + \left ( \frac{11}{24} E_0^{3/2} k_0^{5/2} + \zeta \right) k^{-11/2}\right]^{2/3}.
\end{equation}
When $\zeta <0$ (i.e. the transfer rate too small relatively to the prescribed spectrum at $k_0$), the spectrum  scales like $k^2$ at large wavenumbers, corresponding to an absolute equilibrium regime, classically obtained in the absence of energy flux. Differently, when $\zeta>0 $ (thus the transfer rate relatively too strong), the spectrum vanishes  at a wavenumber $k_* =  k_0 \left( 1 + (11/24) E_0^{3/2}k_0^{5/2}/\zeta\right)^{2/11}$, corresponding to a singularity.

\section*{Acknowledgments}
E. Tassi is gratefully acknowledged for useful discussions.

\bibliographystyle{jpp}
\bibliography{biblio}

\begin{thebibliography}{87}
\expandafter\ifx\csname natexlab\endcsname\relax\def\natexlab#1{#1}\fi
\def\au#1{#1} \def\ed#1{#1} \def\yr#1{#1}\def\at#1{#1}\def\jt#1{\textit{#1}}
  \def\bt#1{#1}\def\bvol#1{\textbf{#1}} \def\vol#1{#1} \def\pg#1{#1}
  \def\publ#1{#1}\def\arxiv#1{#1}\def\org#1{#1}\def\st#1{\textit{#1}}

\bibitem[{Abdelhamid} {\em et~al.\/}(2016){Abdelhamid}, {Lingam} \&
  {Mahajan}]{Abdelhamid16}
{\sc \au{{Abdelhamid}, H.~M.}, \au{{Lingam}, M.} \& \au{{Mahajan}, S.~M.}}
  \yr{2016}  \at{Extended {MHD} turbulence and its applications to the solar
  wind}.  \jt{Astrophys. J.}  \bvol{829},  \pg{87}.

\bibitem[{Alexandrova} {\em et~al.\/}(2009){Alexandrova}, {Saur}, {Lacombe},
  {Mangeney}, {Mitchell}, {Schwartz} \& {Robert}]{Alexandrova09}
{\sc \au{{Alexandrova}, O.}, \au{{Saur}, J.}, \au{{Lacombe}, C.},
  \au{{Mangeney}, A.}, \au{{Mitchell}, J.}, \au{{Schwartz}, S.~J.} \&
  \au{{Robert}, P.}} \yr{2009}  \at{{Universality of solar-wind turbulent
  spectrum from {MHD} to electron scales}}.  \jt{Phys. Rev. Lett.}
  \bvol{103}~(16),  \pg{165003}.

\bibitem[{Belcher} \& {Davis}(1971)]{Belcher71}
{\sc \au{{Belcher}, J.~W.} \& \au{{Davis}, Jr., L.}} \yr{1971}
  \at{{Large-amplitude {A}lfv{\'e}n waves in the interplanetary medium, 2}}.
  \jt{J. Geophys. Res.}  \bvol{76},  \pg{3534}.

\bibitem[Benney \& Newell(1969)]{Benney69}
{\sc \au{Benney, D.J.} \& \au{Newell, A.C.}} \yr{1969}  \at{Random wave
  closures}.  \jt{Stud. Appl. Math.}  \bvol{48},  \pg{29--53}.

\bibitem[{Beresnyak}(2014)]{Beresnyak14}
{\sc \au{{Beresnyak}, A.}} \yr{2014}  \at{{Spectra of Strong
  Magnetohydrodynamic turbulence from high-resolution simulations}}.
  \jt{Astrophys. J. Lett.}  \bvol{784},  \pg{L20}.

\bibitem[Beresnyak \& Lazarian(2008)]{Beresnyak08}
{\sc \au{Beresnyak, A.} \& \au{Lazarian, A.}} \yr{2008}  \at{Strong imbalanced
  turbulence}.  \jt{Astrophys. J}  \bvol{682},  \pg{1070--1075}.

\bibitem[Beresnyak \& Lazarian(2009)]{Beresnyak09}
{\sc \au{Beresnyak, A.} \& \au{Lazarian, A.}} \yr{2009}  \at{Structure of
  stationary strong imbalanced turbulence}.  \jt{Astrophys. J}  \bvol{702},
  \pg{460--471}.

\bibitem[Beresnyak \& Lazarian(2010)]{Beresnyak10}
{\sc \au{Beresnyak, A.} \& \au{Lazarian, A.}} \yr{2010}  \at{Scaling laws and
  diffuse locality of balanced and imbalanced magnetohydrodynamic turbulence}.
  \jt{Astrophys. J. Lett.}  \bvol{722},  \pg{L110--L113}.

\bibitem[{Bian} \& {Tsiklauri}(2009)]{Bian09}
{\sc \au{{Bian}, N.~H.} \& \au{{Tsiklauri}, D.}} \yr{2009}  \at{{Compressible
  {H}all magnetohydrodynamics in a strong magnetic field}}.  \jt{Phys. Plasmas}
   \bvol{16}~(6),  \pg{064503}.

\bibitem[Biskamp {\em et~al.\/}(1999)Biskamp, Schwarz, Zeiler, Celani \&
  Drake]{Biskamp99}
{\sc \au{Biskamp, D.}, \au{Schwarz, E.}, \au{Zeiler, A.}, \au{Celani, A.} \&
  \au{Drake, J.~F.}} \yr{1999}  \at{Electron magnetohydrodynamic turbulence}.
  \jt{Phys. Plasmas}  \bvol{6},  \pg{751--758}.

\bibitem[Boldyrev {\em et~al.\/}(2013)Boldyrev, Horaites, Xia \& Perez]{BHXP13}
{\sc \au{Boldyrev, S.}, \au{Horaites, K.}, \au{Xia, Q.} \& \au{Perez, J.C.}}
  \yr{2013}  \at{Toward a theory of astrophysical plasma turbulence at
  subproton scales}.  \jt{Astrophys. J.}  \bvol{777},  \pg{41}.

\bibitem[{Boldyrev} \& {Perez}(2012)]{Boldyrev12}
{\sc \au{{Boldyrev}, S.} \& \au{{Perez}, J.~C.}} \yr{2012}  \at{{Spectrum of
  kinetic-Alfv{\'e}n turbulence}}.  \jt{Astrophys. J. Lett.}  \bvol{758},
  \pg{L44}.

\bibitem[Brizard(1992)]{Bri92}
{\sc \au{Brizard, A.}} \yr{1992}  \at{{Nonlinear gyrofluid description of
  turbulent magnetized plasmas}}.  \jt{Phys. Fluids B}  \bvol{4},
  \pg{1213--1228}.

\bibitem[Bruno \& Carbone(2013)]{Bruno13}
{\sc \au{Bruno, R.} \& \au{Carbone, V.}} \yr{2013}  \at{The solar wind as a
  turbulence laboratory}.  \jt{Living Rev. Solar Phys.}  \bvol{10},  \pg{2}.

\bibitem[Bruno \& Carbone(2016)]{BC16}
{\sc \au{Bruno, R.} \& \au{Carbone, V.}} \yr{2016} {\em Turbulence in the Solar
  Wind\/},  \st{Lectures Notes in Physics},  \vol{vol. 928}.  \publ{Springer}.

\bibitem[{Carbone} {\em et~al.\/}(2009){Carbone}, {Marino}, {Sorriso-Valvo},
  {Noullez} \& {Bruno}]{Carbone09}
{\sc \au{{Carbone}, V.}, \au{{Marino}, R.}, \au{{Sorriso-Valvo}, L.},
  \au{{Noullez}, A.} \& \au{{Bruno}, R.}} \yr{2009}  \at{{Scaling laws of
  turbulence and heating of fast solar wind: the role of density
  fluctuations}}.  \jt{Phys. Rev. Lett.}  \pg{p. 061102}.

\bibitem[Cerri {\em et~al.\/}(2017)Cerri, Servidio \& Califano]{Cerri17}
{\sc \au{Cerri, S.~S.}, \au{Servidio, S.} \& \au{Califano, F.}} \yr{2017}
  \at{Kinetic cascade in solar-wind turbulence: {3D3V} hybrid-kinetic
  simulations with electron inertia}.  \jt{Astrophys. J. Lett.}  \bvol{846},
  \pg{L18}.

\bibitem[Chandran(2008)]{Chandran08}
{\sc \au{Chandran, B.~D.~G.}} \yr{2008}  \at{Strong anisotropic {MHD}
  turbulence with cross helicity}.  \jt{Astrophys. J.}  \bvol{685},
  \pg{646--658}.

\bibitem[Chen(2016)]{Chen16}
{\sc \au{Chen, C.~H.~.K.}} \yr{2016}  \at{Recent progress in astrophysical
  plasma turbulence for solar wind observations}.  \jt{J. Plasma Phys.}
  \bvol{82},  \pg{535820602}.

\bibitem[Chen \& Boldyrev(2017)]{Chen-Bold17}
{\sc \au{Chen, C.~H.~K.} \& \au{Boldyrev, S.}} \yr{2017}  \at{Nature of kinetic
  scale turbulence in the {E}arth's magnetosheath}.  \jt{Astrophys. J.}
  \bvol{842},  \pg{122}.

\bibitem[Cho \& Kim(2016)]{Cho16}
{\sc \au{Cho, J.} \& \au{Kim, H.}} \yr{2016}  \at{Spectral evolution of helical
  electron magnetohydrodynamics turbulence}.  \jt{J. Geophys. Res. Space Phys.}
   \bvol{121},  \pg{6157--6167}.

\bibitem[{Clark} {\em et~al.\/}(2009){Clark}, {Rubinstein} \&
  {Weinstock}]{Clark09}
{\sc \au{{Clark}, T.~T.}, \au{{Rubinstein}, R.} \& \au{{Weinstock}, J.}}
  \yr{2009}  \at{{Reassessment of the classical turbulence closures: the Leith
  diffusion model}}.  \jt{J. Turbulence}  \bvol{10},  \pg{35}.

\bibitem[{Connaughton} \& {Nazarenko}(2004)]{CN04}
{\sc \au{{Connaughton}, C.} \& \au{{Nazarenko}, S.}} \yr{2004}  \at{{Warm
  cascades and anomalous scaling in a diffusion model of turbulence}}.
  \jt{Phys. Rev. Lett.}  \bvol{92}~(4),  \pg{044501}.

\bibitem[{Cranmer} \& {van Ballegooijen}(2003)]{Cranmer03}
{\sc \au{{Cranmer}, S.~R.} \& \au{{van Ballegooijen}, A.~A.}} \yr{2003}
  \at{{Alfv{\'e}nic turbulence in the extended solar corona: Kinetic effects
  and proton heating}}.  \jt{Astrophys. J.}  \bvol{594},  \pg{573--591}.

\bibitem[Dyachenko {\em et~al.\/}(1992)Dyachenko, Newell, A.Pushkarev \&
  Zakharov]{Dyachenko92}
{\sc \au{Dyachenko, S.}, \au{Newell, A.~C.}, \au{A.Pushkarev} \& \au{Zakharov,
  V.~E.}} \yr{1992}  \at{Optical turbulence: weak turbulence, condensates and
  collapsing filaments in the nonlinear {S}chr\"odinger equation}.  \jt{Physica
  D}  \bvol{57},  \pg{96--160}.

\bibitem[Fj{\o}rtoft(1953)]{Fjortoft53}
{\sc \au{Fj{\o}rtoft, R.}} \yr{1953}  \at{On changes in the spectral
  distribution of kinetic energy for two-dimensional nondivergent flow}.
  \jt{Tellus}  \bvol{5},  \pg{225}.

\bibitem[{Fox} {\em et~al.\/}(2016){Fox}, {Velli}, {Bale}, {Decker},
  {Driesman}, {Howard}, {Kasper}, {Kinnison}, {Kusterer}, {Lario}, {Lockwood},
  {McComas}, {Raouafi} \& {Szabo}]{Fox16}
{\sc \au{{Fox}, N.~J.}, \au{{Velli}, M.~C.}, \au{{Bale}, S.~D.}, \au{{Decker},
  R.}, \au{{Driesman}, A.}, \au{{Howard}, R.~A.}, \au{{Kasper}, J.~C.},
  \au{{Kinnison}, J.}, \au{{Kusterer}, M.}, \au{{Lario}, D.}, \au{{Lockwood},
  M.~K.}, \au{{McComas}, D.~J.}, \au{{Raouafi}, N.~E.} \& \au{{Szabo}, A.}}
  \yr{2016}  \at{{The Solar Probe Plus mission: humanity's first visit to our
  star}}.  \jt{Space Sci. Rev.}  \bvol{204},  \pg{7--48}.

\bibitem[Galtier(2006)]{Galtier06}
{\sc \au{Galtier, S.}} \yr{2006}  \at{Wave turbulence in incompressible {H}all
  magnetohydrodynamics}.  \jt{J. Plasma Phys.}  \bvol{72},  \pg{721--769}.

\bibitem[Galtier \& Bhattacharjee(2003)]{Galtier03}
{\sc \au{Galtier, S.} \& \au{Bhattacharjee, A.}} \yr{2003}  \at{Anisotropic
  weak whistler wave turbulence in electron magnetohydrodynamics}.  \jt{Phys.
  Plasmas}  \bvol{10},  \pg{3065--3075}.

\bibitem[Galtier \& Buchlin(2010)]{GBuch10}
{\sc \au{Galtier, S.} \& \au{Buchlin, E.}} \yr{2010}  \at{Nonlinear diffusion
  equations for anisotropic magnetohydrodynamic turbulence with cross
  helicity}.  \jt{Astrophys. J.}  \bvol{722},  \pg{1977--1983}.

\bibitem[Galtier \& Meyrand(2015)]{Galtier15}
{\sc \au{Galtier, S.} \& \au{Meyrand, R.}} \yr{2015}  \at{Entanglement of
  helicity and energy in kinetic {A}lfv{\'e}n wave/whistler turbulence}.
  \jt{J. Plasma Phys.}  \bvol{81},  \pg{325810106}.

\bibitem[Galtier {\em et~al.\/}(2002)Galtier, Nazarenko, Newell \&
  Pouquet]{Galtier02}
{\sc \au{Galtier, S.}, \au{Nazarenko, S.~V.}, \au{Newell, A.~C.} \&
  \au{Pouquet, A.}} \yr{2002}  \at{Anisotropic turbulence of shear {A}lfv\'en
  waves}.  \jt{Astrophys. J. Lett.}  \bvol{564},  \pg{L49--L52}.

\bibitem[{Goldreich} \& {Sridhar}(1997)]{GS97}
{\sc \au{{Goldreich}, P.} \& \au{{Sridhar}, S.}} \yr{1997}
  \at{{Magnetohydrodynamic turbulence revisited}}.  \jt{Astrophys. J.}
  \bvol{485},  \pg{680--688}.

\bibitem[Grappin {\em et~al.\/}(1983)Grappin, Pouquet \& L\'eorat]{Grappin83}
{\sc \au{Grappin, R.}, \au{Pouquet, A.} \& \au{L\'eorat, J.}} \yr{1983}
  \at{Dependence of {MHD} turbulence spectra on the velocity-field magnetic
  correlation}.  \jt{Astron. Astrophys.}  \bvol{126},  \pg{51--58}.

\bibitem[{Grasso} {\em et~al.\/}(1999){Grasso}, {Pegoraro}, {Porcelli} \&
  {Califano}]{Grasso99}
{\sc \au{{Grasso}, D.}, \au{{Pegoraro}, F.}, \au{{Porcelli}, F.} \&
  \au{{Califano}, F.}} \yr{1999}  \at{{Hamiltonian magnetic reconnection}}.
  \jt{Plasma Phys. Control. Fusion}  \bvol{41},  \pg{1497--1515}.

\bibitem[{Gro{\v s}elj} {\em et~al.\/}(2018){Gro{\v s}elj}, {Mallet},
  {Loureiro} \& {Jenko}]{Groselj18}
{\sc \au{{Gro{\v s}elj}, D.}, \au{{Mallet}, A.}, \au{{Loureiro}, N.~F.} \&
  \au{{Jenko}, F.}} \yr{2018}  \at{{Fully kinetic simulation of 3D kinetic
  Alfv{\'e}n turbulence}}.  \jt{Phys. Rev. Lett.}  \bvol{120}~(10),
  \pg{105101}.

\bibitem[Howes {\em et~al.\/}(2006)Howes, Cowley, Dorland, Hammett, E.Quataert
  \& Schekochihin]{HCD06}
{\sc \au{Howes, G.~G.}, \au{Cowley, S.~C.}, \au{Dorland, W.}, \au{Hammett,
  G.~W.}, \au{E.Quataert} \& \au{Schekochihin, A.~A.}} \yr{2006}
  \at{Astrophysical gyrokinetics: basic equations and linear theory}.
  \jt{Astrophys. J.}  \bvol{651},  \pg{590--614}.

\bibitem[{Howes} {\em et~al.\/}(2011){Howes}, {Tenbarge} \& {Dorland}]{Howes11}
{\sc \au{{Howes}, G.~G.}, \au{{Tenbarge}, J.~M.} \& \au{{Dorland}, W.}}
  \yr{2011}  \at{{A weakened cascade model for turbulence in astrophysical
  plasmas}}.  \jt{Phys. Plasmas}  \bvol{18}~(10),  \pg{102305--102305}.

\bibitem[Kim \& Cho(2015)]{KimCho15}
{\sc \au{Kim, H.} \& \au{Cho, J.}} \yr{2015}  \at{Inverse cascade in imbalanced
  electron magnetohydrodynamic turbulence}.  \jt{Astrophys. J.}  \bvol{801},
  \pg{75}.

\bibitem[{Kraichnan}(1971)]{Kraichnan71}
{\sc \au{{Kraichnan}, R.~H.}} \yr{1971}  \at{{Inertial-range transfer in two-
  and three-dimensional turbulence}}.  \jt{J. Fluid Mech.}  \bvol{47},
  \pg{525--535}.

\bibitem[Leith(1967)]{Leith67}
{\sc \au{Leith, C.~E.}} \yr{1967}  \at{Diffusion approximation to inertial
  energy transfer in isotropic turbulence}.  \jt{Phys. Fluids}  \bvol{10},
  \pg{1409--1416}.

\bibitem[Lithwick \& Goldreich(2003)]{Lithwick03}
{\sc \au{Lithwick, Y.} \& \au{Goldreich, P.}} \yr{2003}  \at{Imbalanced weak
  magnetohydrodynamic turbulence}.  \jt{Astrophys. J.}  \bvol{582},
  \pg{1220--1240}.

\bibitem[Lithwick {\em et~al.\/}(2007)Lithwick, Goldreich \&
  Sridhar]{Lithwick07}
{\sc \au{Lithwick, Y.}, \au{Goldreich, P.} \& \au{Sridhar, S.}} \yr{2007}
  \at{Imbalanced strong {MHD} turbulence}.  \jt{Astrophys. J.}  \bvol{655},
  \pg{269--274}.

\bibitem[Lucek \& Balogh(1998)]{Lucek98}
{\sc \au{Lucek, E.~A.} \& \au{Balogh, A}} \yr{1998}  \at{The identification and
  characterization of {A}lfv\'enic fluctuations in {Ulysses} data at
  midlatitudes}.  \jt{Astrophys. J.}  \bvol{507},  \pg{984--990}.

\bibitem[Lyutikov(2013)]{Lyutikov13}
{\sc \au{Lyutikov, M.}} \yr{2013}  \at{Electron magnetohydrodynamics:
  {D}ynamics and turbulence}.  \jt{Phys. Rev. E}  \bvol{88},  \pg{053103}.

\bibitem[{MacBride} {\em et~al.\/}(2008){MacBride}, {Smith} \&
  {Forman}]{MacBride08}
{\sc \au{{MacBride}, B.~T.}, \au{{Smith}, C.~W.} \& \au{{Forman}, M.~A.}}
  \yr{2008}  \at{{The turbulent cascade at 1 AU: energy transfer and the
  third-order scaling for {MHD}}}.  \jt{Astrophys. J.}  \bvol{679},
  \pg{1644--1660}.

\bibitem[{Mallet} {\em et~al.\/}(2017){Mallet}, {Schekochihin} \&
  {Chandran}]{Mallet17}
{\sc \au{{Mallet}, A.}, \au{{Schekochihin}, A.~A.} \& \au{{Chandran},
  B.~D.~G.}} \yr{2017}  \at{{Disruption of sheet-like structures in
  Alfv{\'e}nic turbulence by magnetic reconnection}}.  \jt{MNRAS}  \bvol{468},
  \pg{4862--4871}.

\bibitem[{Marino} {\em et~al.\/}(2009){Marino}, {Sorriso-Valvo}, {Carbone},
  {Noullez}, {Bruno} \& {Bavassano}]{Marino09}
{\sc \au{{Marino}, R.}, \au{{Sorriso-Valvo}, L.}, \au{{Carbone}, V.},
  \au{{Noullez}, A.}, \au{{Bruno}, R.} \& \au{{Bavassano}, B.}} \yr{2009}
  \at{{The energy cascade in solar wind {MHD} turbulence}}.  \jt{Earth Moon and
  Planets}  \bvol{104},  \pg{115--119}.

\bibitem[{Marsch} \& {Tu}(1990)]{Marsch-Tu90}
{\sc \au{{Marsch}, E.} \& \au{{Tu}, C.-Y.}} \yr{1990}  \at{{On the radial
  evolution of MHD turbulence in the inner heliosphere}}.  \jt{J. Geophys.
  Res.}  \bvol{95 (A6)},  \pg{8211--8229}.

\bibitem[Matthaeus {\em et~al.\/}(2009)Matthaeus, Oughton \& Zhou]{Matthaeus09}
{\sc \au{Matthaeus, W.~H.}, \au{Oughton, S.} \& \au{Zhou, Y.}} \yr{2009}
  \at{Anisotropic magnetohydrodynamic spectral transfer in the diffusion
  approximation}.  \jt{Phys. Rev. E}  \bvol{79},  \pg{035401(R)}.

\bibitem[Meyrand {\em et~al.\/}(2016)Meyrand, Galtier \& Kiyani]{Meyrand16}
{\sc \au{Meyrand, R.}, \au{Galtier, S.} \& \au{Kiyani, K.~H.}} \yr{2016}
  \at{Direct evidence of the transition from weak to strong magnetohydrodynamic
  turbulence}.  \jt{Phys. Rev. Lett.}  \bvol{116},  \pg{105002}.

\bibitem[Meyrand {\em et~al.\/}(2015)Meyrand, Kiyani \& Galtier]{Meyrand15}
{\sc \au{Meyrand, R.}, \au{Kiyani, K.~H.} \& \au{Galtier, S.}} \yr{2015}
  \at{Weak magnetohydrodynamic turbulence and intermittency}.  \jt{J. Fluid
  Mech.}  \bvol{770},  \pg{R1}.

\bibitem[Meyrand {\em et~al.\/}(2018)Meyrand, Kiyani, G\"urcan \&
  Galtier]{MeyrandX18}
{\sc \au{Meyrand, R.}, \au{Kiyani, K.~H.}, \au{G\"urcan, \"O.~D.} \&
  \au{Galtier, S.}} \yr{2018}  \at{Coexistence of weak and strong wave
  turbulence in incompressible {H}all magnetohydrodynamics}.  \jt{Phys. Rev. X}
   \bvol{8},  \pg{031066}.

\bibitem[Moya {\em et~al.\/}(2015)Moya, Pinto, Vi\~{n}as, Sibeck, Kurth,
  Hospodarsky \& Wygant]{Moya15}
{\sc \au{Moya, P.~S.}, \au{Pinto, V.~A.}, \au{Vi\~{n}as, A.~F.}, \au{Sibeck,
  D.~G.}, \au{Kurth, W.~S.}, \au{Hospodarsky, G.~B.} \& \au{Wygant, J.~R.}}
  \yr{2015}  \at{Weak kinetic {A}lfv\'en waves turbulence during the 14
  {N}ovember {2012} geomagnetic storm: Van {A}llen probes observations}.
  \jt{J. Geophys. Res.}  \bvol{120 (A7)},  \pg{5504--5523}.

\bibitem[Nazarenko(2011)]{Nazarenko11}
{\sc \au{Nazarenko, S.}} \yr{2011} {\em Wave turbulence\/},  \st{Lectures Notes
  in Physics},  \vol{vol. 825}.  \publ{Springer}.

\bibitem[Newell {\em et~al.\/}(2001)Newell, Nazarenko \& Biven]{Newell01}
{\sc \au{Newell, A.~C.}, \au{Nazarenko, S.} \& \au{Biven, L.}} \yr{2001}
  \at{Wave turbulence and intermittency}.  \jt{Physica D}  \bvol{152-153},
  \pg{520--550}.

\bibitem[{Ng} \& {Bhattacharjee}(1996)]{Ng96}
{\sc \au{{Ng}, C.~S.} \& \au{{Bhattacharjee}, A.}} \yr{1996}  \at{{Interaction
  of shear-{A}lfv\'en wave packets: implication for weak magnetohydrodynamic
  turbulence in astrophysical plasmas}}.  \jt{Astrophys. J.}  \bvol{465},
  \pg{845--854}.

\bibitem[Passot {\em et~al.\/}(2018)Passot, Sulem \& Tassi]{PST18}
{\sc \au{Passot, T.}, \au{Sulem, P.L.} \& \au{Tassi, E.}} \yr{2018}
  \at{Gyrofluid modeling and phenomenology of low-$\beta_e$ {A}lfv\'en wave
  turbulence}.  \jt{Phys. Plasmas}  \bvol{25},  \pg{042107}.

\bibitem[Passot \& Sulem(2015)]{PS15}
{\sc \au{Passot, T.} \& \au{Sulem, P.~L.}} \yr{2015}  \at{A model for the
  non-universal power law of the solar wind sub-ion-scale magnetic spectrum}.
  \jt{Astrophys. J. Lett.}  \bvol{812},  \pg{L37}.

\bibitem[Passot {\em et~al.\/}(2017)Passot, Sulem \& Tassi]{PST17}
{\sc \au{Passot, T.}, \au{Sulem, P.~L} \& \au{Tassi, E.}} \yr{2017}
  \at{Electron-scale reduced fluid models with gyroviscous effects}.  \jt{J.
  Plasma Phys.}  \bvol{83},  \pg{715830402}.

\bibitem[Perez \& Boldyrev(2009)]{Perez-Boldyrev09}
{\sc \au{Perez, J.~C.} \& \au{Boldyrev, S.}} \yr{2009}  \at{Role of
  cross-helicity in magnetohydrodynamic turbulence}.  \jt{Phys. Rev. Lett.}
  \bvol{102},  \pg{025003}.

\bibitem[Perez \& Chandran(2013)]{Perez13}
{\sc \au{Perez, J.~C.} \& \au{Chandran, B.~D.~G.}} \yr{2013}  \at{Direct
  numerical simulations of reflection-driven, reduced magnetohydrodynamic
  turbulence from the sun to the {A}lfv\'en critical point}.  \jt{Astrophys.
  J.}  \bvol{776}~(2),  \pg{124}.

\bibitem[Perez {\em et~al.\/}(2012)Perez, Mason, Boldyrev \& Cattaneo]{Perez12}
{\sc \au{Perez, J.~C.}, \au{Mason, J.}, \au{Boldyrev, S.} \& \au{Cattaneo, F.}}
  \yr{2012}  \at{On the energy spectrum of strong magnetohydrodynamic
  turbulence}.  \jt{Phys. Rev. X}  \bvol{2},  \pg{041005}.

\bibitem[Podesta(2013)]{Podesta13}
{\sc \au{Podesta, J.~J.}} \yr{2013}  \at{Evidence of kinetic {A}lfv\'en waves
  in the solar wind at 1 {AU}}.  \jt{Solar Phys.}  \bvol{286},  \pg{529--548}.

\bibitem[Podesta \& Bhattacharjee(2010)]{Podesta10}
{\sc \au{Podesta, J.~J.} \& \au{Bhattacharjee, A.}} \yr{2010}  \at{Theory of
  incompressible magnetohydrodynamic turbulence with scale-dependent alignment
  and cross-helicity}.  \jt{Astrophys. J.}  \bvol{718},  \pg{11511157}.

\bibitem[{Podesta} \& {Borovsky}(2010)]{Podesta10b}
{\sc \au{{Podesta}, J.~J.} \& \au{{Borovsky}, J.~E.}} \yr{2010}  \at{{Scale
  invariance of normalized cross-helicity throughout the inertial range of
  solar wind turbulence}}.  \jt{Phys. Plasmas}  \bvol{17}~(11),  \pg{112905}.

\bibitem[{Pouquet} {\em et~al.\/}(1976){Pouquet}, {Frisch} \&
  {Leorat}]{Pouquet76}
{\sc \au{{Pouquet}, A.}, \au{{Frisch}, U.} \& \au{{Leorat}, J.}} \yr{1976}
  \at{{Strong MHD helical turbulence and the nonlinear dynamo effect}}.  \jt{J.
  Fluid Mech.}  \bvol{77},  \pg{321--354}.

\bibitem[Roberts {\em et~al.\/}(1987)Roberts, Goldstein, Klein \&
  Matthaeus]{Roberts87}
{\sc \au{Roberts, D.~A.}, \au{Goldstein, M.~L.}, \au{Klein, L.~W.} \&
  \au{Matthaeus, W.~H.}} \yr{1987}  \at{Origin and evolution of fluctuations in
  the solar wind: Helios observations and {H}elios-{V}oyager comparisons}.
  \jt{J. Geophys. Res.: Space Physics}  \bvol{92}~(A11),  \pg{12023--12035}.

\bibitem[Roytershteyn {\em et~al.\/}(2018)Roytershteyn, Boldyrev, Delzanno,
  Chen, Grošelj \& Loureiro]{Royter18}
{\sc \au{Roytershteyn, V.}, \au{Boldyrev, S.}, \au{Delzanno, G.~L.}, \au{Chen,
  C.~H.~K.}, \au{Grošelj, D.} \& \au{Loureiro, N.~F.}} \yr{2018}
  \at{Numerical study of inertial kinetic-{A}lfv\'en turbulence}.
  \jt{arXiv:1810.12428v1 [physics.plasm-ph]} .

\bibitem[Sahraoui {\em et~al.\/}(2010)Sahraoui, Goldstein, Belmont, Canu \&
  Rezeau]{Sahraoui10}
{\sc \au{Sahraoui, F.}, \au{Goldstein, M.~L.}, \au{Belmont, G.}, \au{Canu, P.}
  \& \au{Rezeau, L.}} \yr{2010}  \at{Three dimensional anisotropic $k$ spectra
  of turbulence at subproton scales in the solar wind}.  \jt{Phys. Rev. Lett.}
  \bvol{105},  \pg{131101}.

\bibitem[Sahraoui {\em et~al.\/}(2009)Sahraoui, Goldstein, Robert \&
  Khotyaintsev]{SGRK09}
{\sc \au{Sahraoui, F.}, \au{Goldstein, M.~L.}, \au{Robert, P.} \&
  \au{Khotyaintsev, Y.~U.}} \yr{2009}  \at{Evidence of a cascade and
  dissipation of solar-wind turbulence at the electron gyroscale}.  \jt{Phys.
  Rev. Lett.}  \bvol{102},  \pg{231102}.

\bibitem[Salem {\em et~al.\/}(2012)Salem, Howes, Sundkvist, Bale, Chaston, Chen
  \& Mozer]{Salem12}
{\sc \au{Salem, C.~S.}, \au{Howes, G.~G.}, \au{Sundkvist, D.}, \au{Bale,
  S.~D.}, \au{Chaston, C.~C.}, \au{Chen, C. H.~K.} \& \au{Mozer, F.~S.}}
  \yr{2012}  \at{Identification of kinetic {A}lfv\'en wave turbulence in the
  solar wind}.  \jt{Astrophys. J. Lett.}  \bvol{745},  \pg{L9}.

\bibitem[{Schekochihin} {\em et~al.\/}(2009){Schekochihin}, {Cowley},
  {Dorland}, {Hammett}, {Howes}, {Quataert} \& {Tatsuno}]{Schekochihin09}
{\sc \au{{Schekochihin}, A.~A.}, \au{{Cowley}, S.~C.}, \au{{Dorland}, W.},
  \au{{Hammett}, G.~W.}, \au{{Howes}, G.~G.}, \au{{Quataert}, E.} \&
  \au{{Tatsuno}, T.}} \yr{2009}  \at{{Astrophysical gyrokinetics: kinetic and
  fluid turbulent cascades in magnetized weakly collisional plasmas}}.
  \jt{Astrophys. J. Suppl.}  \bvol{182},  \pg{310--377}.

\bibitem[Schekochihin {\em et~al.\/}(2012)Schekochihin, Nazarenko \&
  Yousef]{Scheko12}
{\sc \au{Schekochihin, A.~A.}, \au{Nazarenko, S.~V.} \& \au{Yousef, T.~A.}}
  \yr{2012}  \at{Weak {A}lfv\'en-wave turbulence revisited}.  \jt{Phys. Rev. E}
   \bvol{85},  \pg{036406}.

\bibitem[Schep {\em et~al.\/}(1994)Schep, Pegoraro \& Kuvshinov]{Schep94}
{\sc \au{Schep, T.~J.}, \au{Pegoraro, F.} \& \au{Kuvshinov, B.~N.}} \yr{1994}
  \at{Generalized two-fluid theory of nonlinear magnetic structures}.
  \jt{Phys. Plasmas}  \pg{pp. 2843--2852}.

\bibitem[{Stansby} {\em et~al.\/}(2019){Stansby}, {Horbury} \&
  {Matteini}]{Stansby19}
{\sc \au{{Stansby}, D.}, \au{{Horbury}, T.~S.} \& \au{{Matteini}, L.}}
  \yr{2019}  \at{{Diagnosing solar wind origins using in situ measurements in
  the inner heliosphere}}.  \jt{MNRAS}  \bvol{482},  \pg{1706--1714}.

\bibitem[Sulem {\em et~al.\/}(2016)Sulem, Passot, Laveder \&
  Borgogno]{Sulem2016}
{\sc \au{Sulem, P.~L.}, \au{Passot, T.}, \au{Laveder, D.} \& \au{Borgogno, D.}}
  \yr{2016}  \at{Influence of the nonlinearity parameter on the solar wind
  sub-ion magnetic energy spectrum: {FLR}-{L}andau fluid simulations}.
  \jt{Astrophys. J.}  \bvol{818},  \pg{66}.

\bibitem[Tassi(2017)]{Tassi17}
{\sc \au{Tassi, E.}} \yr{2017}  \at{Hamiltonian closures in fluid models for
  plasmas}.  \jt{Euro. Phys. J. D}  \bvol{71},  \pg{269}.

\bibitem[Tassi {\em et~al.\/}(2016)Tassi, Sulem \& Passot]{TSP16}
{\sc \au{Tassi, E.}, \au{Sulem, P.~L} \& \au{Passot, T.}} \yr{2016}
  \at{Reduced models accounting for parallel magnetic perturbations: gyrofluid
  and finite {L}armor radius{-L}andau fluid approaches}.  \jt{J. Plasma Phys.}
  \bvol{82},  \pg{705820601}.

\bibitem[{Told} {\em et~al.\/}(2015){Told}, {Jenko}, {TenBarge}, {Howes} \&
  {Hammett}]{Told15}
{\sc \au{{Told}, D.}, \au{{Jenko}, F.}, \au{{TenBarge}, J.~M.}, \au{{Howes},
  G.~G.} \& \au{{Hammett}, G.~W.}} \yr{2015}  \at{{Multiscale Nature of the
  Dissipation Range in Gyrokinetic Simulations of Alfv{\'e}nic Turbulence}}.
  \jt{Phys. Rev. Lett.}  \bvol{115}~(2),  \pg{025003}.

\bibitem[Tronko {\em et~al.\/}(2013)Tronko, Nazarenko \& Galtier]{Tronko13}
{\sc \au{Tronko, N.}, \au{Nazarenko, S.~V.} \& \au{Galtier, S.}} \yr{2013}
  \at{Weak turbulence in two-dimensional magnetohydrodynamics}.  \jt{Phys. Rev.
  E}  \bvol{87},  \pg{033103}.

\bibitem[Tu {\em et~al.\/}(1990)Tu, March \& Rausenbauer]{Tu90}
{\sc \au{Tu, C.~Y.}, \au{March, E.} \& \au{Rausenbauer, H.}} \yr{1990}  \at{The
  dependence of {MHD} turbulence spectra on the inner solar wind stream
  structure near solar minimum}.  \jt{Geophys. Res. Lett.}  \bvol{17},
  \pg{283--286}.

\bibitem[Tu {\em et~al.\/}(1989)Tu, Marsch \& Thieme]{Tu89}
{\sc \au{Tu, C.-Y.}, \au{Marsch, E.} \& \au{Thieme, K.~M.}} \yr{1989}
  \at{Basic properties of solar wind {MHD} turbulence near 0.3 {AU} analyzed by
  means of elsasser variables}.  \jt{J. Geophys. Res.}  \bvol{94 (A9)},
  \pg{11739--11759}.

\bibitem[{Voitenko} \& {De Keyser}(2016)]{Voitenko16}
{\sc \au{{Voitenko}, Y.} \& \au{{De Keyser}, J.}} \yr{2016}  \at{{MHD}-kinetic
  transition in imbalanced {A}lfv\'enic turbulence}.  \jt{Astrophys. J. Lett.}
  \bvol{832},  \pg{L20 (4pp)}.

\bibitem[Voitenko(1998)]{Voitenko98a}
{\sc \au{Voitenko, Y.~M.}} \yr{1998}  \at{Three-wave coupling and parametric
  decay of kinetic {A}lfv\'en waves}.  \jt{J. Plasma Phys.}  \bvol{60},
  \pg{497--514}.

\bibitem[Wicks {\em et~al.\/}(2013)Wicks, Roberts, Mallet, Schekochihin,
  Horbury \& Chen]{Wicks13}
{\sc \au{Wicks, R.~T.}, \au{Roberts, D.~A.}, \au{Mallet, A.}, \au{Schekochihin,
  A.~A.}, \au{Horbury, T.~S.} \& \au{Chen, C. H.~K.}} \yr{2013}
  \at{Correlations at large scales and the onset of turbulence in the fast
  solar wind}.  \jt{Astrophys. J.}  \bvol{778},  \pg{177}.

\bibitem[Zhou \& Matthaeus(1990)]{Zhou90}
{\sc \au{Zhou, Ye.} \& \au{Matthaeus, W.~H.}} \yr{1990}  \at{Models of inertial
  range spectra of interplanetary magnetohydrodynamic turbulence}.  \jt{J.
  Geophys. Res.}  \bvol{95(A9)},  \pg{14881--11892}.

\end{thebibliography}

\end{document}